Diplomarbeit

Charakterisierung eines Messplatzes
zur Vermessung von Brennpunkten
reflektiver Optiken
für die konzentrierende Photovoltaik

Vorgelegt von Manuel Frick

am 04.06.2013

Prof. Dr. Elizabeth von Hauff

Fakultät für Mathematik und Physik

Albert-Ludwigs Universität Freiburg

# Inhalt

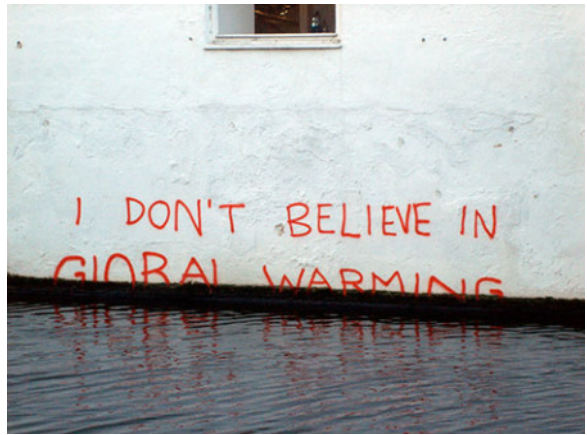
Bansky

# 1. Einleitung

Im Laufe der letzten Jahrzehnte fand die These eines anthropogenen Klimawandels in der wissenschaftlichen Gemeinde wachsenden Zuspruch [IPC95 p. 5] [IPC07 p. 72]. Gleichzeitig wurde jedoch mit den ernüchternden Ergebnissen diverser Klimagipfel klar, dass nicht allein auf politischen Willen gesetzt werden kann, um alternative Energiekonzepte umzusetzen. Während es mit dem aktuellen technologischen Stand mit Hilfe von Speichertechnologien schon jetzt realistisch erscheint unseren Energiebedarf vollständig aus erneuerbaren Energiequellen zu decken [Fra13], scheint der Wille zu fehlen, für dieses Ziel einen Anstieg der Strompreise zu akzeptieren und größere Investitionen vorzunehmen. Daher führt der Weg der angewandten Forschung im Bereich der solaren Energiesysteme unausweichlich über den Versuch niedrige Stromgestehungskosten[1] zu erreichen, um private Investoren und politische Entscheidungsträger zum Ausbau erneuerbarer Energiequellen zu bewegen.

Einen Ansatz um die Stromgestehungskosten für Solarstrom zu reduzieren stellt die konzentrierende Photovoltaik (engl. concentrating photovoltaics, CPV) dar. Grundidee der CPV ist es, teure Solarzellenfläche einzusparen, indem mit einer billigen Optik das Sonnenlicht auf die Solarzelle gebündelt wird. Als Optiken können dabei sowohl Linsen, wie auch Spiegel zum Einsatz kommen.

Um die Stromgestehungskosten niedrig zu halten, sind die marktreifen Produkte, aufgrund der Konzipierung und der Produktionsverfahren, oft weit entfernt vom technischen Optimum. Im Bereich der Optiken für die CPV ist es daher von großer Bedeutung zu analysieren, wie sich die Abweichungen vom Optimum auf die optischen Eigenschaften auswirken und abzuwägen, ob sie zugunsten geringerer Produktionskosten in Kauf genommen werden können. Um die Trennung von nicht-idealen Verhaltensweisen der Proben von Einflüssen der Messapparatur zu ermöglichen, ist eine sehr genaue Kenntnis der Messaufbauten die zu dieser Analyse beitragen nötig.

Ziel dieser Arbeit ist es, einen Messplatz zu charakterisieren, mit dem die Intensität im Brennpunkt reflektiver Optiken für die CPV ortsaufgelöst aufgenommen werden kann. Ein solcher Aufbau erlaubt es, die Auswirkungen der Abweichungen der Optiken von ihrem Optimum zu analysieren.

---

[1] Dieser Begriff bezeichnet die Kosten pro Energieeinheit, die bei der Erzeugung von elektrischem Strom auftreten.

Im Kapitel „Motivation" werden die Vor- und Nachteile von CPV gegenüber Silizium-Flatpanels vorgestellt. Weiter werden reflektive Optiken und Fresnellinsen, die aktuell die am meisten verbreitete Optik in der CPV darstellen, gegenübergestellt.

Im Kapitel „Vorstellung des Messplatzes" geht es um die Anforderungen an einen Messplatz zur Bestimmung der Konzentrationsmatrizen im Brennpunkt reflektiver Optiken für die CPV sowie um mehrere mögliche Ansätze für einen solchen Aufbau und um die genauere Ausgestaltung des bei uns aufgebauten Reflektormessplatzes (RMP).

Da für die CPV Verluste im Prozentbereich relevant sind, ist eine detaillierte systematische Untersuchung der Einzelkomponenten nötig. Im Kapitel „Charakterisierung der Einzelkomponenten" wird genauer auf die Auswirkungen auf das Messergebnis der einzelnen Komponenten des RMPs eingegangen. Dies geschieht auf experimentellem und wo nötig auf theoretischem Weg.

Im Kapitel „Charakterisierung des Gesamtaufbaus" wird die Zusammensetzung der Konzentrationsmatrix anhand der Ergebnisse des vorherigen Kapitels beschrieben. Weiter werden die Ergebnisse der Kalibrierung sowie, zur Illustrierung der späteren Funktionsweise, der Vermessung einer Probe vorgestellt.

Die Arbeit endet mit einer Zusammenfassung der durchgeführten Messungen und ihrer Ergebnisse sowie dem Ausblick über anstehenden Vermessungen und Weiterentwicklungen.

# 2. Motivation

## 2.1. Wieso konzentrierende Photovoltaik?

Das Konzept der konzentrierenden Photovoltaik (CPV) ist es, ein optisches System zu benutzen, um das Sonnenlicht auf eine kleine Fläche zu konzentrieren und somit Solarzellenfläche einzusparen. Durch diese Zellflächeneinsparung wird es finanziell möglich die teureren, aber auch effizienteren Mehrfachsolarzellen einzusetzen.

Bei Solarzellen mit einer einzigen Bandlücke, im Folgenden einfache Solarzellen genannt, ist der Teil des Sonnenspektrums verloren, dessen Photonenenergie geringer als die Bandlücke der Zelle ist. Jedoch ist es auch nicht unbedingt vorteilhaft eine niedrigere Bandlücke zu wählen, denn ist die Energie des Photons höher als die Bandlücke, geht der energetische Unterschied zwischen beiden aufgrund von Thermalisierung mit der Umgebung verloren. Prinzip der Mehrfachzellen ist es, mehrere einfache Zellen mit unterschiedlichen Bandlücken zu „stapeln". Dabei trifft das Licht zunächst auf die Zelle mit der größten Bandlücke, wodurch weniger überschüssige Energie verloren geht. Die Photonen, deren Energie nicht reicht, können durch eine der darunter liegenden Zellen mit einer niedrigeren Bandlücke absorbiert werden, solange ihre Energie mindestens so hoch ist wie die Bandlücke der letzten Zelle.

Dadurch erreichen Mehrfachsolarzellen Wirkungsgradrekorde von bis zu 44%[2] [NREL], gegen 27,6%[3] [Gre12] für Siliziumeinfachzellen unter Konzentration, bzw. 25,0% ohne Konzentration [Gre12]. Beeindruckender ist jedoch die Differenz, wenn statt auf die Rekordzellen auf die kommerziell üblichen Wirkungsgrade geschaut wird. Hier kommen Mehrfachsolarzellen auf 36%[4] [AZU12] bis über 40%[5] [Sol13] gegenüber 5-7% (amorphes Silizium) und 14-17% (monokristallines Silizium) bei Siliziumzellen [sol13]. Auch die theoretische Höchstgrenze für den Wirkungsgrad der Mehrfachsolarzellen liegt mit 86,8% für unendliche viele gestapelte Zellen bzw. 63,9% für drei gestapelte Zellen weit über der Höchstgrenze von ca. 31% bei Siliziumeinfachzellen [McE p. 73]. Die Merhfachsolarzellen bieten aus dieser Sicht also ein größeres Optimierungspotential. So wurde der Rekordwirkungsgrad innerhalb der letzten 10 Jahre bei Mehrfachsolarzellen um 10,5% erhöht, während er bei Silizium nur um 0,8% anstieg [Gre12] [Gre03].

Trotz dieser Zahlen sind die Stromgestehungskosten bei Panelen aus Mehrfachsolarzellen ohne konzentrierende Optik aufgrund ihrer höheren Produktionskosten höher als bei Siliziumflachpanelen. Doch in Kombination mit einer konzentrierenden Optik und einem Tracker, der die Zelle im Laufe des Tages der Sonne nachführt, werden Mehrfachzellen konkurrenzfähig.

---

[2] Dieser Weltrekord wurde im November 2012 bei einer 947-fachen Konzentration durch eine III-V-Zelle von Solar Junction aufgestellt. Dies wurde durch das National Renewable Energy Laboratory verifiziert.
[3] Dieser Weltrekord wurde im November 2004 bei einer 92-fachen Konzentration durch eine monokristalline Si-Zelle von Amonix aufgestellt. Dies wurde durch das National Renewable Energy Laboratory verifiziert.
[4] 3C40C von AzurSpace. Zwischen 36,3% bei 1000-facher Konzentration in Version Glass 10x10mm und 39,0% bei 500-facher Konzentration in Version Air 5x5mm.
[5] Herstellerangaben zur SJMJ-3 von SolarJunction. "The SJMJ-3 is a multi-junction solar cell available in varying sizes and configurations with mean production efficiency exceeding 40% at 1000 suns concentration and Standard Test Conditions (STC)." [Sol13]

Es sollen im Folgenden einige Vor- und Nachteile der CPV gegenüber Siliziumflachpanelen aufgeführt werden, ohne den Anspruch zu erheben über eine Überlegenheit eines der beiden Systeme zu urteilen, die ohnehin stark vom Standort des Solarkraftwerkes abhängt.

Die höhere Komplexität von CPV-Systemen, die die Konkurrenzfähigkeit erst erlaubt, wird ihnen jedoch in vielen Hinsichten auch zum Verhängnis. So besteht der Gesamtwirkungsgrad des Systems nicht nur aus dem der Solarzelle, sondern wird z.B. durch Verluste durch ungenaues Fokussieren oder Tracking und durch einen geringeren nutzbaren Anteil des einfallenden Lichtes weiter verringert. Aktuell liegt der Wirkungsgradrekord für ein gesamtes Modul daher mit 34,9% [AFP13] deutlich unter dem der alleinigen Zelle.

Bei der Optik kommt es zu Verlusten durch Abbildungsfehler, thermische Effekte, Abbildungen neben die Zelle aufgrund zu großer Einfallswinkel, Reflexionen und Verschattungen. Diese Effekte werden jedoch im nächsten Abschnitt ausführlicher erläutert und daher hier nicht weiter ausgeführt.

Es ist entscheidend, dass ein möglichst großer Anteil des einfallenden Lichtes als Spot auf die Zelle fällt. Jedoch ist diese Fokussierung abhängig von der Einfallsrichtung des Lichts und der Anteil der Direktstrahlung[6] wandert somit im Laufe des Tages und des Jahres. Da dies der größte Teil der einfallenden Strahlungsenergie ist, kann CPV nur in Kombination mit einer Nachführung der Optik, die den Einfallswinkel der Sonnenstrahlung auf die Optik konstant hält, funktionieren. Hier ist ebenfalls die Genauigkeit der Nachführung für den Wirkungsgrad des Moduls von Bedeutung. Der Tracker ist daher ein nicht vernachlässigbarer Kostenfaktor bei der Herstellung, Montage und Wartung des Systems. Allerdings führt die Nachführung ebenfalls dazu, dass die erzeugte Leistung im Laufe des Tages schneller auf ihr Maximum ansteigt und dort länger bleibt, wodurch an wolkenfreien Tagen weniger Schwankungen im Stromnetz ausgeglichen werden müssen.

Doch auch bei guter Nachführung erreicht ein größerer Anteil des auf das Modul treffenden Lichtes nie die Zelle, denn im Gegensatz zu Flachpanelen können CPV-Module kein Streulicht umwandeln, da dieses nicht auf die Zelle fokussiert wird. CPV ist also nicht in denselben geographischen Zonen attraktiv wie Flachpanele. Während bei Flachpanelen die Globalstrahlung (GHI)[7] die entscheidende Rolle spielt, ist für die CPV nur die Direktnormalstrahlung von Bedeutung. Zusätzlich zum Abstand zur Sonne spielt also die Bewölkung und der Lichtweg durch die Atmosphäre eine Rolle (siehe Abbildung 1).

Ist die Oberfläche der Optik verschmutzt, so sorgen Schmutzpartikel, neben Verschattung, dafür, dass ein Teil der Direktstrahlung gestreut wird und somit ebenfalls für die CPV verloren geht. Dies bedeutet, dass CPV-Module zur Bewahrung ihrer Effizienz einer häufigeren Reinigung bedürfen als Flachpanele, was einerseits zusätzlichen Wartungskosten entspricht, andererseits je nach Reinigungsverfahren und Standort auch aufgrund des Wasserverbrauchs kritisch ist.

---

[6] Def.: „the amount of solar radiation from the direction of the sun." [NRE13]
[7] $GHI = DNI \cdot \cos(Z) + DHI$, wobei DHI die Diffusstrahlung, DNI die Direktnormalstrahlung und Z der Winkel zwischen Sonne und Zenith sind. [NRE13]

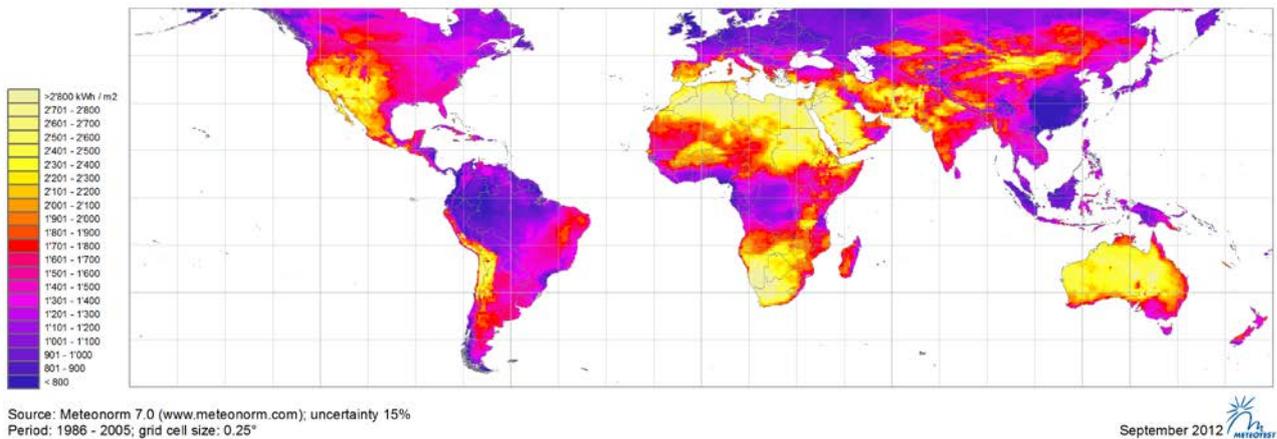

**Abbildung 1: Direktnormalstrahlung in der Welt (© METEOTEST; basierend auf www.meteonorm.com)**

Dafür scheint, dank kleiner Zellfläche, bei derzeitiger Verfügbarkeit der Rohstoffe, die Energetic Payback Time, die benötigte Zeit bis das System seine Produktionsenergie erzeugt hat, bei CPV Systemen geringer zu sein als bei Flachpanelen[8] [ISE12] [McE p. 1106]. Bezüglich der in Anspruch genommenen Fläche verhält sich die CPV neutral: Zwar ist bei gegebener elektrischer Leistung die Fläche der benötigten Module bei einem CPV-Kraftwerk aufgrund der höheren Wirkungsgrade geringer, doch benötigen die Module Freifläche zur Rotation, sodass letztendlich kaum oder kein Flächengewinn resultiert. Die Hersteller von CPV-Modulen argumentieren jedoch, dass diese Freifläche weiterhin für die Agrarwirtschaft benutzt werden kann, da dort Verschattung nicht dauerhaft ist.

In vielen Regionen der Welt ist also schon jetzt CPV aus Kosten- und Umweltperspektive günstiger als geläufige Silizium-Flachpanele. Es ist daher von Bedeutung CPV-Module weiter zu entwickeln, sodass für jeden Standort das jeweils optimale System hochentwickelt zur Verfügung steht.

## 2.2. Wieso reflektive Optiken?

Unter den CPV-Systemen sind Systeme mit Fresnellinsen aus PMMA oder Silikon auf Glas der Standard. Seit 8 Jahren ist das Fraunhofer ISE in der Lage den Brennpunkt solcher Linsen zu vermessen. Doch gibt es neben Fresnellinsen auch Systeme mit reflektiven Optiken.

Ausschlaggebend für die produzierte Strommenge pro Zelle und die Stromgestehungskosten ist, neben den oben genannten Faktoren, der Konzentrationsfaktor der Optik. Ist der Konzentrationsfaktor bei konstanter Apertur der Optik niedriger, muss eine größere Solarzelle verwendet werden, was die Kosten erhöht, oder es geht ein Teil des Lichtes verloren. Genau dieser Konzentrationsfaktor leidet jedoch bei Fresnellinsen an Effekten, die bei üblichen reflektiven Optiken (Parabolspiegel, Cassegrainspiegel) geringer ausfallen oder gar nicht vorkommen.

So treffen beispielsweise aufgrund von Totalreflexion an den steilen Flanken der Fresnellinse, den sogenannten Störflanken, Lichtstrahlen neben die Solarzelle.

---

[8] In Südeuropa weniger als ein Jahr, gegen ca. 1 Jahr bei Si-Flachpanelen. [ISE12]

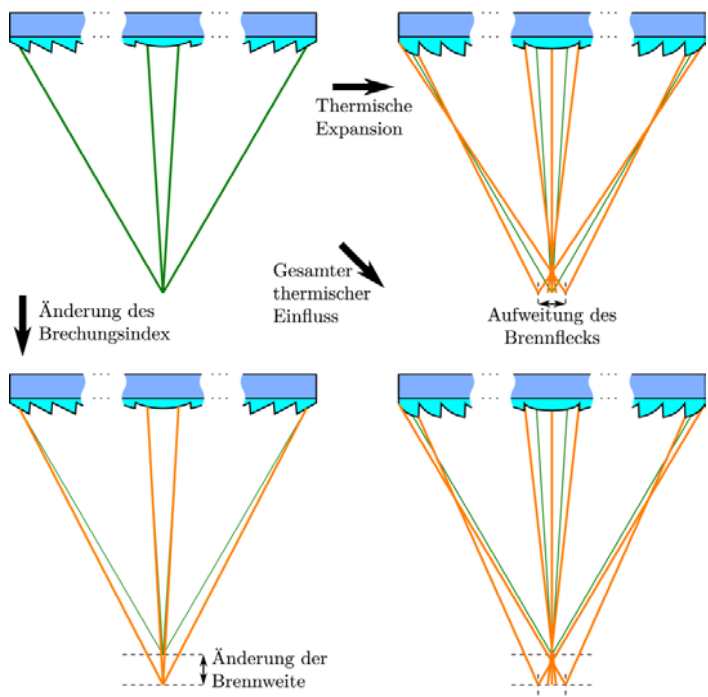

Abbildung 2: Thermische Verformung einer Silikon auf Glas Fresnellinse (aus [Hor13], Abb. 8.1)

Weiter führt die chromatische Aberration, aufgrund der Dispersion des Brechungsindexes der Linse, zu einer „Ausschmierung" des resultierenden Brennpunktes der Linse. Denn wenn die Solarzelle in die Brennebene einer bestimmten Wellenlänge gesetzt wird, sitzt sie bezüglich aller anderen Wellenlängen off-focus.

Dieselbe Auswirkung hat auch die thermische Ausdehnung der Fresnellinsen. So wird bei PMMA-Fresnellinsen durch die thermische Ausdehnung senkrecht zur optischen Achse die Brennweite vergrößert. Bei Silikon auf Glas ist eine Ausdehnung der gesamten Linse senkrecht zur optischen Achse aufgrund der geringen Ausdehnung von Glas nicht signifikant, doch führt die Ausdehnung des Silikons zu einer Wölbung der einzelnen Wirkflanken, was Winkelfehler je nach Austrittsort aus der Flanke verursacht [alH10] (siehe Abbildung 2). Bei beiden Versionen (PMMA und Silikon auf Glas) führt darüber hinaus die Änderung des Brechungsindexes aufgrund der Dichteschwankung zu einer Änderung der Brennweite. [Hor13]

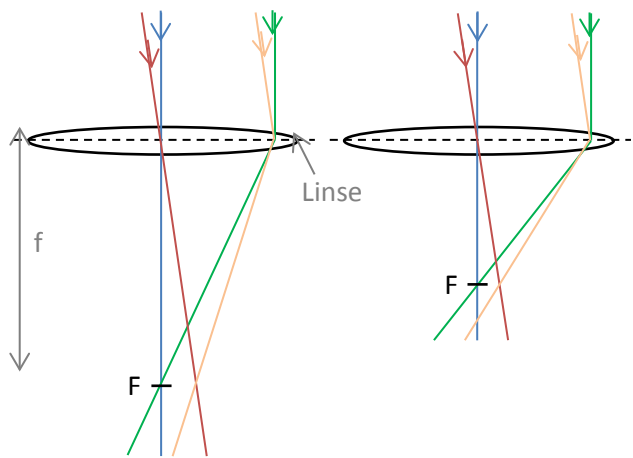

Abbildung 3: Abhängigkeit des Durchmessers des Brennpunkts von der Brennweite

Letztendlich haben reflektive Optiken typischerweise kürzere Brennweiten als Fresnellinsen[9], wodurch, wie aus Abbildung 3 klar wird, ein geringerer Brennpunktdurchmesser aufgrund der Sonnendivergenz zustande kommt[10]. Unter der vereinfachenden Annahme der abbildenden und der Gauß'schen Optik beträgt dieser $f \cdot \tan \alpha$ wobei $f$ die Brennweite und $\alpha$ der Divergenzwinkel sind.

Wir können also anhand von reflektiven Optiken höhere Konzentrationen erreichen. Doch haben solche Optiken nicht nur Vorteile gegenüber Fresnellinsen.

---

[9] Bei Fresnellinsen werden bei kurzen Brennweiten die Wirkflanken steil, woraus hohe Fresnelreflexionen resultieren.
[10] Die Brennpunktdurchmesser, die aus der Sonnendivergenz resultieren, sind viel größer als die durch die Fourieroptik vorhergesagten unteren Grenzen, sodass die aus letzterer folgende Verbreiterung des Brennpunkts bei längerer Brennweite aufgrund eines kleineren Einfallswinkelspektrums keine Rolle spielt.

So ist durch die geringere Brennweite ebenfalls die Tiefenschärfe geringer (siehe Abbildung 4). Betrachten wir den Radius r(z) des Lichtkegels, so gilt

$$r(z) = f \cdot \tan(\alpha) + \left( \frac{a/2 \pm f \cdot \tan(\alpha)}{f} \cdot |z - f| \right) \quad (1)$$

wobei „+" bei $z > f$ (rote Strahlen) und „-" bei $z < f$ (orange Strahlen) gilt. Eine gegebene Defokussierung um das Streckenelement $|z - F|$ wirkt sich demnach unter Annahme der paraxialen Optik bei einer kurzbrennweitigen Optik stärker auf den Durchmesser des Spots auf der Zelle aus.

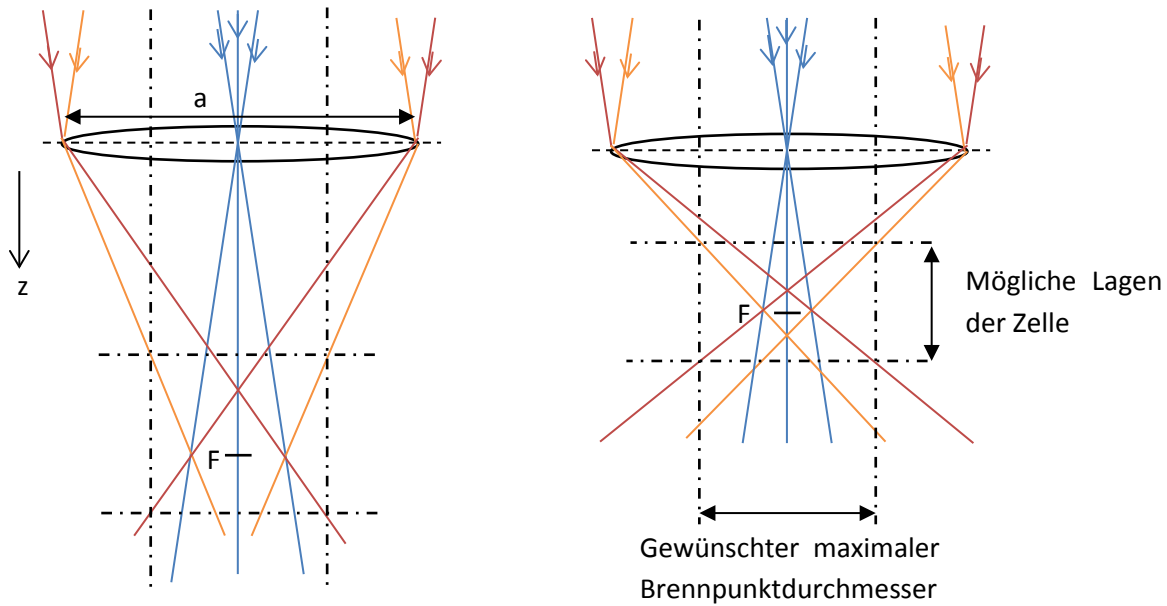

Abbildung 4: Geringere Tiefenschärfe bei Optiken mit kurzer Brennweite (rechts) bei Bestrahlung mit divergentem Licht.

Wichtiger sind jedoch die Problematiken der Kühlung der Solarzelle sowie die nötige hohe Formtreue der Spiegel.

Während bei CPV-Systemen mit Fresnellinsen (oder mit komplexeren reflektiven Systemen wie einem Cassegrainsystem) problemlos an der Rückseite der Solarzelle ein thermischer Leiter zur passiven Kühlung angebracht werden kann, würde dieser bei reflektiven Optiken, deren Brennpunkt sich auf der Seite der Optik befindet, von der auch das Licht einfällt, Verschattung verursachen. Hier sind also komplexere Kühlungsmethoden, bzw. eine Beschränkung der Konzentration nötig.

Letztendlich wirken sich Oberflächenfehler der Optiken (z.B. aus Abformungsfehlern, thermischen Verzerrungen) bei reflektiven Optiken stärker aus als bei refraktiven Optiken. Ausführlichere Rechnungen zum Thema wurden in [Hor13], Kapitel 4 durchgeführt. Wir wollen hier nur grob auf das Thema eingehen.

Besitzt ein Spiegel lokal eine um den Winkel α zur Sollgeometrie verkippte Oberfläche, so verändert sich der Winkel zwischen einem einfallenden Strahl und der Oberflächennormale um ebendiesen Winkel. Damit ändert sich der Ausfallswinkel um 2α: Einmal aufgrund des geänderten Einfallswinkels und einmal aufgrund der verdrehten Oberflächennormale (siehe Abbildung 5 oben).

Bei refraktiven Optiken wird das Licht an Grenzflächen nach dem Snelliusschen Gesetz gebrochen. Sei $\beta_i$ der Einfallswinkel, $\beta_o$ der Austrittswinkel, $\alpha$ ein kleiner Fehler auf die Neigung der Grenzfläche und $\alpha'$ die aus dem Fehler der Grenzfläche resultierende Verkippung des Austrittsstrahls. Es gilt dann

$$n_i \cdot \sin(\beta_i + \alpha) = n_o \sin(\beta_o + \alpha') \tag{2}$$

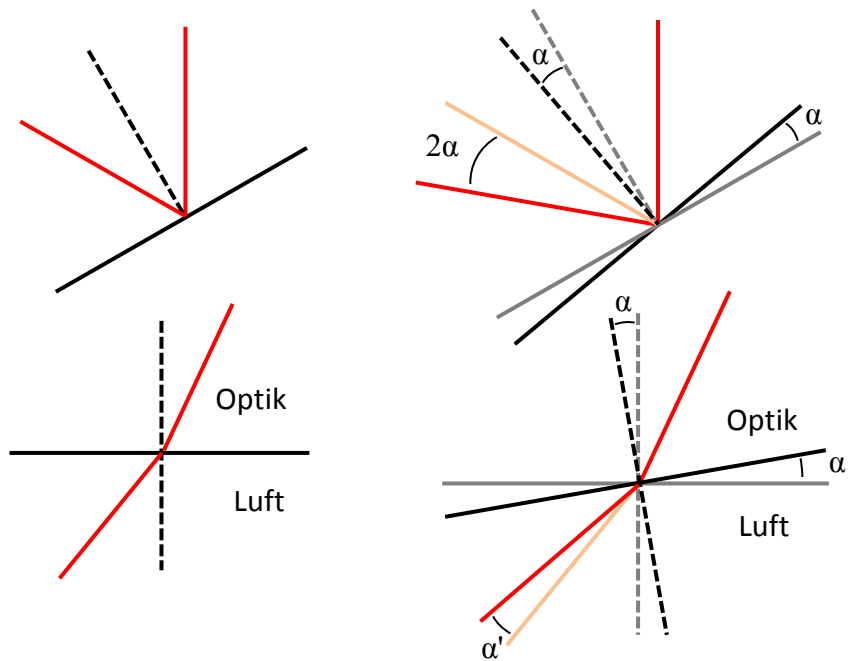

**Abbildung 5: Auswirkung von Oberflächenfehlern bei reflektiven und refraktiven Optiken**

Unter der Annahme, dass sowohl $\alpha$ wie auch $\alpha'$ sehr klein sind, können in der obigen Formel beide Sinusterme taylorentwickelt und mit Snellius-Gesetz substituiert werden. Daraus folgt

$$n_i \cdot \alpha \cdot \cos\beta_i = n_o \cdot \alpha' \cdot \cos\beta_o \tag{3}$$

$$\Leftrightarrow \frac{\alpha'}{\alpha} = \frac{n_i/n_o \cdot \cos\beta_i}{\cos\left(\sin^{-1}(n_i/n_o \cdot \sin\beta_i)\right)} \tag{4}$$

Betrachten wir bei Fresnellinsen den Austritt aus dem Silikon, also die Stelle, an der einerseits die Oberflächenfehler am höchsten sind, andererseits sich die Fehler am stärksten auswirken, so gilt[11] $n_i \approx 1{,}4$ und $n_o \approx 1$ bei 589nm. Für diese Brechungsindizes ist $\alpha'$ bis zu Einfallswinkeln von 0.63 rad kleiner als $2\alpha$ (siehe Abbildung 6). Solche hohen Winkel werden bei Fresnellinsen typischerweise nur für die Wirkflanken am äußersten Rand der Linse erreicht.

---

[11] Silikone können sehr unterschiedliche Brechungsindizes haben. Für Fresnellinsen wird jedoch oft Silikon mit einem Brechungsindex nahe dem von Glas benutzt. Die benutzten Werte stammen aus [Hor13], S.164.

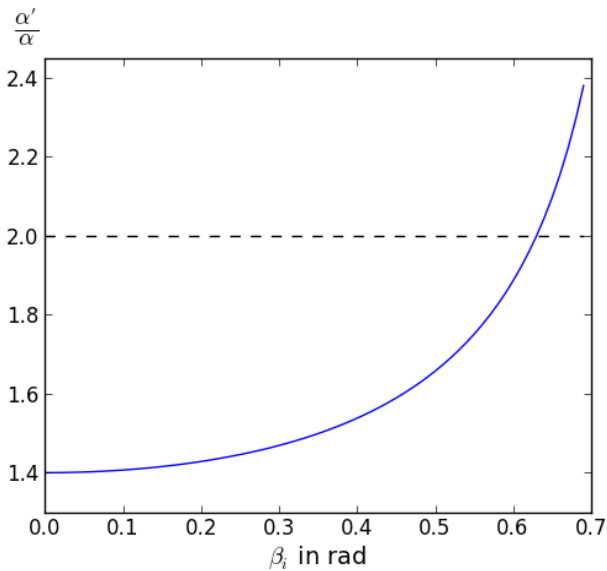

**Abbildung 6: Fortpflanzung eines Oberflächenfehlers auf den Ausgangsstrahl in Abhängigkeit des Einfallwinkels bei einer SOG-Fresnellinse, bei $n_{Silikon} = 1,4$**

Ähnlich verhält es sich, wenn die Oberfläche lokal versetzt ist, wobei also dieselbe Neigung beibehalten, die Fläche jedoch verschoben wird. Anhand von Abbildung 7 sieht man, dass ein solcher Versatz $d$ der Oberfläche bei reflektiven Optiken zu einem Versatz $2d \cdot \sin \beta_i$ des reflektierten Strahls entlang der Oberfläche führt, während derselbe Versatz der Oberfläche bei einer refraktiven Optik einen Versatz $x = d \cdot (\sin \beta_o - \tan \beta_i \cdot \cos \beta_o)$ des Austrittstrahls verursacht. Für den obigen Brechungsindex von 1,4 für die Optik und durch Einsetzen des Snellius'schen Brechungsgesetzes für $\sin \beta_o$ sieht man, dass der Versatz bei refraktiven Optiken geringer ist.

Wir sehen also, dass sich in den meisten Fällen ein Fehler der Oberfläche der Optik bei reflektiven Optiken stärker auf den Ausgangsstrahl auswirkt als bei refraktiven Optiken und somit ein kleiner Fehler der Optik zu einer größeren Verschmierung des Brennpunktes führt. Gelingt es jedoch die Oberflächenfehler bei vernünftigen Produktionskosten der Spiegel klein zu halten, so können dank kürzerer Brennweite, geringeren thermischen Effekten und Abwesenheit von chromatischen Aberrationen höhere Konzentrationen und somit letztendlich geringere Stromgestehungskosten erreicht werden.

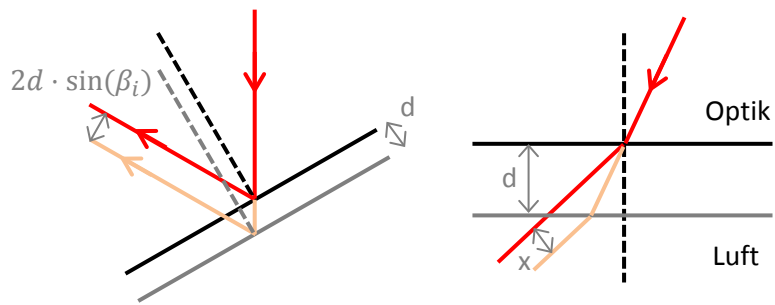

**Abbildung 7: Auswirkung eines Oberflächenversatzes bei reflektiven und refraktiven Optiken**

# 3. Vorstellung des Messplatzes

## 3.1. Anforderungen an den Messplatz

Um die in der CPV zum Einsatz kommenden Optiken zu verbessern, müssen die im vorherigen Kapitel genannten Fehler genau untersucht werden. So können Modelle für diese Effekte aufgestellt und überprüft werden, die es ermöglichen abzuwägen, ob der aus einer Korrektur der Fehler resultierende Gewinn im Wirkungsgrad die eventuellen Mehrkosten bei der Herstellung rechtfertigt.

Wir haben gesehen, dass bei reflektiven Optiken die Abweichungen der Sollgeometrie entscheidend sind. Weiter ist der Einfluss der Rauheit der Spiegeloberfläche bedeutend. Dies sind die Fehler, die anhand des hier charakterisierten Messplatzes untersucht werden sollen.

Um ein Verständnis für die Effekte zu bekommen und die Modelle zu überprüfen, reicht es nicht mit einer Fotodiode die im Brennpunkt auftreffende Leistung zu messen. Wir müssen die Auswirkungen der Fehler auf die Intensitätsverteilung, dort wo die Solarzelle sitzen würde, verstehen und in der Lage sein, diese quantitativ zu beschreiben. Dies wird anhand der Konzentrationsmatrix getan. Diese gibt für jeden Punkt in der Brennebene das Verhältnis zwischen der Bestrahlungsstärke an diesem Punkt und an der Eintrittsapertur des Systems an. Sie gibt also für jeden Punkt an, um welchen Faktor das auf ihn treffende Licht im Vergleich zur direkten Bestrahlung ohne Optik konzentriert wurde. In der Konzentrationsmatrix sind noch keine geometrischen Betrachtungen zum Flächenverhältnis der Eintrittsapertur und der Zellfläche enthalten. Die Effizienz des Systems, bei der diese Größen mitberücksichtigt sind, kann jedoch daraus berechnet werden. An der Konzentrationsmatrix kann beispielsweise ein „Ausschmieren" des Brennpunktes im Vergleich zur idealen Optik, also ein flacherer Abfall der Intensität zum Rand der Zelle hin, Verschattungen an einzelnen Stellen der Zelle oder Verformungen des Brennpunktes beobachtet werden. Anhand von Modellen kann dann versucht werden diese Effekte zu simulieren.

Ziel unseres Messplatzes ist es daher, Brennpunkte punktfokussierender Spiegelsysteme, deren Brennpunkte sonnenseitig liegen, ortsaufgelöst darzustellen. Dabei sollen einerseits eine Wiederhol- und Vergleichbarkeit gegeben, andererseits Rückschlüsse zum späteren Einsatz unter der Sonne möglich sein.

Aufgrund der Wetterabhängigkeit, der Verschmutzung und dem örtlich und zeitlich schwankenden Anteil an Streulicht ist ein Test im Außenbereich bezüglich der Wiederhol- und Vergleichbarkeit nicht ideal. Wir müssen also im Labor Bedingungen schaffen, die Aussagen zum Einsatz im Außenbereich erlauben.

Während bezüglich der örtlichen Homogenität der Strahlung und der Strahldivergenz der Lichtquelle an diesem Messplatz bereits ähnliche Bedingungen wie unter freiem Himmel geschaffen wurden, werden die Optiken jedoch mit einer LED, also unter quasi-monochromatischen Bedingungen gemessen. Dies stellt zwar eine Abweichung der späteren Einsatzbedingungen dar, vereinfacht dafür die Gestaltung der Lichtquelle sowie die Interpretation der Messergebnisse. Ohnehin wäre es kaum möglich, einen Detektor zu finden, dessen spektrale Empfindlichkeit der der später eingesetzten Solarzellen entspricht. Somit wäre die über die Wellenlängen gewichtete integrierte Strahlungsstärke am Messplatz nicht vergleichbar mit der für die Solarzelle entscheidenden Bestrahlungstärke.

Messergebnisse an diesem Messplatz spiegeln also nicht zwangsweise direkt das Verhalten im Außenbereich wider. Jedoch können, durch den Vergleich von Proben unter monochromatischer Beleuchtung, die einzelnen Einflüsse analysiert werden. Das Verständnis für diese Effekte sollte dann eine Simulation des Verhaltens unter der Sonne erlauben.

## 3.2. Konzept und Aufbau

Wir wollen also die Intesitätsverteilung, genauer gesagt die Konzentrationsmatrix, dort betrachten wo die Solarzelle platziert wäre. Am einfachsten wäre es, dort einen CCD zu montieren. Allerdings muss dieser dort befestigt, verkabelt und gekühlt oder eine gesamte CCD-Kamera dort montiert werden. Da sich jedoch bei einem einfach reflektierenden Spiegel die Zelle selbst und somit alle hier genannten Bauteile im Strahlengang befinden, würde dies eine zu große Verschattung verursachen.

Dabei sind Verschattungen kritisch, da sie nicht nur dafür sorgen, dass weniger Licht im Brennpunkt ankommt, sondern ganze Bereiche der Spiegel bei der Messung keinen Einfluss haben. Ist nun beispielsweise der mittlere Bereich eines Parabolspiegels besserer optischer Qualität als der Rand, so sinkt die gemessene Güte des Spiegels durch eine Verschattung des zentralen Bereichs. Es ist demnach wichtig die Verschattung so gering wie möglich zu halten, um diese für jedes System auf ein sinnvolles Maß, das heißt auf die im späteren CPV-System durch die Zelle, Kühlung und Halterung verursachte Verschattung, erhöhen zu können.

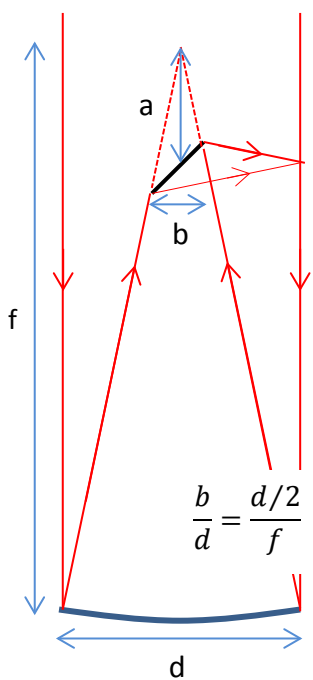

**Abbildung 8:** Alternatives Konzept ohne Diffusor

$$\frac{b}{d} = \frac{d/2}{f}$$

Bei Spiegeln, deren Brennweite groß ist verglichen mit ihrem Durchmesser bzw. ihrer Kantenlänge, könnte auch das zusammenlaufende Strahlenbündel nach dem Spiegel anhand eines Umlenkspiegels aus dem Strahlengang des einfallenden Lichtes heraus geführt werden, sodass eine Kamera ohne Verschattung angebracht werden kann. Tolerieren wir jedoch eine Verschattung von 10% der Spiegelfläche, so müssten die Brennweiten der vermessbaren Spiegel immer noch über anderthalb Mal so groß wie deren Kantenlänge sein(siehe Abbildung 8). Dieses Messverfahren würde daher eine zu starke Einschränkung der vermessbaren Spiegel darstellen.

Das Grundkonzept dieses Messplatzes wurde durch Tobias Schmid vor Beginn dieser Arbeit basierend auf [Ant03] entworfen und im Mai 2013 auf der „9. International Conference on Concentrator Photovoltaic Systems" vorgestellt. Das zugehörige Paper ist im Anhang zu finden. Die Idee ist dabei, dass die Solarzelle im Brennpunkt durch eine streuende Folie, einen sogenannten Diffusor, ersetzt wird. Dieser wird dann über einen Umlenkspiegel anhand einer Kamera mit Objektiv betrachtet. Der Diffusor strahlt auf seiner Rückseite von jedem beleuchteten Punkt aus proportional zur eintreffenden Lichtstärke ab, sodass wir dadurch die ortsaufgelöste Bestrahlungsstärke in der Ebene des Diffusors erhalten. Auch hier ist Verschattung durch den Diffusor und den Umlenkspiegel sowie deren Halterung vorhanden. Diese beträgt jedoch nur 31,2 mm$^2$, was bei dem Spiegel, den wir zum Kalibrieren einsetzen, 6,6% der Fläche entspricht.

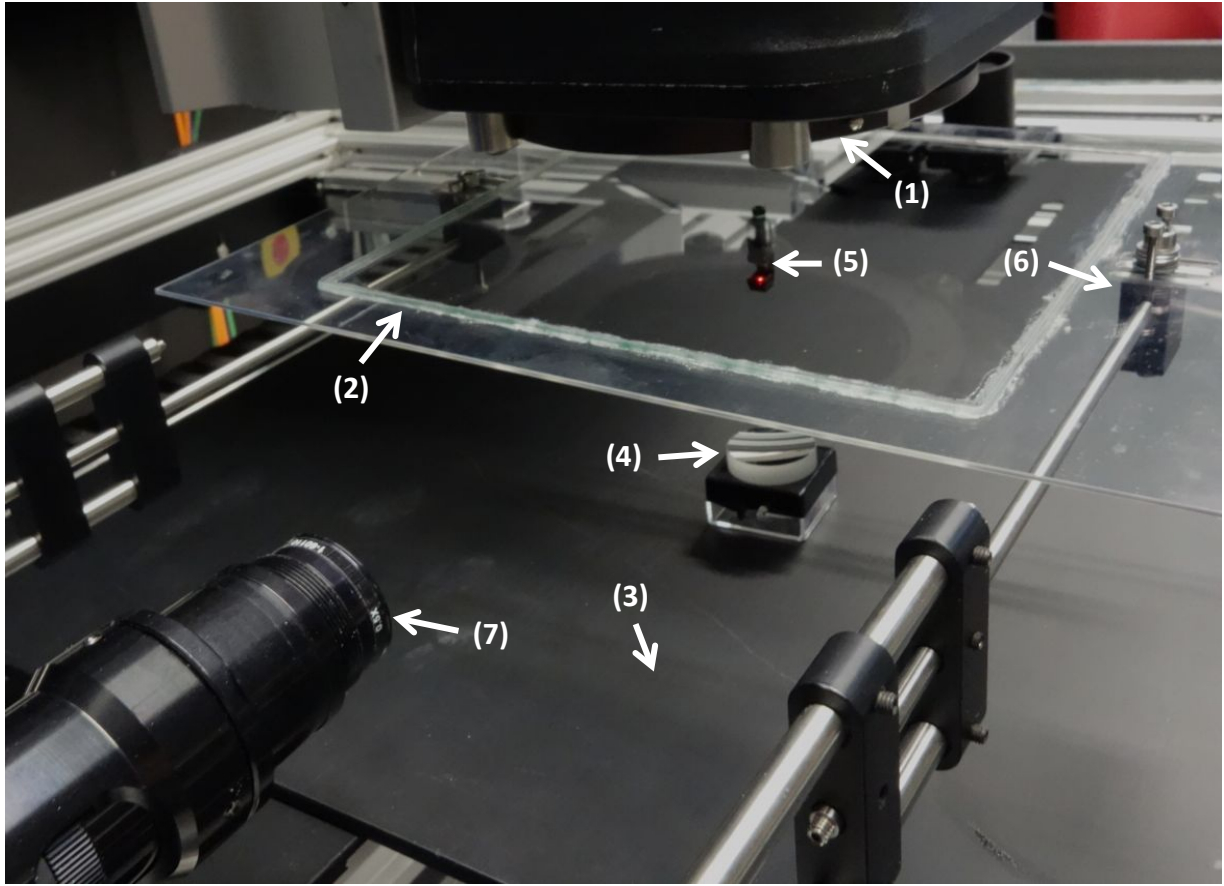

**Abbildung 9: Aufbau des RMP. (1) Lichtquelle, (2) Glasplatte, (3) Auflageplatte, (4) Kalibrierspiegel, (5) Umlenkspiegel- und Diffusorhalter, (6) Justage, (7) Objektiv.**

Der genaue Aufbau ist in Abbildung 9 zu sehen. Die Lichtquelle (1) emittiert rotes Licht, dessen Divergenz der direkten Solarstrahlung entspricht. Dieses Licht durchquert eine schräggestellte Glasplatte (2), an der der Umlenkspiegel- und Diffusorhalter (5) montiert ist. Die Schrägstellung der Glasplatte dient der Vermeidung von Streulicht, einerseits durch Rückreflexionen in die Lichtquelle beim ersten Auftreffen der Strahlen auf die Platte, andererseits durch Spiegelung von Strahlen, die zuvor durch den Diffusor reflektiert wurden (siehe dazu Abbildung 10). Die Glasplatte ist über Regelblöcke (6) an der Kamera befestigt. Hier können der Abstand zur Kamera und somit die Scharfstellung des Bildes, der vertikale Abstand zwischen Glasplatte und Objektiv sowie zwei Verkippungen eingestellt werden. Die dritte Drehachse ist durch die Befestigung des Halters (5) an der Glasplatte gegeben. Um unnötiges Streulicht durch streifenden Einfall am Umlenkspiegel- und Diffusorhalter zu vermeiden und die Divergenz der Strahlen nahe der Halterung nicht zu modifizieren, sind die Seitenwände des Halters leicht angefast. Dieser Halter ist in Abbildung 11 näher zu sehen.

Die Regelblöcke sowie der Umlenkspiegel- und Diffusorhalter wurden im Rahmen dieser Arbeit in Zusammenarbeit mit T. Schmid entworfen.

Auf eine Auflageplatte (3), deren Orthogonalität zum einfallenden Licht vor den Messungen nachjustiert wird, können verschiedene Spiegel aufgelegt werden. In dieser Arbeit werden ein sphärischer $\lambda/4$ Spiegel mit einer Brennweite von (50,80±1,02)mm und einem Durchmesser von (24,50+0/-0,3)mm, im folgenden Kalibrierspiegel (4) genannt, ein silberbeschichteter $\lambda/4$ Planspiegel, im folgenden Hellbildspiegel genannt und konkave Spiegel aus industrieller

Serienproduktion mit unbekannter Oberflächengüte, im folgenden Probe genannt, benutzt[12]. Diese Spiegel reflektieren das Licht durch die Öffnung im Umlenkspiegel- und Diffusorhalter auf den Diffusor. Die Form dieser Öffnung ist so gestaltet, dass möglichst große Einfallswinkel ermöglicht werden. Der Diffusor wird, damit er sich nicht wölbt, zwischen zwei Unterlegscheiben mit einer eckigen Öffnung der Größe 3x3mm eingespannt. Auch hier soll die Form der Unterlegscheiben möglichst große Einfallswinkel erlauben.

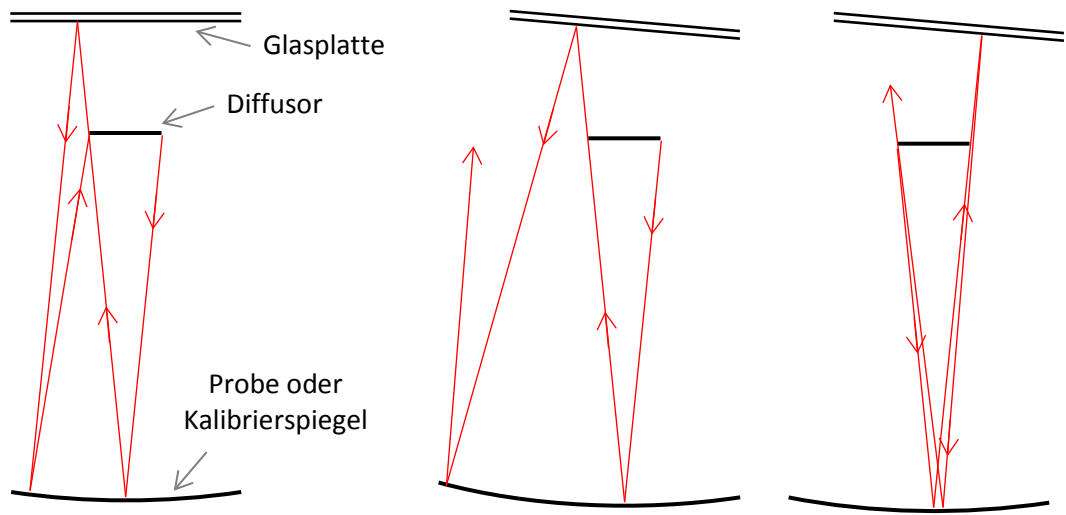

**Abbildung 10: Verkippung der Glasplatte zur Vermeidung von Rückreflektionen des am Diffusor gestreuten Lichts auf diesen.**

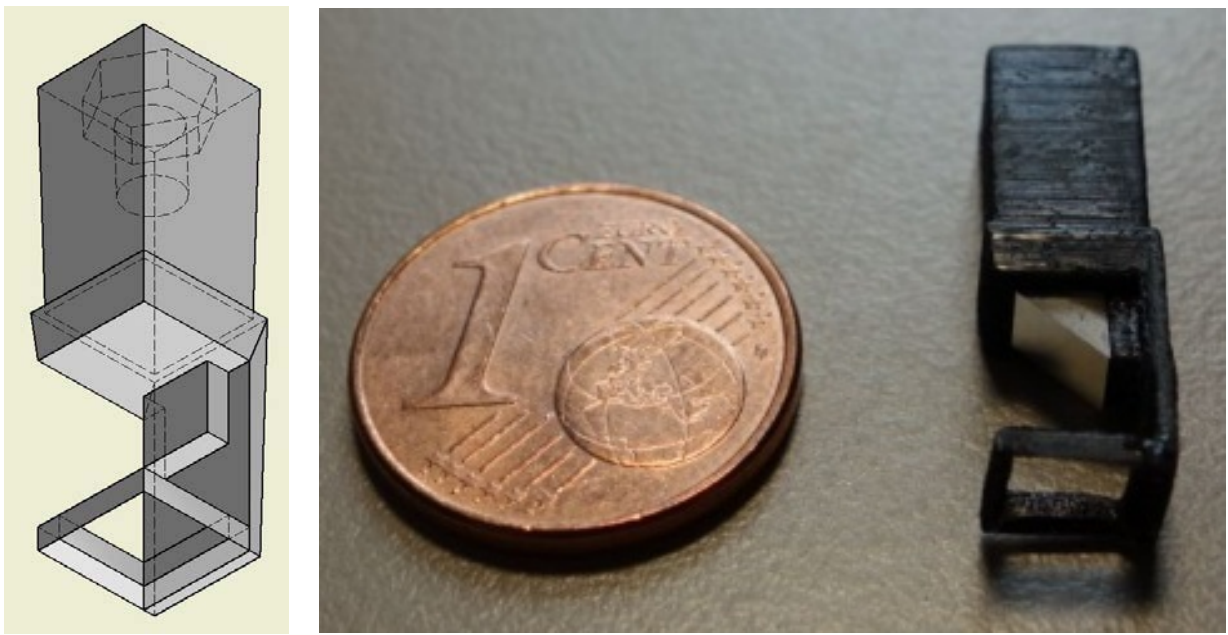

**Abbildung 11: Umlenkspiegel- und Diffusorhalter. Links CAD-Zeichnung ohne Spiegel. Rechts der 3D-gedruckte Halter im Vergleich mit einem 1-cent-Stück. Der Diffusor wird auf die quadratische Öffnung an der Unterseite aufgelegt.**

---

[12] Herstellerangaben

Ein Teil des vom Diffusor gestreuten Lichtes wird nun durch einen $\lambda/8$ Spiegel mit spiegelnder Fläche von 3x4.2mm, den sogenannten Umlenkspiegel, zur Kamera gelenkt. Dabei ist die Entfernung zwischen Diffusor und Umlenkspiegel von Bedeutung. Wäre der Umlenkspiegel zu nah an dem Diffusor, so würden unter bestimmten Winkeln gestreute Strahlen unter so kleinen Winkeln zur Oberflächennormale des Umlenkspiegels auftreffen, dass die reflektierten Strahlen wieder auf dem Diffusor auftreffen und somit das Bild verfälschen würden (siehe Abbildung 12). Daher muss die Unterkante des Umlenkspiegels mindestens dieselbe Entfernung vom Diffusor haben, wie die am weitesten entfernte Kante des Diffusors von der Rückwand des Halters (beide Abstände in Abbildung 12 eingezeichnet).

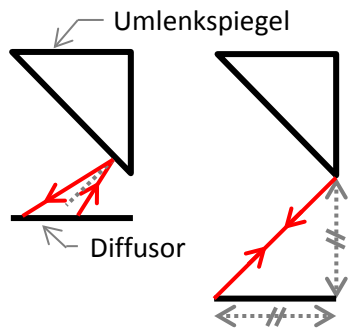

Abbildung 12: Wahl des Abstandes zwischen Umlenkspiegel und Diffusor. Links Rückreflektionen auf den Diffusor bei zu kleinem Abstand. Rechts minimaler Abstand

Die Objektweite des Objektivs der Kamera (7) ist schließlich so gewählt, dass es außerhalb des auf die Probe, Hellbild- oder Kalibrierspiegel einfallenden Strahlenbündels liegt und somit keine Verschattung verursacht. Weiter ist der vertikale Abstand der Glasplatte zum Objektiv so gewählt, dass kein Streulicht direkt vom Diffusor oder der Glasplatte oder vom Diffusor über die Glasplatte auf den CCD gelangen kann.

## 3.3. Messprinzip

Mit dem Aufbau aus Kapitel 3.2 ist es möglich die Konzentrationsmatrix zu ermitteln. Dazu werden mit der Kamera zwei Aufnahmen gemacht. Beim eigentlichen Messbild wird die Probe bzw. der Kalibrierspiegel eingelegt, der das einfallende Licht auf den Diffusor fokussiert. Dagegen wird beim sogenannten Hellbild dieser Spiegel durch den leicht gekippten und seitlich versetzten Hellbildspiegel ersetzt. Dieser lenkt somit das einfallende Licht unfokussiert auf den Diffusor um. Der Strahlengang für Mess- und Hellbild ist respektiv in Abbildung 14 und Abbildung 13 dargestellt.

Bilden wir pixelweise das Verhältnis zwischen Mess- und Hellbild, so bekommen wir für jeden Pixel den Konzentrationsfaktor zwischen Mess- und Hellbild. Dies ist jedoch noch nicht die Konzentrationsmatrix der vermessenen konzentrierenden Optik, denn die einzelnen Komponenten des Messplatzes beeinflussen das Messergebnis. Diese Einflüsse werden im folgenden Kapitel untersucht.

Mit Hilfe eines Raytracing-Programms wird der Strahlengang für die jeweilige konzentrierende Optik simuliert. Die Ergebnisse der Charakterisierungen der Einzelelemente ermöglichen dabei die Eigenschaften der verschiedenen Komponenten im Strahlengang bei der Simulation vorzugeben.

Zur Überprüfung der Zuverlässigkeit und der Vollständigkeit der Simulation wird der bekannte Kalibrierspiegel vermessen und mit der Simulation verglichen.

Stimmt dies überein, können Proben gemessen und entsprechend ihrer Sollgeometrie simuliert werden. Treten hier Abweichungen auf, kommen diese von Abweichungen der Probe von der Sollgeometrie, die dann untersucht werden können.

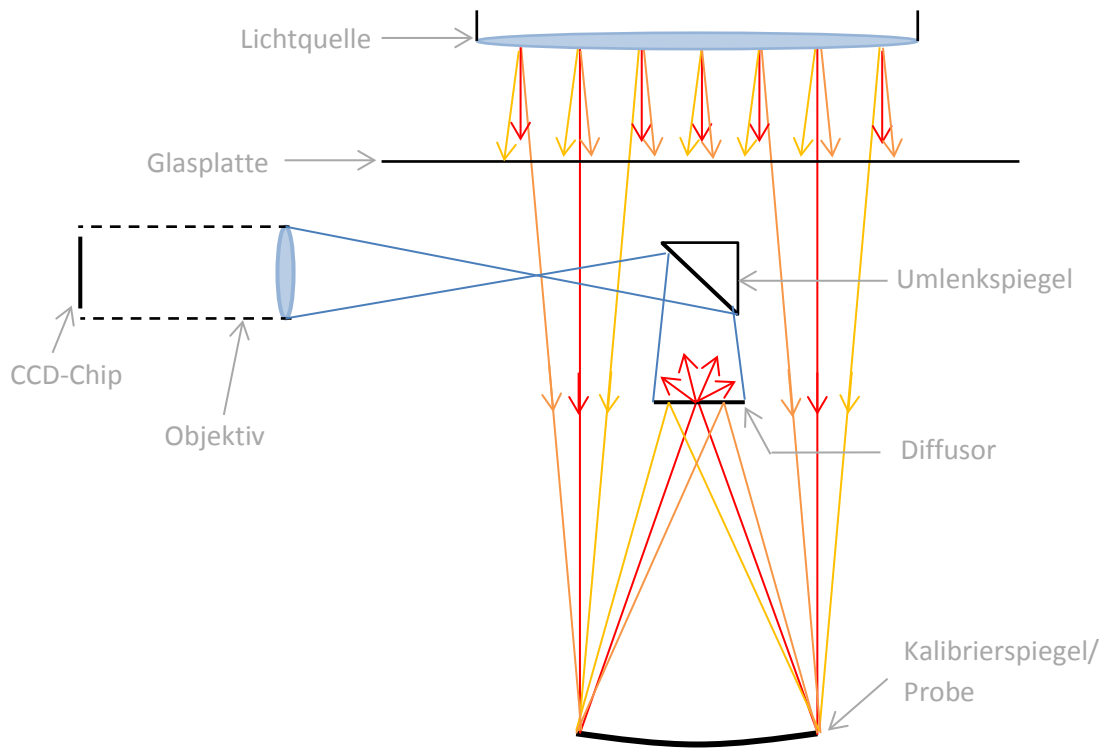

**Abbildung 14:** Strahlengang am RMP für das Messbild mit Kalibrierspiegel oder Probe. Die Divergenz der Strahlen und die Größe des Diffusors wurden hier zur Sichtbarkeit erhöht. Die Lichtquelle leuchtet homogen mit sonnenähnlicher Divergenz.

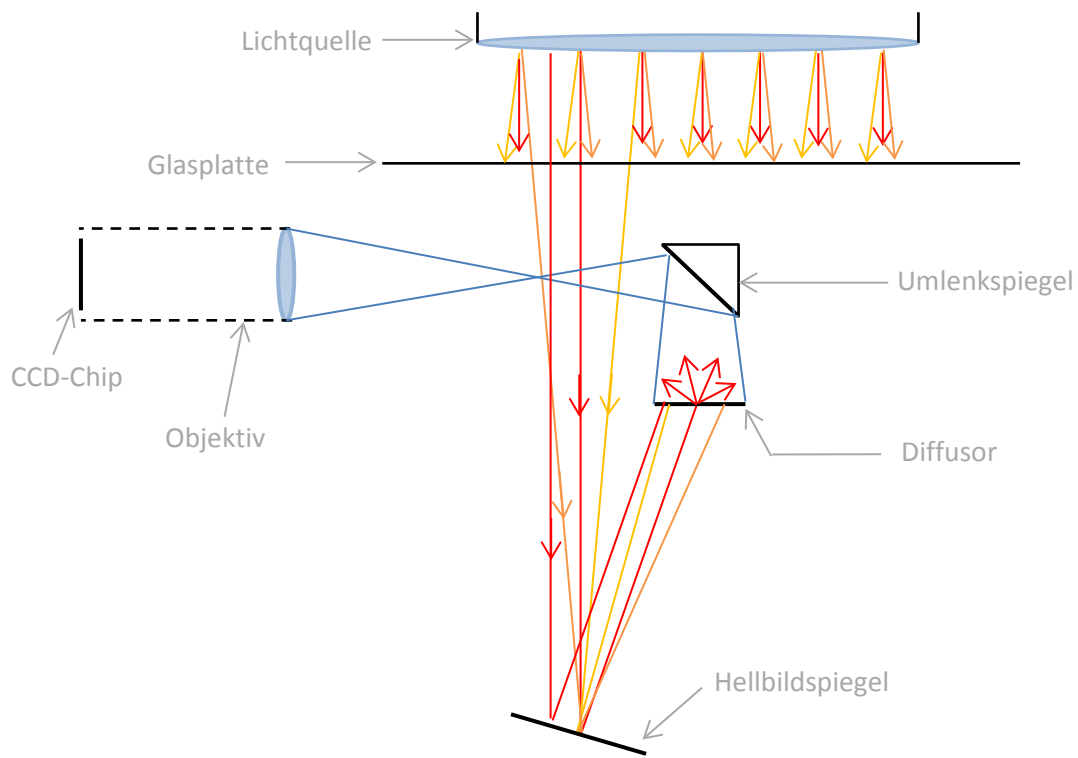

**Abbildung 13:** Strahlengang am RMP für das Hellbild. Die Divergenz der Strahlen und die Größe des Diffusors wurden hier zur Sichtbarkeit erhöht. Die Lichtquelle leuchtet homogen mit sonnenähnlicher Divergenz. Der Diffusor wird dadurch ebenfalls homogen mit dieser Divergenz beleuchtet.

# 4. Charakterisierung der Einzelkomponenten

## 4.1. Kamera

Aufgrund der benötigten Ortsauflösung kommt der Kamera eine besonders wichtige Bedeutung zu. Jeder ihrer Fehler und jedes nicht-ideale Verhalten wirken sich nicht nur auf die späteren Messbilder des RMPs aus, sondern auch auf die meisten der im Rahmen dieser Arbeit durchgeführten Experimente zur Charakterisierung anderer Komponenten. Daher sollen zu allererst ihre Funktionsweise und die Fehler, die durch die Kamera zustande kommen, näher betrachtet werden.

### 4.1.1. Funktionsweise und Fehlerquellen

Im RMP und bei jedem Experiment mit Kamera im Rahmen dieser Arbeit wird eine vom Hersteller für wissenschaftliche Zwecke vorgesehene monochromatische Kamera eingesetzt. Jedem der 1600x1200 Pixel des CCDs wird ein Zahlenwert zwischen 0 und $2^{14}$ – ein sogenannter Grauwert – zugeordnet, der mit der auftreffenden Photonenzahl steigt. Wir werden die Einheit dieser Werte im Folgenden als Graustufen bezeichnen.

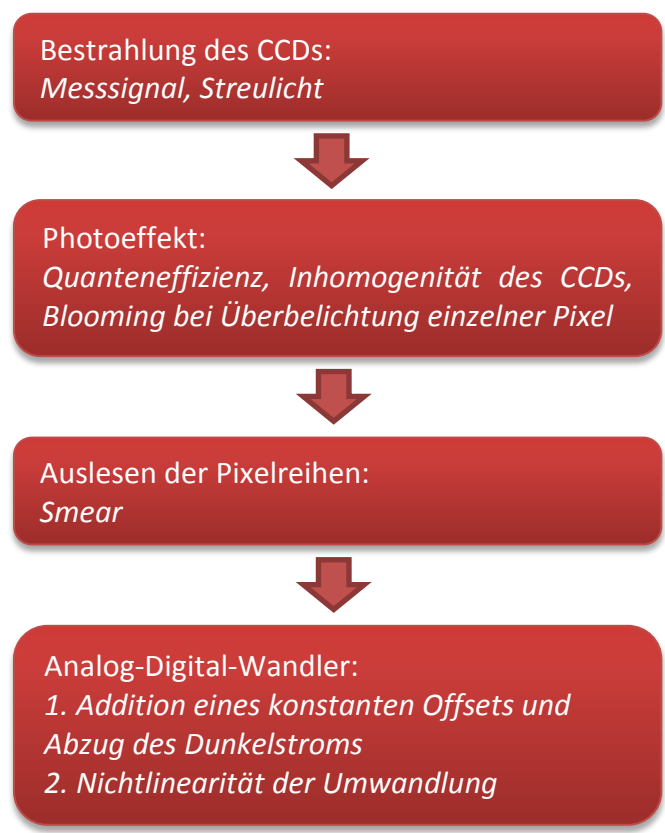

**Abbildung 15: Verlauf der Bildentstehung mit jeweiligen Effekten**

Ziel von Messungen mit der Kamera ist es Verhältnisse von Bestrahlungsstärken aufzunehmen. Dabei können sich diese Verhältnisse auf die Bestrahlungsstärken am Ort unterschiedlicher Pixel einer selben Aufnahme beziehen oder von Pixeln unterschiedlicher Aufnahmen die mit kurzem Zeitabstand relativ zu Schwankungen der Bestrahlung aufgenommen wurden. Da diese quantitativ benötigt werden muss der Grauwert eines Pixels proportional zur auf diesen Pixel treffenden Strahlungsenergie sein. Dies gilt jedoch nur eingeschränkt. Wir wollen daher zunächst die genauere Funktionsweise der Kamera betrachten, die in Abbildung 15 schematisch dargestellt ist.

Bei der Bestrahlung des CCDs mit der gewünschten Beleuchtung und dem Streulicht wird aufgrund des Photoeffekts ein Teil der Photonen absorbiert und freie Elektronen erzeugt. Dabei hängt die im Durchschnitt nötige Anzahl an auftreffenden Photonen um ein Elektron freizusetzen, die sogenannte Quanteneffizienz, von der Wellenlänge der auftreffenden Photonen ab. Außerdem ist es wichtig, an welchem Ort auf dem CCD-Chip die Photonen auftreffen. Diese Inhomogenität des CCD-Chips, die sogenannte photoresponse non-uniformity (PRNU), kann durch Verschmutzungen (Staub, Glaßsplitter von den CCD-Abdeckgläsern, Metalsplitter vom C-Mount-Gewinde), Beschädigungen (Kratzer auf dem CCD-Abdeckglas) und Kristallfehler des Wafers [alS96] starke lokale Variationen sowie durch Variationen der Dotiermenge im Wafer großflächigere Empfindlichkeitsgradienten [Hil03] aufweisen.

Da der Photoeffekt nur mit einer bestimmten Wahrscheinlichkeit stattfindet, entsteht an dieser Stelle ein statistisches Rauschen.

Die einzelnen Pixel können nur eine begrenzte Anzahl an Elektronen fassen, was nicht mit einer einfachen Überbelichtung zu verwechseln ist. Diese Grenze beträgt bei unserer Kamera 40 000 Elektronen pro Pixel, was einer 1,14-fachen Überbelichtung entspricht [Dat]. An jedem Pixel wird, vom Halbleiter durch einen Isolator getrennt, ein positives Potential angelegt. Dadurch entsteht im Halbleiter am Isolator eine Potentialmulde, in der sich Elektronen aus dem Leitungsband ansammeln [Bar75]. Werden während der Belichtungszeit so viele Elektronen in das Leitungsband gehoben, dass die abstoßenden Ladungen der angesammelten Elektronen das angelegte Potential kompensieren, können die Elektronen vom am Nachbarpixel angelegten Potential angezogen werden. Dieser Effekt wird oft als ein Überlaufen des Potentialtopfes in die Nachbarpixel beschrieben. Wenn er stark ausgeprägt ist, kann er zu einem erneuten Überlaufen der Nachbarpixel führen und wird als Blooming bezeichnet. Die von uns eingesetzte Kamera besitzt dabei keine Anti-Blooming-Gates, die den Übergang in Nachbarpixel durch eine Abführung der Überschüssigen Elektronen behindern [Jan01 p. 300f].

Zum Auslesen der Pixel befindet sich zwischen je zwei aktiven Pixelspalten eine abgedeckte Spalte [Dat], in die die Elektronen nach der Belichtungszeit überführt werden, um von dort aus zu einer Zeile transportiert zu werden in der die Pixel ausgelesen und somit geleert werden. Trotz der Abdeckung dieser Spalten gerät etwas Licht an den Halbleiter, was zusammen mit übergetretenen Elektronen aus den Nachbarpixeln ein elektronisches Rauschen verursacht [Hop04 p. 102]. Um dieses zu reduzieren werden die Inhalte der Pixel der Dunkelspalten ständig zeilenweise zyklisch durchgeschoben und jeder Pixel beim Durchgang durch die Auslesezeile geleert. Wird nach der Belichtungszeit der Inhalt der aktiven Pixel in die Pixel der abgedeckten Spalten verschoben, beinhalten diese noch die Elektronen, die sich dort seit ihrem letzten Durchgang durch die Auslesezeile angehäuft haben. Wird ein Pixelinhalt in dieser Zeit oder zwischen dem Befüllen mit den Elektronen aus den aktiven Pixeln und dem Auslesen in einen Pixel geschoben, der einer hohen Bestrahlungsstärke ausgesetzt wird, wird dort trotz Abdeckung eine beachtliche Zahl an Elektronen zugeführt. Dies gibt im Messbild einen senkrechten Streifen über den gesamten CCD sobald ein Pixel der Spalte einer hohen Bestrahlungsstärke ausgesetzt ist. Dieser Effekt wird Smear genannt. Die absolute Menge an Smear hängt nicht von der Belichtungszeit ab, sondern ist proportional zu der Zeit für einen solchen Zyklus. Solange nicht überbelichtet wird, ist jedoch die relative Menge an Smear invers proportional zur Belichtungszeit.

In derselben Zeit in der auch der Smear entsteht sowie während der Belichtungszeit werden aufgrund der thermischen Stöße im gesamten Halbleiter einige Elektronen in das Leitungsband gehoben [Til05 p. 186]. Dies ist der sogenannte Dunkelstrom. Der Dunkelstrom nimmt mit der Temperatur zu, weshalb die Kamera ein Peltier-Element zur Kühlung des CCDs besitzt. Je nach Betriebstemperatur werden zwischen ca. 0,005 (bei 253,15K) und 0,25 (bei 293,15K) Graustufen pro Pixel und pro Sekunde thermisch angeregt [Dat]. Sofern nichts anderes angegeben ist, wurde für Messungen im Rahmen dieser Arbeit der Chip auf 259,15 K gekühlt. Zusätzlich zu dieser Reduzierung besitzt der CCD-Chip, um den Dunkelstrom herauszurechnen, im Randbereich abgedeckte

Pixelspalten und -reihen die nicht zur Aufnahme des Messbildes genutzt werden. Der Mittelwert der Elektronen dieser Pixel dient dann zur Korrektur der Elektronenmengen der aktiven Pixel. Da es jedoch sein kann, dass der Dunkelstrom eines beliebigen Pixels geringer ist als der Mittelwert in den gesonderten Pixeln zur Korrektur, könnten schwach beleuchtete Pixel negative Werte annehmen. Um bei der Konversion zu einem digitalen Signal nicht ein eigenes Bit für das Vorzeichen zu benötigen, wird ein konstanter Offset von circa 100 Graustufen auf das Bild aufaddiert.

Schließlich wird aus den Elektronen durch einen Analog-Digital-Wandler (ADC) ein Grauwert erzeugt. Im Idealfall sollte dieser linear mit der Anzahl der Elektronen steigen. Dies ist jedoch insbesondere bei wenigen Elektronen und nahe der Überbelichtung nicht genau der Fall. Weiter entsteht auch hier wieder ein Rauschen.

Damit nimmt der Grauwert am Pixel px unter Bestrahlung $s(\lambda) + u(\lambda)$ die Form

$$M(px) := \left\{ \left[ \int \left( s(\lambda) + u(\lambda) \right) \cdot \eta(\lambda) \, d\lambda \right](px) \cdot h(px) + sm(px) + off \right\} \cdot l(G) \qquad (5)$$

an, wobei

$s(\lambda)$ : Messsignal
in Graustufen pro Wellenlänge

$u(\lambda)$ : Streulicht
in Graustufen pro Wellenlänge

$off$: Offset in Graustufen

$\eta(\lambda)$ : Quanteneffizienz

$h(px)$ : Inhomogenität

$l(G)$ : Linearität

$sm(px)$ : Smear in Graustufen

und $G$ für Grauwert, $px$ für Pixel steht. Die thermischen Elektronen wurden hier nicht aufgeführt, da sie bereits in der Kamera intern wieder abgezogen werden. Der Smear ist nicht direkt mit der Homogenität des Chips an der Stelle des Pixels zu gewichten. Die dafür verantwortlichen Elektronen werden während des zyklischen Verschiebens der Ladungen angeregt und gehen also in allen Pixeln derselben Spalte in das Leitungsband über. Damit hängt der Smear von der Kristallstruktur und Verschmutzungen sowie von der Bestrahlungsstärke am Ort anderer Pixel derselben Reihe ab. Der Smear besteht somit für den Pixel der Spalte i und Reihe j aus

$$sm(px_{ij}) = \text{const} \cdot \frac{t_{zyklus}}{t_{belichtung}} \cdot \int \sum_k \left( s(\lambda) + u(\lambda) \right)(px_{kj}) \cdot h(px_{kj}) \cdot \eta(\lambda) \, d\lambda \qquad (6)$$

Dabei ist $t_{zyklus}$ die Periode der Ausleerung der Pixel der abgedeckten Reihen. Die Konstante resultiert durch die Abschwächung durch die Abdeckung dieser Reihen. $t_{zyklus}$ beträgt bei uns ca. 0,2s. Gilt $t_{zyklus} \ll t_{belichtung}$, ist also die Lichtintensität schwach, kann der Smear gegenüber dem Bild vernachlässigt werden. Dabei ist anzumerken, dass $s(\lambda) + u(\lambda)$ das Messsignal des Messbildes ist und somit proportional zur Belichtungszeit ist. Der Smear ist also belichtungszeitunabhängig. Der Smear wird im Folgenden als $sm(px)$ zusammengefasst, da er als solcher, ohne ihn in seine Komponenten zu zerlegen, korrigiert werden kann.

### 4.1.2. Signifikanz und Minimierung der Fehlerquellen

Im Idealfall möchten wir in der Lage sein jeden der Terme aus Gleichung 5 heraus zu korrigieren, um das reine Messsignal zu bekommen. Dazu müssen wir zunächst die Kameralinearität betrachten. Sie wurde bereits vom Hersteller vermessen, jedoch nur für Grauwerte über 400 und mit einer relativ geringen Punktdichte. Daher wurde sie im Rahmen dieser Arbeit nochmals vermessen.

*Linearität*

Die Grundidee, um die Linearität der Kamera zu bestimmen, ist es anzunehmen, dass die Anzahl der in der Potentialmulde eingefangenen Elektronen proportional mit der Belichtungszeit steigt. Dies ist bei konstanter Bestrahlungsstärke der Fall, wenn die in das Leitungsband angehobenen Elektronen schnell genug in die Potentialmulde abgeführt werden, um Einschwingprozesse der Elektronenzahl im Leitungsband außerhalb der Potentialmulde aufgrund von Rekombinationen vernachlässigen zu können und wenn die Anzahl der im Laufe der Belichtungszeit angeregten Elektronen klein gegenüber der Gesamtanzahl an Elektronen im Valenzband ist, sodass sich die Anregungsrate nicht merklich im Laufe der Belichtungszeit ändert. Wir nehmen an, dass dies gegeben ist. Abweichungen von einem linearen Anstieg der Grauwerte mit der Belichtungszeit bei konstanter Bestrahlungsstärke resultieren daher aus einer nicht-linearen Konversion der Elektronen in Graustufen.

Um die Linearität der Kamera zu messen wurde mit 109 unterschiedlichen Belichtungszeiten von 0,5ms bis 6300ms dieselbe, näherungsweise homogen ausgeleuchtete, Fläche photographiert. Dabei wurden die Aufnahmen in willkürlicher Reihenfolge durchgeführt, damit sich eventuelle Drifts der LED und der Kamera nur in Form eines statistischen Rauschens auswirken. Für das am längsten belichtete Bild bei dem noch keine Pixel überbelichtet sind[13] und somit mit Sicherheit noch kein Blooming auftritt, wurde festgestellt welche Pixel höchstens um 5 Graustufen vom Mittelwert des Bildes abweichen. Nur diese Pixel (im Folgenden Maske genannt) wurden für die Auswertung aller Bilder benutzt. Die Belichtungszeit eines Großteils der anderen Bilder ist kürzer als die dieses Bildes. Somit stellen die 5 Graustufen eine obere Grenze für die Abweichung der Grauwerte der ausgewählten Pixel dieser Bilder von ihrem Mittelwert dar. Für jedes Bild wurde über die ausgewählten Pixel gemittelt und der Mittelwert gegen die Belichtungszeit aufgetragen (siehe Abbildung 16, links). Wir können sehen, dass die Grauwerte in erster Näherung linear mit der Belichtungszeit, und somit mit der auftreffenden Energie, steigen. Lediglich für Grauwerte unter 500 und über 15000 ist eine Abweichung von der Linearität zu erkennen (Abbildung 16, rechts).

Prinzipiell kann nun anhand dieser Kurve eine Korrektur der Linearität der Messbilder vorgenommen werden. Ziel einer solchen Korrektur wäre pixelweise die Grauwerte zu korrigieren, sodass sie anstatt ihres ursprünglichen Wertes den dazugehörigen Wert auf der roten Linie in Abbildung 16 annehmen. Das kann erreicht werden, indem zunächst bei der Linearitätskurve jeder Zahlenwert des mittleren Grauwertes der Maske durch den Zahlenwert seiner Belichtungszeit geteilt wird und dieses Verhältnis gegen den Grauwert aufgetragen wird. Dann kann jeder Grauwert späterer Messbilder, indem er durch dieses Verhältnis geteilt wird, korrigiert werden.

---

[13] Es handelt sich um die Belichtungszeit von 6077ms

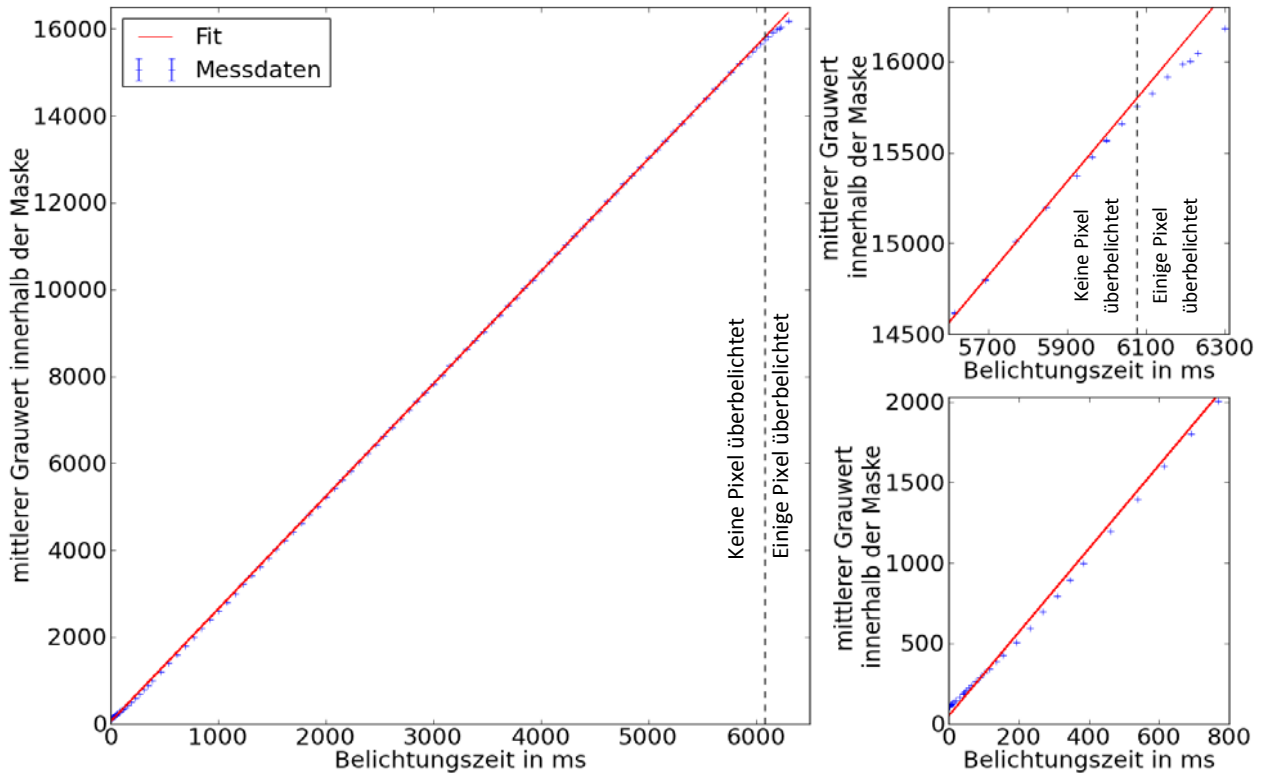

**Abbildung 16:** Linearität der Kamera bei Aufnahmen zu 109 unterschiedlichen Belichtungszeiten. Links: Übersicht. Rechts: bei sehr geringen und sehr hohen Grauwerten.

Die Ergebnisse dieser Messung stimmen allerdings nicht mit den gröberen und auf einen kleineren Grauwertebereich beschränkten Angaben des Herstellers überein (siehe Abbildung 17). Insbesondere weicht die Messung des Herstellers im unteren Bereich schon bei höheren Grauwerten von der Linearität ab und die gemessenen Grauwerte sind im oberen Bereich höher als sie bei einem linearen Verhalten wären.

Daher wurde von T. Schmid die Linearität der Kamera mit einer differentiellen Methode überprüft. Die Grundidee ist dabei, für zwei nahegelegene Belichtungszeiten den CCD so zu bestrahlen, dass auf den Bildern jeder Grauwert annähernd gleich häufig vorkommt. Nun kann für jeden Pixel die Änderung des Grauwertes zwischen den zwei Aufnahmen als Ableitung der Linearität an der Stelle des Grauwertes angesehen werden. Diese Ableitungen können gegen den zugehörigen Grauwert aufgetragen und die Kurve anschließend integriert werden. Auch diese Messung stimmt weder mit der im Rahmen dieser Arbeit durchgeführten direkten Messung, noch mit der Herstellermessung überein. Zwar stimmt hier der qualitative Verlauf der Abweichung von der Linearität mit der Messung durch den Hersteller überein, doch sind die Abweichungen ca. um einen Faktor 7 höher.

Derzeit gibt es für diese Abweichungen keine zufriedenstellende Erklärung. Bei höheren Belichtungszeiten könnte prinzipiell der gebildete Mittelwert durch einen Überlauf aus überbelichteten Nachbarpixeln, also aufgrund von Blooming, erhöht werden. Doch sollte dies dann auch rechts der gestrichelten Linie in Abbildung 16 in Form eines Anstiegs der Steigung der Messkurve zu erkennen sein.

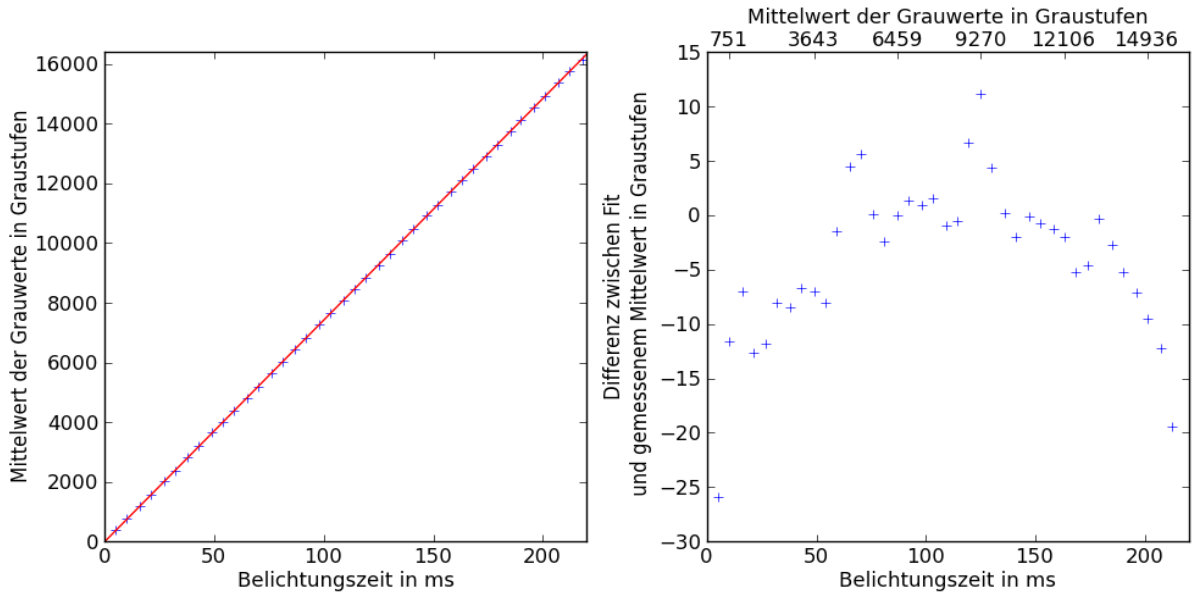

Abbildung 17: Linearität der Kamera gemessen durch den Hersteller. Links: Messung des Herstellers und Fit, rechts: Abweichung der Messpunkte vom Fit

Jedoch verläuft jede dieser 3 Messungen im Bereich zwischen 4000 und 12000 Grauwertstufen in guter Näherung linear mit maximalen Abweichungen der Linearität von ca. 0,2% bei Grauwerten nahe 4000 Graustufen. Wo es möglich ist, reicht es somit die Belichtungszeit so zu wählen, dass alle in die Auswertung einfließenden Pixel Grauwerte in diesem Intervall besitzen.

Im Allgemeinen ist anzumerken, dass die relativen Abweichungen von der Linearität, bis auf den Fall von Grauwerten von unter 500 Graustufen, sehr gering sind. Daher entsteht durch die Linearität nur eine geringe Unsicherheit die einer eventuell fehlerhaften Korrektur vorzuziehen ist.

### Smear

Dank der Fähigkeit der Kamera sehr kurz zu belichten und der Belichtungszeitunabhängigkeit des Smears, kann dieser ohne großen Aufwand herauskorrigiert werden. Dazu wird eine Aufnahme mit der minimal möglichen Belichtungszeit von 500ns durchgeführt. Verglichen mit den bei uns üblichen Belichtungszeiten von 500ms bis zu einer knappen Stunde ist diese Belichtungszeit vernachlässigbar klein, sodass für alle unsere Anwendungen die Näherung, das Smearbild enthalte weder Messsignal noch Streulicht, berechtigt ist. Wir können also das Smearbild durch

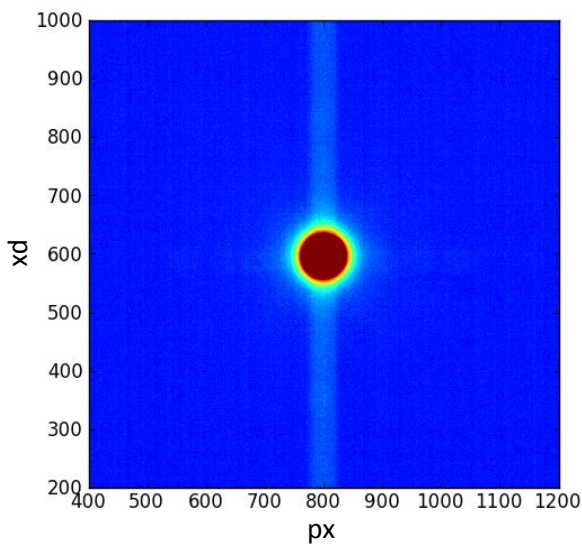

Abbildung 18: Ein Beispiel für Smear

$$M_{sm}(px) = (\,sm(px) + off\,) \cdot l(G) \qquad (7)$$

beschreiben. Vernachlässigt man wie oben begründet die Linearität erhält man wie gewünscht nur noch den Smear und den Offset. Wird vom Messbild (siehe Gleichung 5), nach Vernachlässigung der Linearität, nun die Smear-Aufnahme abgezogen, haben wir nur noch:

$$M'(px) := \left[\int \big(s(\lambda) + u(\lambda)\big) \cdot \eta(\lambda)\ d\lambda\right](px) \cdot h(px) \tag{8}$$

*Photoresponse non-uniformity*

Die PRNU wurde vom Hersteller mitgeteilt und von T. Hornung nachgemessen (siehe [Hor13], Kap. 7.5). Dabei wird der CCD homogen bestrahlt, wodurch sich Sensibilitätsverhältnisse der Pixel untereinander ergeben. Wir können damit die Korrektur vornehmen, indem das Messbild pixelweise durch die PRNU-Aufnahme geteilt wird. Damit dadurch der Betrag des Messbildes im Mittel nicht verändert wird, wird zuvor jeder Grauwert der PRNU-Aufnahme durch den Mittelwert der Grauwerte derselben Aufnahme geteilt. Dies entspricht dem Faktor h(px) in Gleichung 5.

Beide Messungen stimmen, bis auf einen hellen Fleck bei den Herstellerdaten in der Mitte des Bildes, der aus dem Messverfahren resultiert, überein. Aufgrund dieses Flecks wurde für alle Korrekturen die Aufnahme von T. Hornung benutzt.

*Streulicht*

Auch für das Streulicht wird eine eigene Aufnahme gemacht. Dabei wird der direkte Strahlengang durch eine Abschirmung unterbrochen, wobei darauf geachtet wird potentielle Streuquellen analog wie beim Messbild zu bestrahlen. Die Linearität kann wieder vernachlässigt werden, weil die erwarteten Grauwerte niedrig sind und eine leichte Skalierung somit kaum eine Auswirkung hat. Da durch die Abschirmung des direkten Strahlengangs an keiner Stelle des Chips eine hohe Intensität auftrifft, ist hier kein nennenswerter Smear vorhanden. Daher kann die so durchgeführte Untergrundaufnahme durch

$$M_u(px) := \left(\left[\int u(\lambda) \cdot \eta(\lambda)\ d\lambda\right](px) \cdot h(px)\right) + off \tag{9}$$

beschrieben werden. Auch hier ist wieder der Offset vorhanden. Da wir diesen von Messbild zeitgleich mit der Smearkorrektur bereits abgezogen haben, müssen wir diesen dieses Mal von der Untergrundaufnahme abziehen. Wir müssen also den Offset getrennt bestimmen bevor wir auch dieses Bild mit der PRNU korrigieren und anschließend vom Messbild (siehe Gleichung 8) abziehen können.

*Dunkelstrom und Offset*

Da der Dunkelstrom einen thermischen Ursprung hat, kann die Menge an Dunkelstrom reduziert werden indem der CCD-Chip gekühlt wird. Wie bereits erläutert wird der Chip am RMP und für die Charakterisierungen im Rahmen dieser Arbeit bei 259.15K betrieben. Im Grunde übernimmt bereits die Kamera selbst größtenteils die Korrektur des Dunkelstroms, indem sie den mittleren Dunkelstrom von jedem Pixel abzieht. Bei einem Dunkelbild bleiben nur Schwankungen um diesen Mittelwert übrig. Hier besteht die Frage, ob zum Beispiel durch Unregelmäßigkeiten in dem Halbleiter oder durch unterschiedliche Wärmeabführung, bzw. Kühlung manche Pixel bzw. Pixelgruppen systematisch bei allen Aufnahmen einen höheren oder niedrigeren Dunkelstrom aufweisen als andere. Wäre dies der Fall, so würde die Korrektur der Kamera, die auf Pixelreihen am Rand des Chips basiert, je nach Pixel systematisch zu viel oder zu wenig abziehen. Die Frage ist damit, ob die Schwankungen im Dunkelstrom über den CCD statistisch oder systematisch sind. Um dies zu überprüfen werden im Folgenden die Schwankungen in Dunkelstromaufnahmen des Herstellers mit denen in eigenen Aufnahmen verglichen und auf Korrelationen untersucht.

Dabei wird zugrunde gelegt, dass sich die Standardabweichung der pixelweisen Differenz von zwei Bildern bei statistischen Schwankungen der Grauwerte über den Chip von derselben Standardabweichung bei systematischen Schwankungen über den Chip unterscheidet. Sind die Schwankungen rein statistisch, so gilt

$$\sigma_{(Bild1-Bild2)} = \sqrt{\sigma_{Bild1}^2 + \sigma_{Bild2}^2} \tag{10}$$

Sind die Schwankungen dagegen rein systematisch (mit Korrelation 1), so gilt

$$\sigma_{(Bild1-Bild2)} = |\sigma_{Bild1} - \sigma_{Bild2}| \tag{11}$$

Sind die Schwankungen teilweise systematisch, ergibt sich eine Standardabweichung die zwischen beiden Werten liegt. Es sei angemerkt, dass diese Schwankungen nur teilweise ihren Ursprung im Dunkelstrom haben. Dazu kommt noch elektronisches Rauschen beim Auslesen [Dat], welches sich statistisch verhalten sollte. Wir werden im Abschnitt „Statistischer Fehler" nochmals darauf zurückkommen. Daher wird eine vollständige Korrelation nicht erwartet.

In einem dunklen Raum wurden 5 Messbilder mit Deckel auf der Kamera aufgenommen. Wir nehmen daher an, dass alle Anregungen von Elektronen in das Leitungsband thermischen Ursprung haben.

Zunächst werden diese 5 Bilder pixelweise gemittelt. Für das resultierende Bild und für die Aufnahme des Herstellers wird jeweils die Standardabweichung zwischen den Grauwerten des Bildes berechnet. Da der Dunkelstrom temperaturabhängig ist, gilt selbiges für dessen Schwankungen. Diese zwei Standardabweichungen sind daher nicht unbedingt von gleicher Größe. Weiter wird die Herstelleraufnahme pixelweise von der eigenen abgezogen und aus dem resultierenden Bild ebenfalls die Standardabweichung berechnet. Es ergeben sich

$$\sigma_{\overline{exp}} = 2{,}082 \qquad \sigma_{hersteller} = 1{,}018 \qquad \sigma_{\overline{exp}-hersteller} = 2{,}315 \qquad \sqrt{\sigma_{\overline{exp}}^2 + \sigma_{hersteller}^2} = 2{,}318$$

Graustufen. Dabei steht der Index $\overline{exp}$ für das gemittelte eigene Bild. Wir sehen, dass die Abweichungen beinahe statistisch verteilt sind. Die Ungenauigkeit $\sigma/[2 \cdot (N-1)]$ der Berechnung der Standardabweichung [Squ01] ist in der Größenordnung $10^{-6}$ und reicht damit nicht aus die leichte Korrelation zu erklären. Diese könnte aus einer residualen Belichtung resultieren, wodurch in beiden Bildern dieselbe PRNU einspielt, oder aus einem der eingangs genannten Effekte herrühren. Wir behandeln im Folgenden die Schwankungen des Dunkelstroms als rein statistisch.

Zwischen den eigenen Aufnahmen und der Herstelleraufnahme konnte keine bedeutende Korrelation festgestellt werden. Es besteht aber weiterhin die Frage, ob eine solche Korrelation bei kurz hintereinander aufgenommenen Bildern besteht. Dies könnte aufgrund von kurzlebigen Temperaturgradienten im Chip der Fall sein. Um dies zu überprüfen, wurden mit demselben Verfahren die 5 ungemittelten eigenen Aufnahmen untereinander verglichen. Hierzu wurde aus den Einzelaufnahmen jeweils die Standardabweichung gebildet und daraus der Mittelwert $\bar{\sigma}_{Einzelbilder}$ berechnet. Da die 5 Aufnahmen unter selben Bedingungen aufgenommen wurden sind ihre Standardabweichungen ähnlich. Anschließend wurden die Bilder paarweise pixelweise subtrahiert und daraus je die Standardabweichung gebildet. $\bar{\sigma}_{Differenz\,von\,je\,zwei\,Bildern}$ ist der Mittelwert dieser Standardabweichungen. Hier ergeben sich

$$\bar{\sigma}_{Einzelbilder} = 4{,}60 \qquad \bar{\sigma}_{Differenz\,von\,je\,zwei\,Bildern} = 6{,}48 \qquad \sqrt{2} \cdot \bar{\sigma}_{Einzelbilder} = 6{,}50$$

Graustufen. Analog zu oben kann keine nennenswerte Korrelation festgestellt werden.

Der Dunkelstrom schwankt daher in guter Näherung statistisch, wobei die Standardabweichung von der Temperatur des CCD-Chips und der Belichtungszeit abhängt.

Da der Mittelwert des Dunkelstroms über die ca. 48500 dazu bestimmten Pixel des Chips [Dat] in guter Näherung gleich dem Mittelwert des Dunkelstroms über die gesamte aktive Fläche des CCDs sein sollte, bekommen wir bei einem Dunkelbild nur noch die Schwankungen des Dunkelstroms, das elektronische Rauschen und den Offset der in Kapitel 4.1.1 eingeführt wurde. Der Mittelwert der obigen Aufnahmen sollte also dem Offset entsprechen. Dieser beträgt 99,2597 Graustufen. Den so bestimmten Offset können wir also vom Untergrundbild (Gleichung 9) abziehen. Damit ergibt sich

$$M''(px) := M(px) - M_{sm}(px) - M_u(px) + 99{,}2597$$
$$= \left[\int s(\lambda)\cdot\eta(\lambda)\, d\lambda\right](px)\cdot h(px) \quad (12)$$

Dies kann wieder durch die Homogenität geteilt werden, sodass nur noch das Integral über die wellenlängenabhängigen Terme stehen bleibt.

*Quanteneffizienz*

Nach den vorhergegangenen Korrekturen besteht unser Bild nur noch aus dem Integral des Messsignals multipliziert mit der Quanteneffizienz über die Wellenlänge.

Sowohl am RMP, als auch bei allen Experimenten zur Charakterisierung der Komponenten, die anhand der Kamera durchgeführt wurden, wird eine rote Lichtquelle mit Maximum bei (630±0,5)nm und einer FWHM von ca. 13,6nm benutzt (siehe Kapitel 4.3.2). Am RMP wird diese in naher Zukunft durch eine Lichtquelle mit Maximum nahe 623nm und FWHM von etwa 14nm ersetzt[14]. Zwischen den Grenzen der FWHM ändert sich die Quanteneffizienz um 7,6% bzw. 6,5% bei der neuen Lichtquelle. Die Quanteneffizienz kann daher, zumindest auf erste Sicht, nicht vernachlässigt werden.

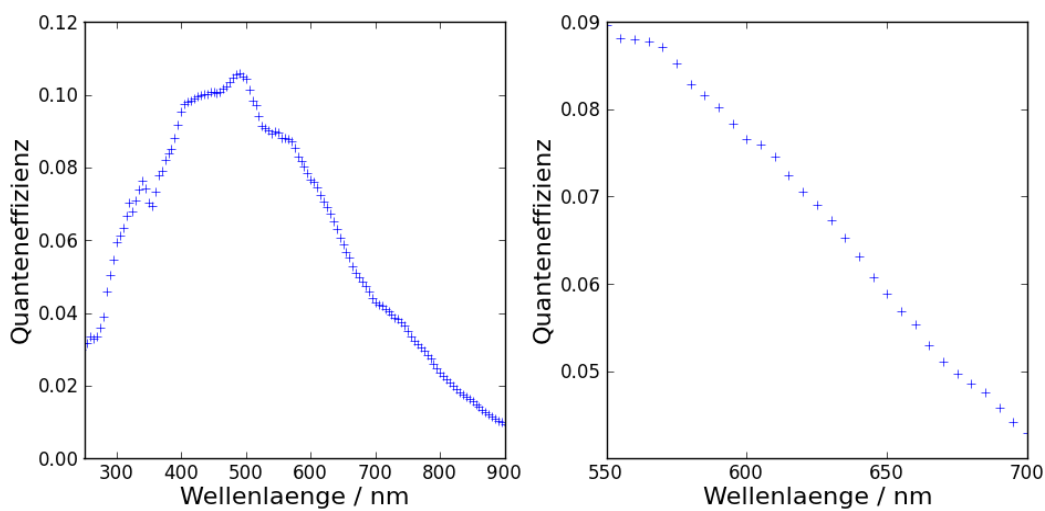

**Abbildung 19: Quanteneffizienz der Kamera in Abhängigkeit der Wellenlänge**

---

[14] Herstellerangaben zur Diode. Aus vertraulichkeitsgründen sind keine näheren Angaben möglich.

Bei unseren Auswertungen wird oft nicht das alleinige Messbild benutzt, sondern dieses durch ein sogenanntes Hellbild geteilt. Das Hellbild ist eine zusätzliche Aufnahme, die möglichst unter selben Bedingungen wie das Messbild durchgeführt wird, außer dass dabei die Probe entfernt wird. Auch sie wird mit einem Untergrundbild und einem Smearbild korrigiert. Wir berechnen somit das Verhältnis

$$\frac{M''_{mess}(px)}{M''_{hell}(px)} = \frac{\int s_{mess}(\lambda) \cdot \eta(\lambda)\, d\lambda}{\int s_{hell}(\lambda) \cdot \eta(\lambda)\, d\lambda} \tag{13}$$

Können wir nun annehmen, dass die wellenlängenabhängigen Anteile von $s_{mess}(\lambda)$ und $s_{hell}(\lambda)$ identisch sind, so kürzen sich beide Integrale über die Wellenlängenabhängigkeiten heraus und es bleibt nur noch das Verhältnis der nicht-wellenlängenabhängigen Anteile stehen.

Diese Annahme ist bei einfachen Experimenten oft in guter Näherung gegeben, da dieselbe Lichtquelle für Hell- und Messbild eingesetzt wird. Am RMP wird jedoch sowohl die Lichtquelle leicht verändert (siehe Kapitel 4.3.2), wie auch bei den 2 Messungen unterschiedliche Spiegel in den Strahlengang eingefügt. Damit sind die Wellenlängenabhängigkeiten nicht mehr dieselben und kürzen sich nicht mehr heraus. Aufgrund der starken Änderung der Quanteneffizienz kann diese Abhängigkeit auch nicht ohne vorherige Untersuchung vernachlässigt werden. Dies ist der Grund dafür, wieso bei der Charakterisierung weiterer Elemente des Messplatzes sowie im Kapitel 5.2 immer wieder Wellenlängenabhängigkeiten untersucht werden.

*Statistischer Fehler*

Die Kamera liefert bei jeder Aufnahme ein Messbild bei dem jedem Pixel ein Grauwert zugewiesen wird. Nun ist jedoch die Frage mit welchem Fehler diese Grauwerte behaftet sind. Es gibt mehrere Fehlerquellen: Die Schwankungen des Dunkelstroms und das elektronische Rauschen, von denen wir bereits im Abschnitt „Dunkelstrom und Offset" gesehen haben, dass sie sich statistisch verhalten, und der Photoeffekt selbst, der mit einer Wahrscheinlichkeit behaftet ist, weshalb auch die Anzahl der angeregten Elektronen bei selber Bestrahlungsstärke schwanken kann.

Beim Photoeffekt handelt es sich um ein Bernouilli-Experiment: Das Elektron kann nur angeregt werden oder nicht angeregt werden. Die Quanteneffizienz für den Photoeffekt liegt bei dem von uns benutzten Wellenlängenbereich ca. bei 0,07. Die Wahrscheinlichkeit für den Übergang eines Elektrons vom Valenz- in das Leitungsband aufgrund des Photoeffekts ist somit gering. Bei fester Bestrahlungsstärke und Belichtungszeit ist die Anzahl der angeregten Elektronen durch eine Wahrscheinlichkeitsverteilung gegeben. Wir erwarten aufgrund der geringen Wahrscheinlichkeit für die Anregung, dass diese durch eine Poisson-Verteilung beschrieben werden kann (siehe [Art65], Kap. 4). Die Standardabweichung der Anzahl angeregter Elektronen sollte daher durch die Wurzel dieser Anzahl gegeben sein. Nun liegt bei Aufnahmen jedoch nicht die Anzahl der Elektronen, sondern nur der Grauwert vor. Hier kommt ein Konversionsfaktor zum Tragen, der laut Hersteller 2,14814 Elektronen für eine Graustufe beträgt [Dat]. In den Grauwerten ist, wie im Abschnitt „Dunkelstrom und Offset" dieses Kapitels bereits erwähnt, noch der künstliche Offset enthalten. Als Offset wurden 99,26 Graustufen bestimmt. Es gilt also

$$\sigma_G^{photoeffekt} = \frac{\sigma_{N_{e^-}}}{2{,}14814} = \frac{\sqrt{N_{e^-}}}{2{,}14814} = \frac{\sqrt{N_G - 99{,}26}}{\sqrt{2{,}14814}} \text{ Graustufen} \tag{14}$$

wobei $N_{e^-}$ die Anzahl an angeregten Elektronen und $N_G$ die Anzahl an Graustufen ist.

Neben den Schwankungen aufgrund des Photoeffekts kommt es noch zu den Schwankungen aufgrund des elektronischen Rauschens und des Dunkelstroms. Die Schwankungen des Dunkelstroms skalieren mit dem Dunkelstrom selbst und hängen damit von der Temperatur und der Belichtungszeit ab. Im Prinzip können diese Schwankungen auf experimentellem oder auf theoretischem Weg bestimmt werden.

Theoretisch wäre es möglich diese in Abhängigkeit der Temperatur genau zu bestimmen. Es handelt sich bei der thermischen Anregung von Elektronen wieder um ein Bernouilli-Experiment mit niedriger Erfolgswahrscheinlichkeit, wodurch auch hier die Standardabweichung der thermischen Anregung durch die Wurzel der Anzahl der Anregungen gegeben ist. Man benötigt ebenfalls wieder den Konversionsfaktor. Die Temperaturabhängigkeit der Besetzungszahlen ist durch die Bolzmannverteilung gegeben [Bah00 p. 73] [15]. Führt man nun die Messung für mehrere Temperaturen durch, kann man die fehlenden Parameter, Gesamtzahl der Elektronen und Entartungsgrade, per Fit bestimmen. Damit bekommen wir für eine feste Belichtungszeit die Anzahl der thermischen Anregungen in Abhängigkeit der Temperatur und daher ebenfalls die Standardabweichung dieser. Die Anzahl der Anregungen steigt, unter Annahme des thermodynamischen Gleichgewichts, linear mit der Belichtungszeit. Damit wäre die Standardabweichung durch den Dunkelstrom in Abhängigkeit der Temperatur und der Belichtungszeit bestimmt. Es müsste bei einer solchen Bestimmung noch das elektronisches Rauschen berücksichtigt werden, das jedoch unabhängig von Belichtungszeit und Temperatur ist und somit durch einen weiteren konstanten Fitparameter beschrieben werden kann.

Am RMP wird jedoch der einfachere Weg gegangen, indem diese Schwankungen bei jeder Änderung der Messbedingungen experimentell bestimmt werden, wie es in Abschnitt „Dunkelstrom und Offset" bereits für die dortigen Bedingungen getan wurde.

Aus der vorherigen Herleitung folgt zusammen mit Gleichung 14, dass die Standardabweichung eines Grauwerts durch

$$\sqrt{\frac{N_G - 99{,}26}{2{,}15} + \sigma^2_{Dunkelstrom+elektronisches\ Rauschen}(T,t)} \qquad (15)$$

gegeben sein sollte. Dies soll hier überprüft werden.

Dazu wurden für 6 Belichtungszeiten jeweils 20 Bilder in einer Messreihe aufgenommen. Die Messung wurde bei einer Chiptemperatur von 1℃ durchgeführt.[16] Für jede dieser Serien an 20 Bildern wurden die Bilder pixelweise gemittelt. Für jedes der 6 gemittelten Bilder wurden die Pixel, deren Grauwerte identisch sind, zusammengefasst. Aus den zwanzig Mal so vielen ungemittelten Grauwerten dieser ausgewählten Pixel wurde dann die Standardabweichung berechnet. So wird jedem Grauwert der in einem gemittelten Bild vorkommt eine Standardabweichung zugewiesen. Die Standardabweichungen wurden gegen den zugehörigen Mittelwert aufgetragen (siehe Abbildung 20).

---

[15] Hierbei wird vernachlässigt, dass die bei der Anregung der Elektronen erzeugten Löcher durch ein Potential abgeführt werden.

[16] Zum Zeitpunkt der Messung war ein Betrieb auf -14℃ aufgrund eines Kameradefekts nicht möglich.

Aufgrund der eingeschränkten Stichprobe zur Berechnung der Standardabweichung sind die so berechneten Standardabweichungen fehlerbehaftet. Dieser Fehler ist durch $\sigma/[2 \cdot (N-1)]$ gegeben, wobei N die Anzahl der zur Berechnung der Standardabweichung benutzten Pixel ist und $\sigma$ der fehlerbehaftete Wert selbst [Squ01 p. 22]. Dieser Fehler ist ebenfalls in Abbildung 20 zu sehen. Grauwerte nahe dem mittleren Grauwert einer Messreihe kommen öfters vor. Daher fließen mehr Grauwerte in die Berechnung der Standardabweichung und diese können genauer bestimmt werden als die der Extremwerte einer Messreihe.

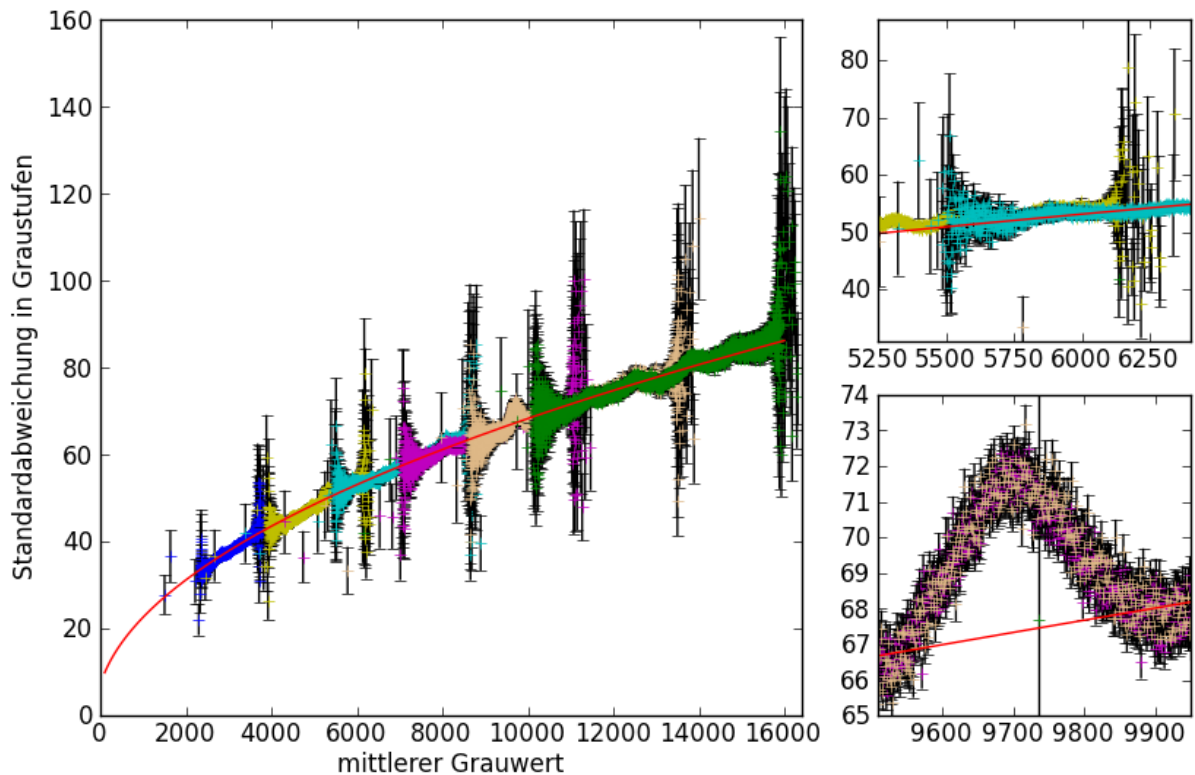

**Abbildung 20: Standardabweichung der Grauwerte von Pixeln bei selber Bestrahlungsstärke in Abhängigkeit des mittleren Grauwerts der Pixel. Die verschiedenen Farben entsprechen den 6 Messreihen. Die Fehler sind durch $\sigma/[2 \cdot (N-1)]$ gegeben [Squ01]. Die rote Linie stellt den Fit dar. links: Übersicht; rechts oben: Detailansicht eines Übergangsbereichs zwischen zwei Messreihen; rechts unten: Detailansicht einer systematischen Abweichung von der Theorie.**

Diese Daten wurden mit der oben theoretisch hergeleiteten Funktion unter Berücksichtigung der Fehler gefittet, wobei der Konversionsfaktor und $\sigma_{Dunkelstrom+elektronisches\ Rauschen}(T,t)$ freigegeben wurden. Daraus ergibt sich ein Konversionsfaktor von 2,164±0,001 Graustufen$^{-1}$ und $\sigma_{Dunkelstrom+elektronisches\ Rauschen}(T,t) = 9,88 \pm 0,06$ Graustufen mit einem $\chi^2/ndf$ von 4,98.

Der Konversionsfaktor deckt sich nicht im Rahmen des Fehlers mit dem vom Hersteller angegebenen Wert von 2,14814. Seine relative Abweichung ist jedoch gering. Der höhere Offset als im Abschnitt „Dunkelstrom und Offset" resultiert vermutlich nicht nur aus der höheren Chiptemperatur. Zusätzliche Schwankungen der Lichtquelle zwischen den Bildern einer Messreihe erhöhen hier diesen Wert. Das $\chi^2/ndf$ weist darauf hin, dass das Modell den Verlauf noch nicht vollkommen beschreibt. So sind für mehrere Grauwerte Abweichungen von dem vorhergesagten Verlauf erkennbar, die nicht durch statistische Schwankungen erklärbar sind. Besonders gut erkennbar ist dies, wie in Abbildung 20, rechts unten zu sehen, für Grauwerte im Bereich um 9700 Graustufen. Der Ursprung dieser

Abweichungen ist nicht bekannt, beträgt jedoch maximal 5 Graustufen und betrifft nur wenige Grauwerte. Daher wurde er nicht weiter untersucht.

Trotz kleiner Abweichungen vom Modell (Gleichung 15) für manche Grauwerte und leicht abweichendem Konversionsfaktor konnte dieses Modell validiert werden. Die obige Formel wird daher bei den folgenden Charakterisierungen und am RMP zur Beschreibung des statistischen Rauschens der Kamera eingesetzt. Der Term $\sigma_{Dunkelstrom+elektronisches\,Rauschen}(T,t)$ wird am RMP durch eine getrennte Aufnahme bestimmt.

*Beugung an den Pixelkanten?*

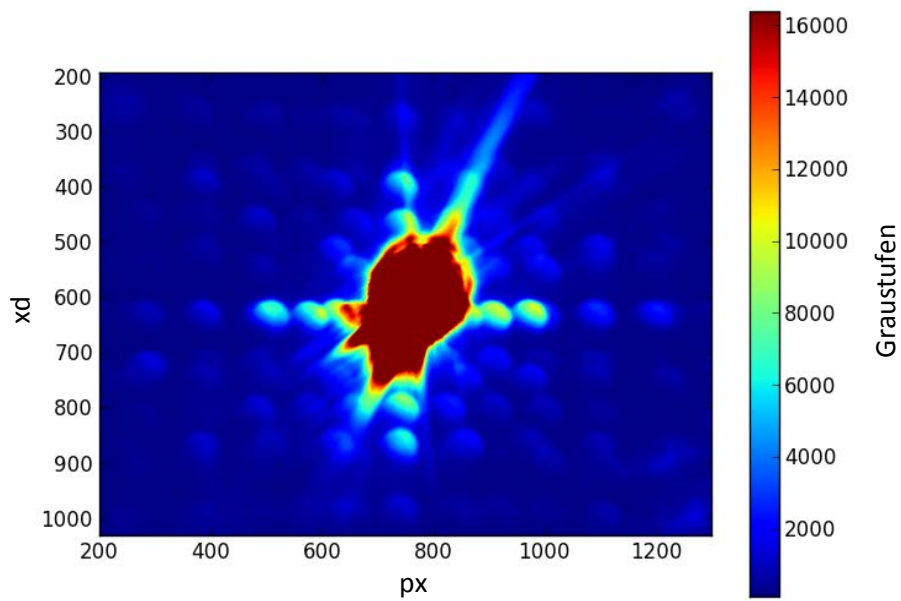

**Abbildung 21: lokale Grauwerterhöhungen bei Bestrahlung des CCDs mit parallelem Licht hoher Intensität**

Trifft an einer Stelle des Chips paralleles Licht hoher Intensität auf, so treten, insbesondere in Richtung der Zeilen und der Spalten des CCDs neben der beleuchteten Stelle lokale Grauwerterhöhungen auf, die in Abbildung 21 gezeigt sind. Hier wurde das Bild ca. 100-fach überbelichtet um diesen Effekt besser sichtbar zu machen. Der Ursprung dieser lokalen Erhöhungen ist nicht endgültig geklärt. Die Anordnung hauptsächlich entlang waagrechten und senkrechten Linien lässt vermuten, dass der CCD mit seiner regelmäßigen Pixelbreite als Reflexionsgitter fungieren könnte. Das Licht würde dann durch die Vorder- oder Rückseite der Abdeckscheibe des Chips oder der Gaszelle in der sich der Chip befindet wieder auf den CCD zurückgespiegelt. Eine grobe Berechnung ergab, dass der Abstand der Beugungsmaxima durch einen solchen Effekt in derselben Größenordnung liegen würde, wie der hier beobachtete Abstand zwischen Stellen mit höherer Bestrahlungsstärke.

Bei einem nicht überbelichteten Bild erreichen diese Überhöhungen bis zu 10 Graustufen außerhalb des Spots selbst. Allerdings ist nicht bekannt wie hoch dieser Effekt im Bereich des Spots selbst ist. Leider ist dieser Effekt ohne ein gutes Verständnis des Vorgangs und einer zeitgleichen Kenntnis der Reflektivitäten der einzelnen Abdeckscheiben, des Einfallwinkels und der Strahldivergenz nicht korrigierbar.

Fassen wir die Effekte der Kamera nochmals zusammen ergibt sich Folgendes:

Das Messbild ist gegeben durch (Gleichung 5)

$$\left\{\left(\left[\int (s(\lambda)+u(\lambda))\cdot\eta(\lambda)\ d\lambda\right](px)\cdot h(px)\right)+sm(px)+off\right\}\cdot l(G)$$

| | |
|---|---|
| **Linearität** $l(G)$ | Es gibt drei sich teils widersprechende Messungen zur Charakterisierung der Linearität. Aus allen resultiert jedoch, dass die Linearität in einem zentralen Grauwertebereich vernachlässigt werden kann. |
| **Offset** *off* | Der Offset wurde durch eine Dunkelmessung bestimmt und kann wo Bedarf ist abgezogen werden. |
| **Smear** *sm(px)* | Am RMP ist die Bestrahlungsstärke so gering, dass der Smear vernachlässigt werden kann (Belichtungszeiten groß gegenüber $t_{zyklus} \approx 0{,}2$s). Wo dies nicht der Fall ist werden Aufnahmen mit 500ns Belichtungszeit zur Korrektur des Smears gemacht. |
| **Homogenität** *h(px)* | Zur Korrektur der Homogenität der CCD Sensitivität wird eine Messung von T. Hornung benutzt. |
| **Streulicht** *u(λ,px)* | Das Streulicht wird durch eine Untergrundsaufnahme korrigiert, bei der eine Abschattung im direkten Strahlengang platziert und die Belichtungszeit gleich wie bei der eigentlichen Aufnahme gewählt wird. |
| **Quanteneffizienz** $\eta(\lambda)$ | Zur Quanteneffizienz sind Herstellerangaben vorhanden. |
| **Beugungserscheinungen** | Außerhalb des Fokus sind die Beugungserscheinungen vernachlässigbar, innerhalb unbekannt. Sie können nicht korrigiert werden. |
| **Statistischer Fehler** $\sigma_s$ und $\sigma_u$ | Der statistische Fehler auf die Grauwerte wurde theoretisch beschrieben und die Theorie experimentell bestätigt. |

## 4.2. Objektiv

Während wir zur Charakterisierung des Umlenkspiegels und des Hellbildspiegels die Kamera allein benutzen, wurde zur Charakterisierung des Diffusors und wird beim RMP die Kamera in Kombination mit einem Objektiv eingesetzt. Wir werden uns hier nicht mit der genauen Zusammensetzung des Objektivs befassen, die nicht bedeutend zum Verständnis beiträgt, sondern uns mit drei Eigenschaften des Objektivs befassen: dem Randlichtabfall, Verzeichnungen und der Point-spread-function. Eine genauere Behandlung dieser Effekte und der Defokussierung ist in [Ghe11] Kapitel 2.3 zu finden.

### 4.2.1. Randlichtabfall

Der Randlichtabfall eines Objektivs kommt daher zustande, dass für jeden Punkt der Objektebene die Eintrittsapertur des Objektivs einen unterschiedlichen Raumwinkel einnimmt. Die Intensität des jeweils zugehörigen Bildpunktes ist das Integral der Strahlstärke des Objektpunktes über den entsprechenden Raumwinkel [Haf03 pp. 176f, 298f]. Damit hängt der Randlichtabfall von mehreren Parametern ab: von der relativen Verteilung der Strahlstärke über den Abstrahlwinkel und vom Raumwinkel, also vom Durchmesser der Eintrittsapertur b und der Gegenstandsweite $s_1$. Dass der Raumwinkel je nach Ort des Objektpunktes ein anderer ist, kann anhand der zweidimensionalen Vereinfachung in Abbildung 22 links verstanden werden. Hier ist der Winkel

$$\epsilon = \tan^{-1}\left(\frac{b/2 - x}{s_1}\right) + \tan^{-1}\left(\frac{b/2 + x}{s_1}\right) \tag{16}$$

und wird maximiert für x=0. Aufgrund des Verlaufs der Strahlstärke in Abhängigkeit des Abstrahlwinkels ist es auch nicht egal welche Abstrahlwinkel auf die Eintrittsapertur treffen. In der Regel sinkt die Strahlstärke mit dem Abstrahlwinkel. Mit steigendem Abstand x des Objektpunktes zur der optischen Achse kommen größere Abstrahlwinkel vor. In Abbildung 22 Mitte ist dies dargestellt. Die Winkelspektren der zwei Punkte unterscheiden sich nur durch die gestrichelten Anteile. Auch dies führt zu einem Abfall der Intensität der Bildpunkte zum Rand des CCDs hin.

Im Extremfall einer quasi-parallelen Bestrahlung, wie es beispielsweise bei den Hellbildaufnahmen zur Charakterisierung des Diffusors der Fall ist, treten beide der eben genannten Effekte bis zu Werten von x nahe b/2 gar nicht auf, da nicht die Eintrittsapertur, sondern die Divergenz der Strahlung den Raumwinkel der auf die Linse trifft beschränkt (siehe Abbildung 22).

Der Randlichtabfall kann also nicht generell angegeben werden, sondern sieht je nach Abstrahlcharakteristik unterschiedlich aus. Wir werden den Randlichtabfall daher an gegebenen Stellen erneut kurz für den jeweiligen Fall diskutieren.

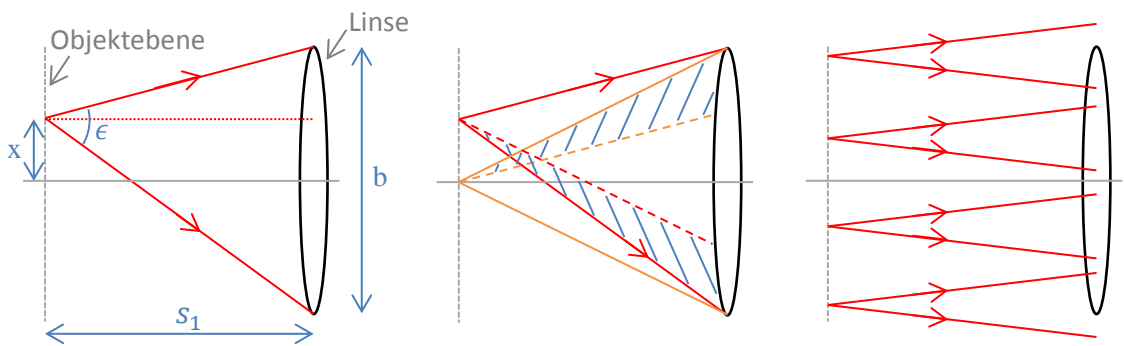

**Abbildung 22: Ursprung des Randlichtabfalls. Links: Änderung des Raumwinkels mit dem Abstand des Objektpunktes zur optischen Achse. Mitte: Unterschiedliche Abstrahlwinkel führen zu unterschiedlichen Strahlstärken. Rechts: Bei schwacher Divergenz tritt der Randlichtabfall nur am äußersten Rand auf.**

### 4.2.2. Verzeichnung

Bei der Abbildung anhand eines Objektivs kann es im Bild zu Gradienten im Abbildungsmaßstab kommen. Diese werden in zwei Kategorien eingeteilt: radiale und tangentiale Verzeichnungen. Bei tangentialer Verzeichnung ist bei selben Abstand zweier Objektpunkte zur optischen Achse der Abbildungsmaßstab unterschiedlich. Sie resultiert hauptsächlich von Zentrierfehlern der Linsen. Da diese Verzeichnungen jedoch typischerweise um eine Größenordnung kleiner sind als radiale Verzeichnungen, werden sie am RMP nicht untersucht werden. [Kar04 p. 52]

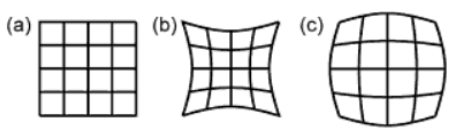

**Abbildung 23: (a) ohne Verzeichnung, (b) kissenförmige Verzeichnung, (c) tonnenförmige Verzeichnung. Quelle: [Küh07]**

Radiale Verzeichnungen sind Änderungen des Abbildungsmaßstabs mit dem Abstand des Objektpunktes zur optischen Achse. Bei mit dem Abstand zur optischen Achse wachsendem Abbildungsmaßstab ist von kissenförmiger Verzeichnung, bei fallendem Abbildungsmaßstab von tonnenförmiger Verzeichnung die Rede (siehe Abbildung 23). Die radiale Verzeichnung entsteht durch das Einbringen von Blenden in den Strahlengang [Ped08 p. 134ff], in Kombination mit der Bildfeldwölbung. Die Bildfeldwölbung ist die Verformung der Bildebene zu einer gekrümmten Fläche aufgrund unterschiedlich starker Brechung von Lichtstrahlen, die unter unterschiedlichen Winkeln zur optischen Achse auf die Linse treffen. [Dem09 p. 291f]

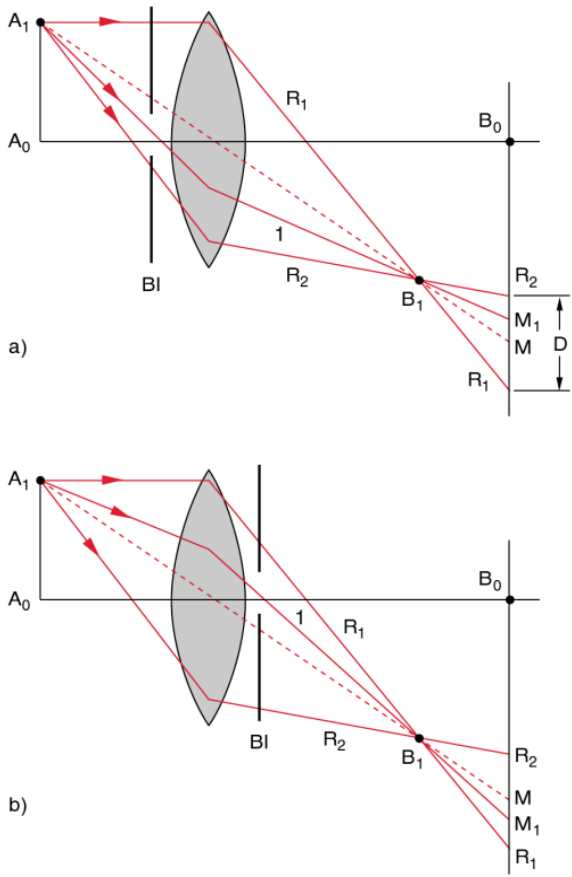

Abbildung 24: Abhängigkeit der Verzeichnung von der Position der Blende. Quelle: [Dem09]

Eine vor oder hinter eine Linse eingebrachte Blende lässt je nach Position Strahlen unter unterschiedlichen Winkeln passieren (siehe Abbildung 24). Wird das Bild trotz Bildfeldwölbung auf einer planen Fläche aufgenommen, wie bei der Kamera auf dem CCD, so sind die Bildpunkte mit steigendem Abstand zur optischen Achse zunehmend defokussiert. Zwischen dem eigentlichen Bildpunkt und dem Sensor breiten sich die Strahlen weiter geradlinig aus. Ihr Auftreffpunkt hängt somit von ihrer Ausbreitungsrichtung und dadurch von der Position der Blende ab.

Mathematisch kann die Position $(x_v, y_v)$ von jedem Bildpunkt eines verzeichneten Bildes in Abhängigkeit der Position $(x, y)$ bei Abwesenheit von Verzeichnung und des Abstands $r$ zur optischen Achse ausgedrückt werden. Dies geschieht in Form der Potenzreihe

$$\begin{pmatrix} x_v \\ y_v \end{pmatrix} = \begin{pmatrix} x \\ y \end{pmatrix} + \sum_k \alpha_i \cdot r^{2k} \begin{pmatrix} x - o_x \\ y - o_y \end{pmatrix} \quad (17)$$

wobei $(o_x, o_y)$ die Position des Schnittpunktes der optischen Achse mit der Bildebene ist. Die ungeraden Potenzen von r können aufgrund der Symmetrie um die optische Achse nicht auftreten. [Bey12 p. 165f] [Bra08]

Durch die Verzeichnung sind die von der Kamera jedem Pixel zugewiesenen Grauwerte dem falschen Ort zugewiesen. Die Verzeichnung kann gemessen werden indem ein quadratisches Gitter fotografiert und die Krümmung der Gitterlinien analysiert wird. Anhand obiger Formel kann so jedem Punkt seine korrigiert Position zugewiesen werden. Um wieder eine quadratische Anordnung an Grauwerten zu bekommen wird anschließend zwischen den in ihrer Position korrigierten gemessenen Grauwerten interpoliert.

Aus Zeitgründen wurde die Vermessung der Verzeichnung noch nicht realisiert wodurch am RMP derzeit die Verzeichnung noch vernachlässigt wird. Um die Verzeichnung zu minimieren wird möglichst die Mitte des CCD-Chips benutzt, die in etwa auf der optischen Achse des Objektivs liegt.

### 4.2.3. Point-Spread-Function

Wird ein Punkt anhand eines Objektivs abgebildet, so ist das Ergebnis nur in erster Näherung wieder ein Punkt. Das Bild des Punkes wird Point-Spread-Function (PSF) genannt und ist die Funktion mit der das Objekt gefaltet werden muss, um das vom Objektiv erzeugte Bild (ohne Verzeichnung und Randlichtabfall) zu bekommen, solange die PSF unabhängig von der Position in der Objektebene ist, was oft in guter Näherung gilt.

Die PSF entsteht aus mehreren verschiedenen Effekten im Objektiv. Einerseits sorgen Abbildungsfehler der Linsen für ein Verwaschen des abgebildeten Punktes [Ped08 p. 747]. Andererseits sorgt die beschränkende Aperturblende im Objektiv für weitere Erscheinungen. Diese sind zum einen Beugungseffekte, aber vor allem ein Beschnitt $g(\vec{k})$ der Fouriertransformierten des Objekts $\mathcal{F}\{f(\vec{r})\}$. Dadurch ergibt sich laut Abbe'scher Bildentstehung für das Bild $\mathcal{F}^{-1}\{\mathcal{F}\{f(\vec{r})\} \cdot g(\vec{k})\}$ (siehe [Gro05], Kap.21). Da das Objektiv rund ist, handelt es sich bei $g(\vec{k})$ um eine Kreisblende und somit bei der Fouriertransformierten um eine Airy-Funktion.

Es gibt viele Näherungsmodelle zur PSF-Berechnung; es seien insbesondere die Modelle von Born & Wolf und das simulated defocus Modell genannt, die beide Defokussierungen berücksichtigen [Kir12]. Uns fehlen wichtige Informationen, wie der Durchmesser der Aperturblende und die zugehörige Brennweite, um die Berechnungen nach diesen Verfahren vorzunehmen. Da die Breite der PSF eher gering ist verglichen mit der Strahlaufweitung durch den Diffusor (siehe Kapitel 4.6.1), die ebenfalls durch eine Faltung beschrieben wird, ist der genaue Verlauf der PSF von nachgeordneter Bedeutung und es wird kein großer Fehler gemacht, indem die PSF durch eine einfache zweidimensionale Gaußverteilung angenähert wird.

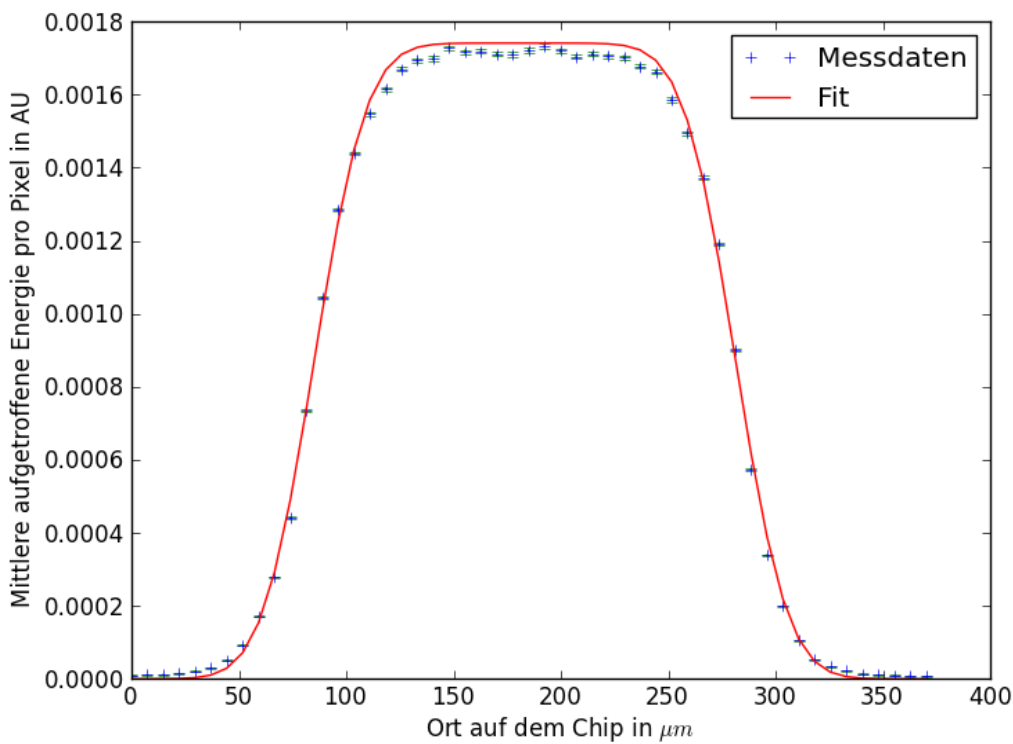

**Abbildung 25: Schnitt durch den zweidimensionalen Fit der PSF**

Zur Bestimmung der PSF wurde daher ein Pinhole mit Durchmesser 100µm anhand einer LED, homogenisiert durch einen Diffusor, beleuchtet und fotografiert. Um das statistische Rauschen zu minimieren wurden 5 Aufnahmen gemittelt. Das Pinhole kann durch eine circ-Funktion, also eine rotationssymmetrische Rechteckfunktion, beschrieben werden. Um die Breite der PSF zu bestimmen wurde ein Fit unter Berücksichtigung des statistischen Fehlers der Kamera durchgeführt, wobei als Fitfunktion die Faltung der zweidimensionalen Gauß-Funktion und der circ-Funktion benutzt wurde. Dabei wurden die Standardabweichung des Gauß sowie aufgrund der Ungenauigkeit des Radiuses der Blende der Radius der circ-funktion als Fitparameter freigegeben. Ein Schnitt ist in Abbildung 25 zu sehen. Aufgrund der Annahmen zur PSF beschreibt diese Funktion nicht perfekt den Verlauf des Messbildes, jedoch für unseren Zweck mit einer ausreichenden Genauigkeit. Aus dem Fit ergibt sich für den Gauß eine Standardabweichung von 9,53±0,05µm, bzw. eine FWHM von 22,4±0,1µm.

## 4.3. Lichtquelle

Die Lichtquelle ist derzeit im Umbau und die später am RMP verbaute Lichtquelle ist noch nicht fertiggestellt. Sie wird im Rahmen des Umbaus getrennt charakterisiert. Dabei werden für den RMP die Homogenität der Beleuchtung, die thermische Stabilität, die Strahldivergenz und das Spektrum der Beleuchtung von Bedeutung sein.

Da jedoch bereits vor der Installation der neuen Lichtquelle erste Messungen am RMP durchgeführt werden, wurden einige Eigenschaften der bisherigen Lichtquelle vermessen. Die bisherige, wie auch die neue Lichtquelle, wurden mit dem Ziel konzipiert, eine möglichst homogene Beleuchtung zu liefern und dabei eine sonnenähnliche Divergenz aufzuweisen. Diese Eigenschaften der alten Lichtquelle wurden bereits zuvor verifiziert und im Rahmen dieser Arbeit nicht nochmals überprüft. Dagegen ist dem Spektrum der Lichtquelle bei ihrer bisherigen Nutzung keine große Bedeutung zugekommen. Weiter wurden bei Messungen, die unter dieser Beleuchtung stattgefunden haben, viel kürzere Belichtungszeiten eingesetzt als aktuell am RMP, sodass andere Aspekte der thermischen Stabilität interessant waren. Es soll daher auf diese zwei Punkte eingegangen werden.

### 4.3.1. Thermische Stabilität

Die thermische Stabilität der Lichtquelle kann durch die Kombination von zwei Effekten beschrieben werden: ein langsamer systematischer Drift und ein schnelles statistisches Schwanken der Intensität der Lichtquelle.

Um dies zu messen wurde die Kamera direkt mit der Lichtquelle des RMPs bei maximaler Bestromung beleuchtet und unter den gewöhnlichen Einstellungen der Kamera alle 16 Sekunden eine Aufnahme gemacht. Vor dem ersten Bild wurde ca. 20 Minuten gewartet, was in etwa der Situation am RMP entspricht. Die Grauwerte jedes der Bilder wurden anschließend aufsummiert und, wie in Abbildung 26 zu sehen, gegen den jeweiligen Zeitpunkt ihrer Aufnahme aufgetragen.

Der Drift der Lichtquelle hat keine bekannte Form und ein mögliches physikalisch motiviertes Modell für seine Beschreibung ist nicht von Interesse. Um die Effekte des Drifts und der statistischen Schwankungen zu entkoppeln, wird ein Fit mit einem Polynom durchgeführt. Das Ergebnis ist ebenfalls in Abbildung 26 zu sehen. Es wird dann die Standardabweichung aus den Abweichungen zu diesem Fit berechnet. Dabei ist die Anzahl der Ordnungen beim Fit von Bedeutung: Bei zu vielen Ordnungen würde der Fit dem statistischen Rauschen folgen. Hierbei handelt es sich um kein exaktes Verfahren, da die Anzahl der benutzten Ordnungen mit bloßem Auge bestimmt werden muss. Eine quantitative Bestimmung bei der die Zahl der Ordnungen so optimiert werden würde, dass $\chi^2/ndf$ möglichst nahe 1 liegt ist nicht möglich. Sie würde zur Berechnung von $\chi^2$ die Standardabweichung benötigen, die wir zu ermitteln versuchen. Trotz dieser Approximation bleibt dieses Verfahren genauer, als das Bilden einer Standardabweichung unter Vernachlässigung des Drifts in einem kleinen Zeitintervall. Anhand dieses Verfahrens bekommen wir eine Standardabweichung von $(1{,}627\pm0{,}007)\cdot10^{-4}$ %. Dies ist auch ohne zeitliche Mittelung weitaus geringer als die Genauigkeit die für den RMP angestrebt wird und kann vernachlässigt werden.

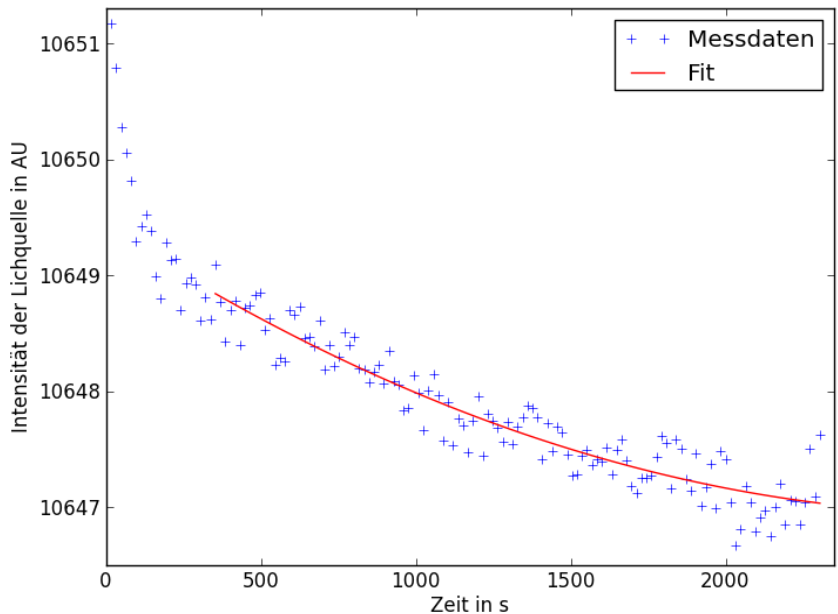

Abbildung 26: thermische Stabilität der aktuellen Lichtquelle am RMP

Neben den statistischen Schwankungen treten systematische Zu- oder Abnahmen der Intensität mit der Laufzeit und der Raumtemperatur auf. Mittelfristig soll bei der neuen Lichtquelle die Intensität der Lichtquelle während der Belichtungszeit mitgemessen werden. Die Integration der Strahlungsleistung während der Belichtungszeit ergibt einen zur Bestrahlungsstärke proportionalen Wert. Somit werden Drifts der Lichtquelle keine Bedeutung mehr zukommen, da die Bilder durch dieses Integral dividiert werden können. Aufgrund der geringen Lichtstärke der bisherigen Lichtquelle beträgt die Belichtungszeit beim RMP ca. 40s für das Messbild, bzw. 40min für das Hellbild. Aus Abbildung 26 kann die Größenordnung des Drifts abgelesen werden. Dieser beträgt 0,04% über 40min. Die genaue Zahl hängt jedoch von Temperaturschwankungen im Raum und der Bestromung der LED ab. Bei nicht übermäßig von der hier durchgeführten Messung abweichenden Bedingungen[17] verursacht demnach auch der Drift der Lichtquelle keinen nennenswerten Fehler.

### 4.3.2. Spektrum

Da diverse Komponenten des RMPs wellenlängenabhängige Eigenschaften besitzen, ist es von Bedeutung zu wissen welches Spektrum die Lichtquelle besitzt. Daher wurde dieses anhand eines Spektrometers vermessen.

In der Lichtquelle wird die Lichtverteilung durch zwei Bauteile homogenisiert. Die zweite örtliche Homogenisierung findet anhand einer Diffusorscheibe statt. Da dieser die Strahlungsleistung in alle

---

[17] Die Raumtemperatur betrug ca. 296K

Raumrichtungen verteilt, lässt er die (homogenisierte) Bestrahlungsstärke am Ausgang der Lichtquelle sinken. Bei den Hellbildaufnahmen, also für Aufnahmen ohne konzentrierende Optik, wären mit Diffusorscheibe ca. 2 Tage Belichtungszeit erforderlich. Die Diffusorscheibe wird daher für das Hellbild ausgebaut, wodurch die Belichtungszeit auf 40min reduziert wird. Dabei war die Annahme, dass die Lichtquelle weiterhin als homogen angesehen werden kann. Dies ist letztendlich nicht der Fall, worauf wir im Kapitel 5.3 zurückkommen werden.

Da diese Diffusorscheibe das Wellenlängenspektrum der Lichtquelle verändern könnte, wurden die Spektren beider zum Einsatz kommenden Aufbauten der Lichtquelle vermessen. Um Rauschen und Abhängigkeiten des Spektrums vom Einkoppelwinkel in den Lichtleiter des Spektrometers zu mitteln, wurde der Lichtleiter 10 Mal neu ausgerichtet und je 10 Messungen des Spektrums vorgenommen. Die gemittelten normierten Spektren sind in Abbildung 27 zu sehen. Das Maximum liegt jeweils bei 630nm $\pm$ 0,5nm, die FWHM beträgt mit Diffusorscheibe 13,7$\pm$0,1nm, ohne Diffusorscheibe 13,5$\pm$0,1nm. Beide relative Verläufe der Spektren unterscheiden sich nur leicht, sodass kein großer Fehler gemacht werden würde, wenn sie in ihrem normierten Verlauf als identisch angenommen werden würde. Da jedoch beide relative Verläufe der Spektren präzise gemessen wurden, ist an dieser Stelle eine solche Näherung nicht nötig und im Folgenden werden beide Spektren weiterhin als unterschiedlich behandelt.

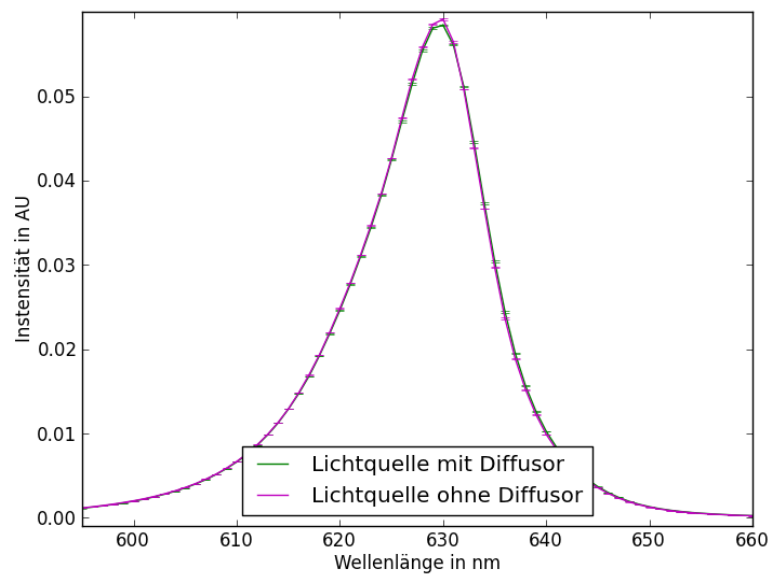

Abbildung 27: normiertes Wellenlängenspektrum der Lichtquelle mit und ohne Diffusor

Neben den relativen Verläufen brauchen wir noch die Amplitude der Spektren, bzw. das Amplitudenverhältnis der zwei Spektren. Deshalb wurde von T. Schmid anhand der Kamera ein Maß für die Abschwächung der Intensität durch den Diffusor bestimmt. Hier wurde die Kamera mit und ohne Diffusorscheibe beleuchtet und je mehrere Aufnahmen gemacht, die gemittelt wurden. Alle Grauwerte eines Bildes wurden anschließen aufsummiert. Das Intensitätsverhältnis der Lichtquelle ohne Diffusorscheibe zur Lichtquelle mit Diffusorscheibe, gewichtet mit der Quanteneffizienz der Kamera und über die Wellenlängen integriert, beträgt 65,53.

## 4.4. Spiegel

### 4.4.1. Einfallswinkelabhängigkeit der Reflektivität

Die benutzten Spiegel sind bis auf die Proben Multilayerspiegel. Die Beschichtung besteht aus einer metallischen Schicht mit komplexem Brechungsindex und mehreren dielektrischen Schichten. Trifft ein Lichtstrahl auf eine der Grenzflächen, so gilt das Snellius'sche Brechungsgesetz. Aus den Maxwellgleichungen an dieser Grenzfläche lassen sich dann die Fresnelsche Formeln

$$\left(\frac{E_{0_r}}{E_{0_i}}\right)_s = \frac{N_1 \cdot \cos\alpha - \frac{\mu_{r_1}}{\mu_{r_2}}\sqrt{N_2^2 - N_1^2 \cdot \sin^2\alpha}}{N_1 \cdot \cos\alpha + \frac{\mu_{r_1}}{\mu_{r_2}}\sqrt{N_2^2 - N_1^2 \cdot \sin^2\alpha}} \tag{18}$$

Und

$$\left(\frac{E_{0_r}}{E_{0_i}}\right)_p = \frac{N_2^2 \cdot \frac{\mu_{r_1}}{\mu_{r_2}} \cdot \cos\alpha - N_1 \cdot \sqrt{N_2^2 - N_1^2 \cdot \sin^2\alpha}}{N_2^2 \cdot \frac{\mu_{r_1}}{\mu_{r_2}} \cdot \cos\alpha + N_1 \cdot \sqrt{N_2^2 - N_1^2 \cdot \sin^2\alpha}} \tag{19}$$

herleiten. Dabei stehen die Indizes s und p für die Polarisationsrichtungen des einfallenden Lichtstrahls bezüglich zur Einfallsebene, r für „reflected" und i für „incident". $\mu_{r_j}$ ist jeweils die Permeabilität, $N_j$ der (komplexe) Brechungsindex der Schicht j, $E_0$ die Amplitude der elektrischen Feldstärke und α der Einfallswinkel.

Die Reflektivität ergibt sich für das am RMP benutzte unpolarisierte Licht

$$R_{unpolarisiert} = \frac{1}{2}\left[\left|\left(\frac{E_{0_r}}{E_{0_i}}\right)_p\right|^2 + \left|\left(\frac{E_{0_r}}{E_{0_i}}\right)_s\right|^2\right] \tag{20}$$

Damit ist die Reflektivität einfallswinkelabhängig. Wir müssen also die Einfallswinkelabhängigkeit berücksichtigen. (Herleitung und Formeln nach [Gro05 pp. 62-75])

Für den Kalibrierspiegel wurde die Reflektivitäten vom Hersteller berechnet und uns mitgeteilt. Sie ist in Abbildung 28 eingezeichnet. Dabei ist die Anzahl der berechneten Punkte sehr gering. Da die Einfallswinkel auf den Kalibriersspiegel (unter Benutzung der doppelten Brennweite für den Radius des sphärischen Kalibrierspiegels) maximal 0,12rad betragen, der Abstand zwischen zwei Punkten aber über 0,4 rad beträgt, müssen wir interpolieren. Prinzipiell könnten die genauen Fresnel'schen Formeln zur Interpolation mit Gleichung 20 benutzt werden, was jedoch einen großen Rechenaufwand bei der Berücksichtigung der Einfallswinkelabhängigkeit bei der späteren Simulation des RMPs (siehe Kapitel 5.3 und 5.4) bedeutet. Wir sehen anhand der berechneten Punkte, dass die Reflektivität in etwa quadratisch abnimmt (siehe fit in grün). Dieser Fit beschreibt schlecht den für uns interessanten Bereich. Daher sollte eine quadratische Interpolation zwischen den zwei ersten Punkten eine bessere Approximation liefern als der Fit und als eine lineare Interpolation. Dies wurde von T. Schmid mit der roten Kurve getan und das Ergebnis wird am RMP zur Korrektur der Winkelabhängigkeit benutzt. Innerhalb der 0,12 rad ändert sich die Reflektivität um 0,18 Prozentpunkte.

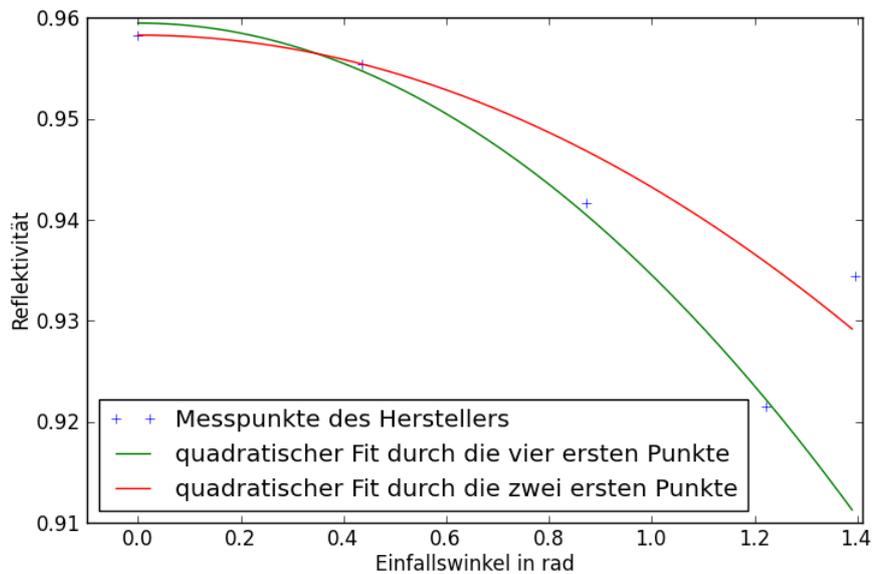

Abbildung 28: Einfallswinkelabhängigkeit der Reflektivität des Kalibrierspiegels

Der Umlenkspiegel besitzt dieselbe Beschichtung wie der Kalibrierspiegel. Allerdings kommen hier Winkel bis $\pi/4$ rad vor. Hier ändert sich die Reflektivität um ca. einen Prozentpunkt. Strahlt der Diffusor zur Betrachtung des Brennpunktes bei Hell- und Messbild mit derselben Streuwinkelverteilung ab, so kürzt sich die Winkelabhängigkeit des Spiegels heraus. Wir werden auf diese Annahme in Kapitel 5.1 zurückkommen.

Beim Hellbildspiegel wird ein einziger Einfallswinkel benutzt. Die in Abbildung 29 aufgetragene spektrale Reflektivität wurde vom Hersteller relativ vermessen und auf einen für diesen Einfallswinkel vom Hersteller berechneten Wert normiert.

Für die Proben muss eine extra Messung durchgeführt werden oder anhand von Literatur und Herstellerangaben zur Beschichtung die Einfallswinkelabhängigkeit berechnet werden.

### 4.4.2. Spektrale Abhängigkeit der Reflektivität

Aus den Fresnel'schen Formeln aus Kapitel 4.4.1 geht aufgrund der Dispersion des Brechungsindex hervor, dass die Reflektivität eines Spiegels von der einfallenden Wellenlänge abhängt. Dies bedeutet, dass sich am RMP das Spektrum der Beleuchtung der einzelnen Komponenten im Laufe der Reflexionen verändert. Erstens hängt damit der reflektierte Anteil des einfallenden Lichts von dessen Spektrum ab und somit davon, welche optischen Komponenten im Strahlengang vor dem Spiegel das Wellenlängenspektrum beeinflussen. So ist der Anteil des am Umlenkspiegel reflektierten Lichts für das Messbild und für das Hellbild nicht exakt derselbe, da die weiter vorne im Strahlengang vorhandenen Hellbild- und Kalibrierspiegel bzw. die Probe das Spektrum des Lichtes unterschiedlich beeinflussen. Zweitens ist das Wellenlängenspektrum beim Auftreffen auf den CCD unterschiedlich, wodurch die Kamera, aufgrund der Wellenlängenabhängigkeit der Quanteneffizienz, bei selber auftreffender Bestrahlungsstärke eine unterschiedliche Anzahl an Grauwerten anzeigt. Daher ist es wichtig die spektrale Abhängigkeit der Reflektivität genau zu kennen.

Für den Umlenkspiegel sowie den Kalibrierspiegel sind diese Eigenschaften durch die Hersteller berechnet, bzw. für den Hellbildspiegel vermessen worden, und sind in Abbildung 29 zu sehen. Die Messung für den Hellbildspiegel wurde auf einen simulierten Wert normiert. Die Plausibilität der Angaben wurde durch die Vermessung der integralen Reflektivität, wie in Kapitel 4.4.4 erläutert, überprüft.

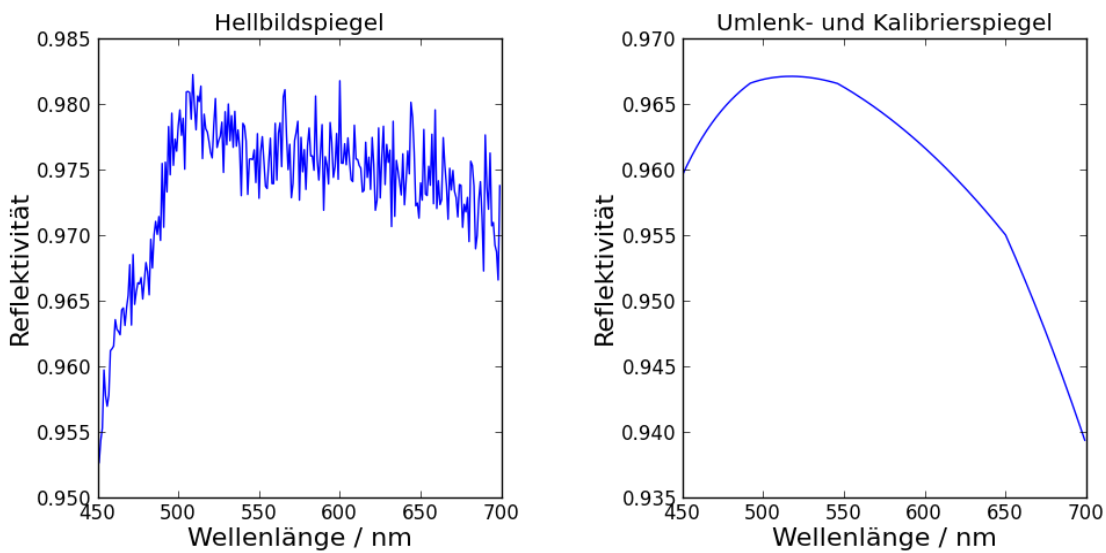

**Abbildung 29: Wellenlängenabhängige Reflektivität des Hellbild-, Umlenk- und Kalibrierspiegels. Links: Vom Hersteller gemessene Reflektivität des Hellbildspiegels, normiert auf einen simulierten Wert für den benötigten Einfallswinkel. Rechts: Vom Hersteller simulierte Reflektivität bei senkrechtem Einfall.**

Wir benötigen jedoch auch die Reflektivitäten der Proben. Diese sind in der Regel dem Hersteller selbst nicht genauer bekannt und wurden daher testweise an zwei Proben vermessen. Dazu wurde der Spiegel in einer sogenannten Ulbrichtkugel, einer mit diffus streuendem Material hoher Reflektivität ausgekleideten Kugel, leicht verdreht montiert. Der Aufbau ist in Abbildung 31 zu sehen. Dort wurde die Probe mit einem parallelen Lichtbündel bestrahlt. Dabei muss darauf geachtet werden, dass keine Strahlen neben die Probe direkt auf das streuende Material treffen. Der Spiegel reflektiert das Licht auf die streuende Innenverkleidung der Kugel. Nach einigen Streuungen innerhalb der Ulbrichtkugel trifft ein Teil des Lichtes auf den Lichtleiter und wird vom Spektrometer analysiert. Der Untergrund wird durch eine Aufnahme mit abgedeckter Lichtquelle abgezogen. Dies ist notwendig, da das Streulicht sowohl auf den Spiegel, wie auch direkt auf die streuende Fläche treffen kann und somit das Ergebnis verfälscht. Das Spektrum des detektierten Lichtes besteht also aus der Multiplikation des Spektrums der Lichtquelle, der wellenlängenabhängigen Reflektivität der Probe und der der Kugel. Um das Messsignal von der Reflektivität der Kugel und des Spektrums der Lichtquelle zu bereinigen, wird zusätzlich eine Messung mit einer Diffusorplatte, deren Reflexionsverhalten bekannt ist, durchgeführt. Diese wird ebenfalls des Untergrunds bereinigt. Teilt man nun das untergrundbereinigte Signal der Aufnahme mit Probe durch das mit Diffusorplatte, so bekommt man das Verhältnis zwischen der wellenlängenabhängigen Reflexion des Spiegels und der der Diffusorplatte in Abhängigkeit der Wellenlänge. Die Reflektivität der Diffusorplatte in Abhängigkeit der Wellenlänge ist bekannt und durch Multiplikation mit selbiger bekommen wir die Reflektivität der Probe.

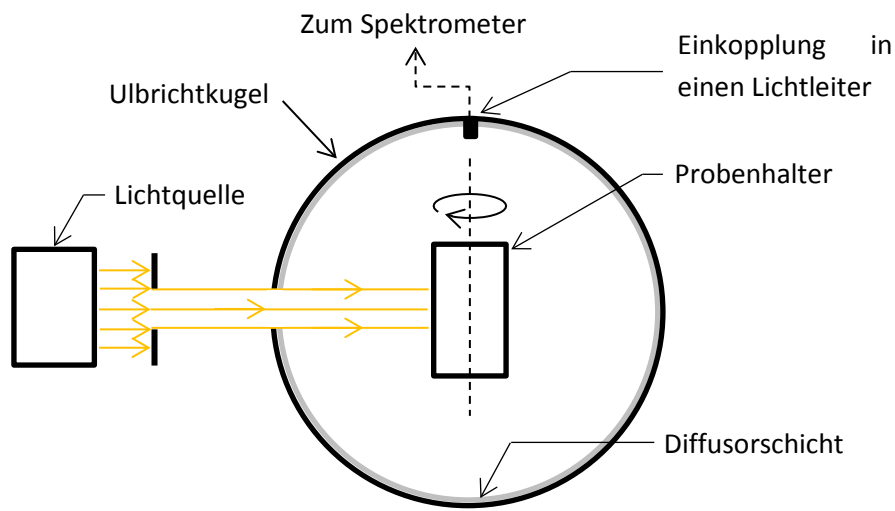

Abbildung 31: Aufbau zur Messung der wellenlängenabhängigen Reflektivität

Damit hätten wir im Prinzip die Reflektivität der Probe bestimmt. Allerdings haben Wiederholungsmessungen einer selben Probe gezeigt, dass die Lichtquelle um einige Prozent in dem Zeitraum zwischen mehreren Messungen, also auch zwischen der Messung der Probe und der Diffusorplatte, schwankt (siehe Abbildung 30). Somit stimmt zwar der Verlauf der Reflektivität, jedoch kann die Reflektivität für jede Wellenlänge um den selben Faktor skaliert sein. Der Skalierungsfaktor entspricht dem Verhältnis der Strahlungsleistungen der Lichtquelle bei Messung der Probe und der Diffusorplatte. In Abbildung 32 sind die Messergebnisse der zwei Proben zu sehen.

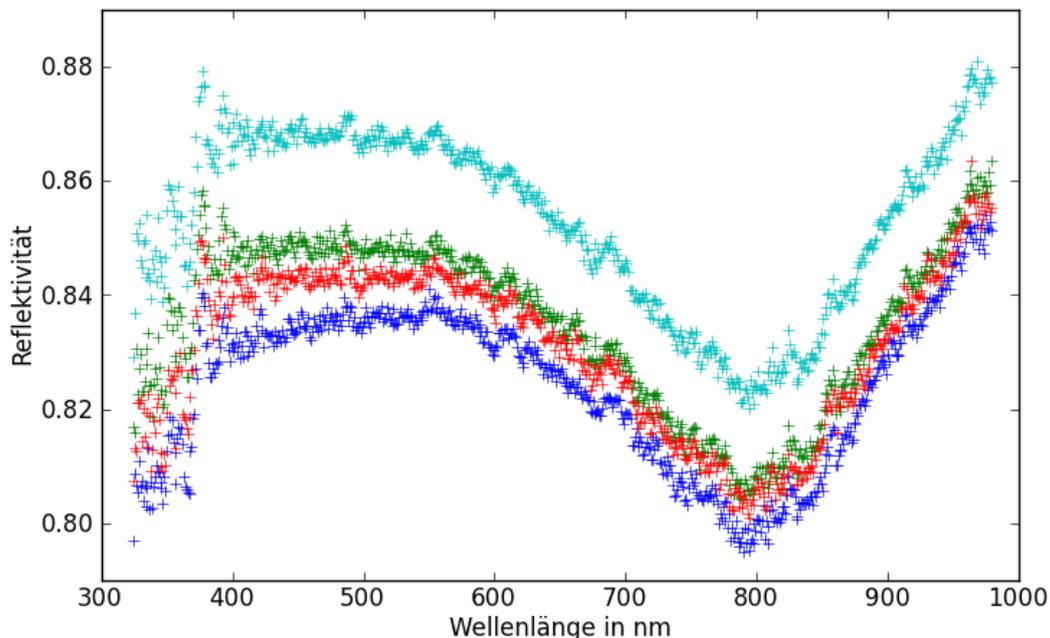

Abbildung 30: Wiederholbarkeit der Messungen von Reflektivitäten an der Ulbrichtkugel

Auch wenn das genaue Wellenlängenspektrum der Reflektivität der Probe und die Dicke der Beschichtung vom Hersteller nicht angegeben werden kann, wissen wir, dass die Beschichtung auf Aluminium basiert. Zur Überprüfung der Plausibilität des Messergebnisses kann dieses mit einer externen Messung der Reflektivität eines Aluminiumspiegels verglichen werden. Letztere ist in Abbildung 33 zu sehen. Eine qualitative Übereinstimmung besteht.

Wir können somit für den Hellbild-, Kalibrier- und Umlenkspiegel die Herstellerangaben benutzen und für die Proben jeweils, wie oben erläutert, den relativen Verlauf der Wellenlängenabhängigkeit der Reflektivität an der Ulbrichtkugel messen.

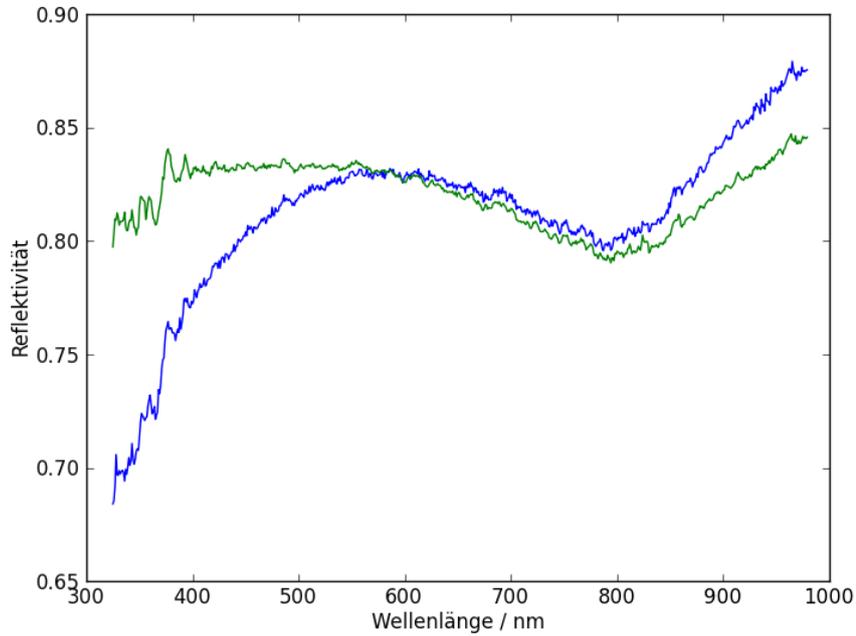

**Abbildung 32: Dispersion der Reflektivität zweier unterschiedlicher Proben. Die Werte können aufgrund von Schwankungen der Lichtquelle skaliert sein.**

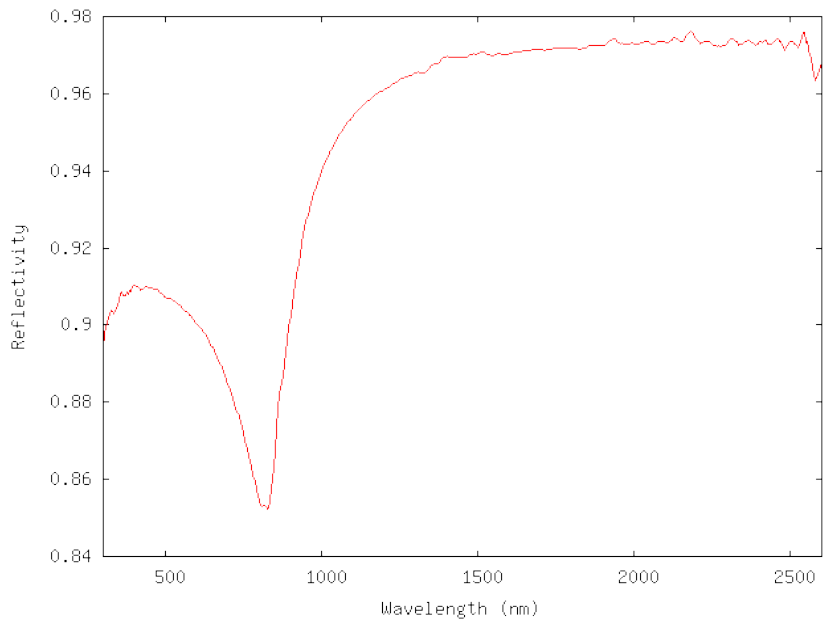

**Abbildung 33: Reflektivität der Alumiumbeschichtung des Primärspiegels des Subaru-Teleskops. Quelle: [NAO13]**

### 4.4.3. Strahlaufweitung

Bei der Reflexion an einem Spiegel wird das Licht nur in erster Näherung entsprechend dem Reflexionsgesetz abgestrahlt. Aufgrund der Oberflächenrauheit des Substrats und der Beschichtung und geometrischer Oberflächenfehler des Substrats des Spiegels, entsteht ein Abstrahlprofil mit einem Maximum in der durch das Reflexionsgesetz vorhergesagten Richtung, das jedoch eine gewisse Breite im Winkelraum besitzt.

Wird am RMP das Licht am Hellbildspiegel, am Kalibrierspiegel oder am Umlenkspiegel reflektiert, kommt es, für jeden Strahl gemäß der idealen geometrischen Optik, zu einer Faltung des Austrittswinkels. Die Auswirkung dieser Fehler wird in Kapitel 5.4 nochmals genauer behandelt. An dieser Stelle ist lediglich wichtig, dass diese Fehler eine ähnliche Auswirkung haben, wie die die wir später bei den Proben bestimmen wollen. Daher ist es notwendig diese Effekte zu vermessen um sie nicht später fälschlicherweise den Proben zuzuschreiben. Wir erwarten jedoch aufgrund der Herstellerangaben (siehe Kapitel 3.2), dass alle Spiegel im Versuchsaufbau, mit Ausnahme der Proben, so glatt und formtreu sind, dass die Strahlaufweitung durch diese Spiegel vollständig vernachlässigt werden kann.

Um dies zu überprüfen wurde der in Abbildung 34 gezeichnete Aufbau benutzt. An einem Photogoniometer wurde auf einem drehbaren Arm eine Lichtquelle montiert, bestehend aus einer roten LED, einem Diffusor und einem Pinhole. Auf denselben Arm wurde eine Linse mit Brennweite 160mm fixiert, sodass der Abstand zum Pinhole möglichst genau der Brennweite der Linse entspricht. Das durch die Linse parallelisierte Licht besitzt aufgrund des endlichen Durchmessers der Blende in der Lichtquelle eine Divergenz von 0,9mrad. Direkt hinter der Linse wird das Lichtbündel mit einer Blende auf den gewünschten Durchmesser von 2mm reduziert. Der drehbare Arm rotiert um einen Probehalter dessen Winkel ebenfalls einstellbar ist und in den bei Bedarf ein Spiegel eingelegt werden kann. Enthält der Halter keinen Spiegel durchqueren, ihn die Lichtstrahlen unbeeinflusst. Hinter dem Probenhalter ist eine Linse mit Brennweite f=200mm montiert, die das Licht auf den CCD-Chip fokussiert.

Wie aus Abbildung 35 klar wird, beobachten wir auf dem CCD direkt den Winkelraum und somit die Intensitätsverteilung der Reflexion im Winkel $\alpha$ um die durch das Reflexionsgesetz vorhergesagte Richtung. Dabei gilt $x = f \cdot \tan(\alpha) \approx f \cdot \alpha$ für kleine $\alpha$, wobei x der Abstand auf dem CCD zu dem Auftreffpunkt laut geometrischer Optik ist. Die Auflösung aufgrund der Pixelgröße beträgt 0,04 mrad. Durch den Vergleich einer Aufnahme ohne Probe (Hellbild) und einer Aufnahme mit Probe (Messbild) kann einerseits die Strahlaufweitung in Form einer Faltung des Hellbildes beobachtet werden, andererseits die Reflektivität des Spiegels für dieses Wellenlängenspektrum berechnet werden.

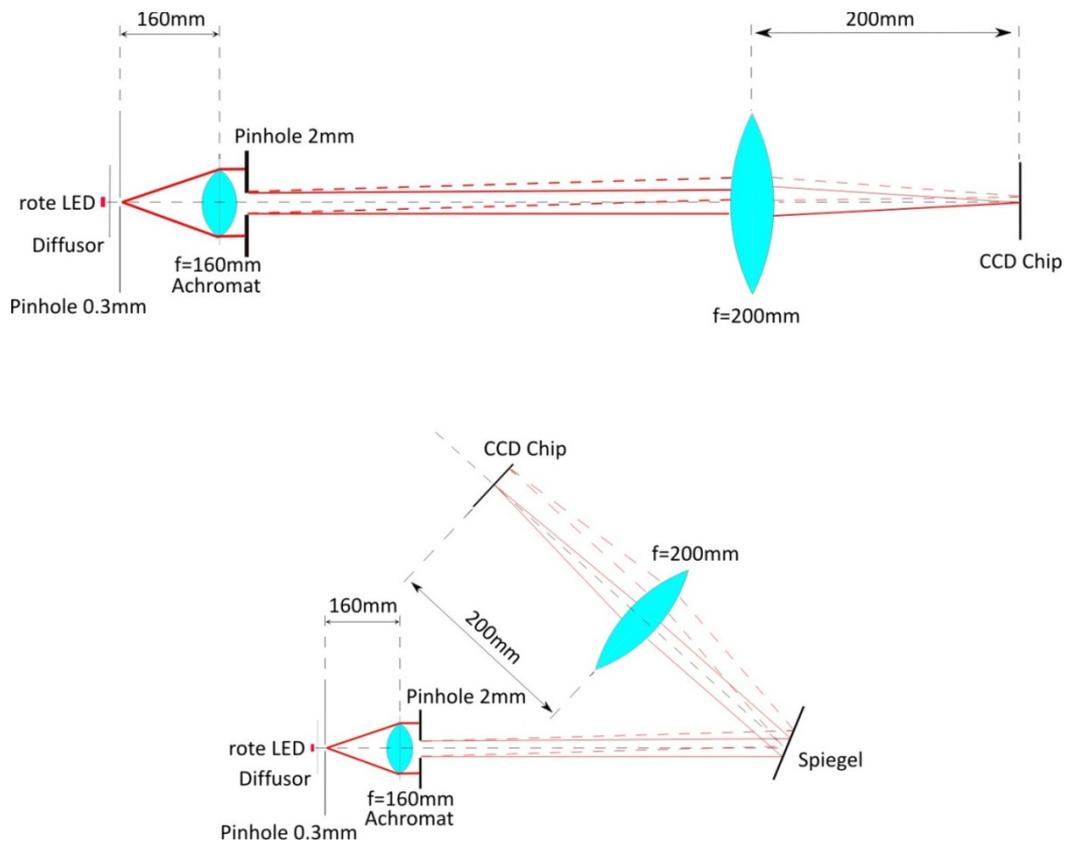

Abbildung 34: Aufbau zur Vermessung der Strahlaufweitung durch Spiegel am Goniometer. Oben: Position zur Aufnahme des Hellbildes. Unten: Position zur Aufnahme des Messbildes. (Zeichnung aus einem internen Bericht von T. Schmid)

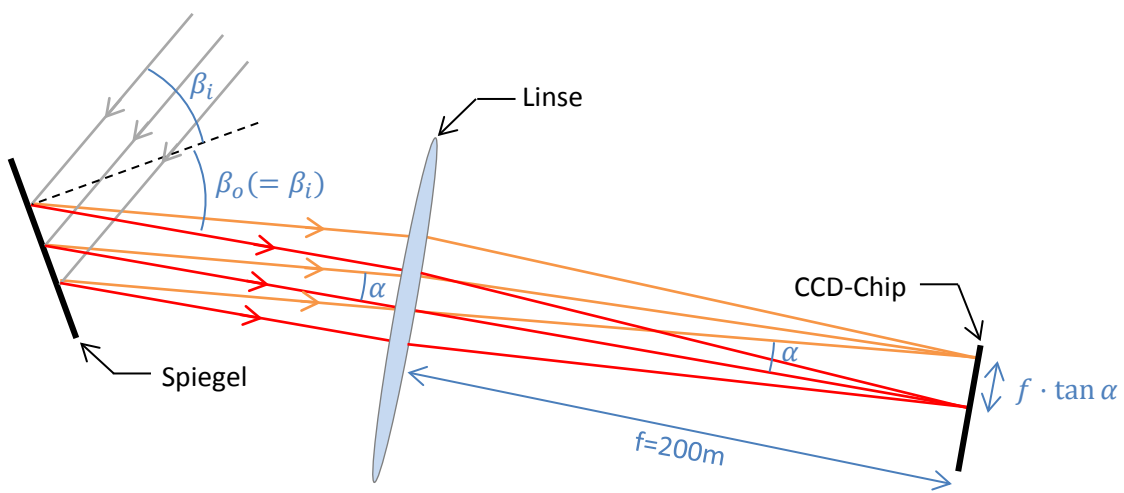

Abbildung 35: Fouriertransformation durch die Linse: Auf dem CCD-Chip wird die Reflektionswinkelverteilung sichtbar

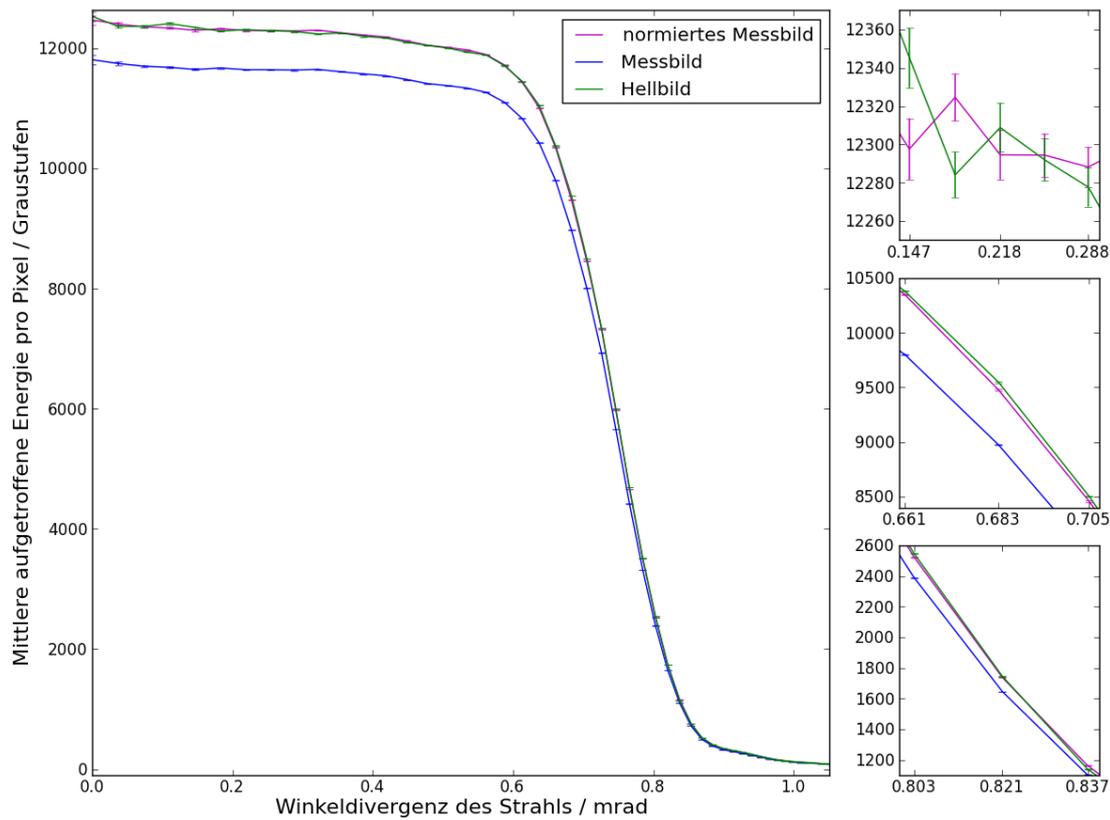

Abbildung 36: Radiale Darstellung des Winkelspektrums der reflektierten Strahlen am Umlenkspiegel, beim Hellbild aufgrund der Divergenz der Beleuchtung, beim Messbild zusätzlich aufgrund der Strahlaufweitung. Die Fehlerbalken resultieren aus dem statistischen Rauschen der Kamera. Links: Gesamtansicht. Rechts: Nahansichten

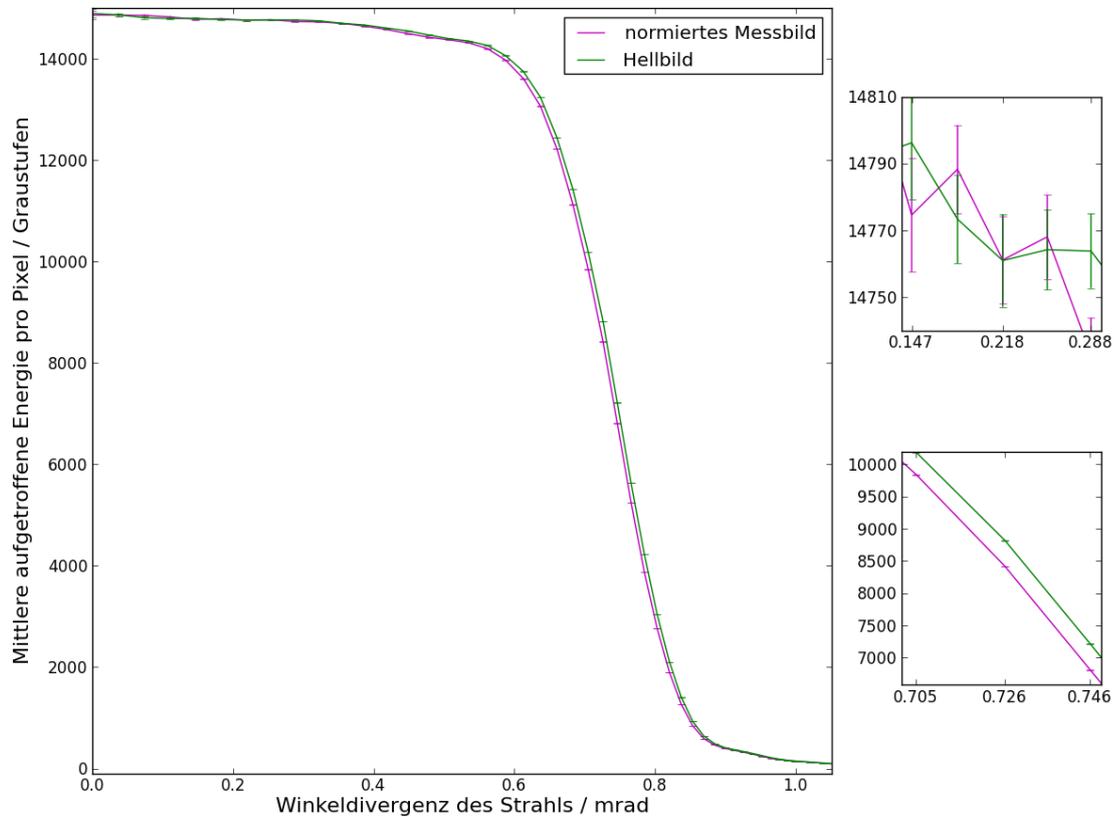

Abbildung 37: Radiale Darstellung des Winkelspektrums der reflektierten Strahlen an der ersten Position des Hellbildspiegels, beim Hellbild aufgrund der Divergenz der Beleuchtung, beim Messbild zusätzlich aufgrund der Strahlaufweitung. Die Fehlerbalken resultieren aus dem statistischen Rauschen der Kamera. Links: Gesamtansicht. Rechts: Nahansichten

Um die Strahlaufweitung zu bestimmen, wurden die Messbilder zunächst wie in Kapitel 4.1.2 erläutert korrigiert. Anschließend wurde jeweils für konzentrische Ringe um den Mittelpunkt des Spots der Mittelwert gebildet, wodurch eine Radialverteilung entsteht. Die Radialverteilung des Hellbildes hat aufgrund der Divergenz des einfallenden Lichtes eine endliche Breite. Die Radialverteilungen wurden für den Umlenkspiegel und den Hellbildspiegel gebildet. Der Kalibrierspiegel ist durch seine Krümmung nicht mit dem Verfahren zu vermessen. Er besitzt jedoch laut Hersteller dieselbe Beschichtung wie der Umlenkspiegel. Daher werden die Eigenschaften des Umlenkspiegels für den Kalibrier- und den Umlenkspiegel benutzt. Für den Hellbildspiegel wurde die Homogenität der Oberfläche und der Beschichtung überprüft, indem an drei verschiedenen Stellen gemessen wurde. Der Umlenkspiegel wurde aufgrund seiner geringen Größe von 3x4,2 mm nur an einer Stelle vermessen. Die Radialverteilung des Umlenkspiegels sowie eine der Radialverteilungen des Hellbildspiegels sind in Abbildung 36 und Abbildung 37 zu sehen. Die Radialverteilungen für die zwei anderen Stellen auf dem Hellbildspiegel sind im Anhang zu finden.

Aufgrund der begrenzten Reflektivität des Spiegels nimmt die Radialverteilung des Messbildes niedrigere Werte an als die des Hellbildes. In der Nähe des Zentrums der Lichtverteilung ändert eine schmale Faltung nichts an der Intensität, da dort ein Plateau vorzufinden ist. Diese Stelle kann daher zur Normierung benutzt werden, wodurch die Reflektivität herausgerechnet wird.

Vergleichen wir nun die Verläufe der Radialverteilung für Hell und Messbild, ergibt sich zunächst, dass beide Verläufe sich sehr ähneln und nur um einige Standardabweichungen des statistischen Rauschen der Kamera unterscheiden. Zugleich wird jedoch, aufgrund von Differenzen von über 3 Standardabweichungen und systematisch niedriger liegender Radialverteilung des Messbildes, klar, dass die Abweichungen nicht durch die Kamera zustande kommen. Letztere Bemerkung schließt ebenfalls eine Strahlaufweitung durch den Spiegel als Ursache aus.

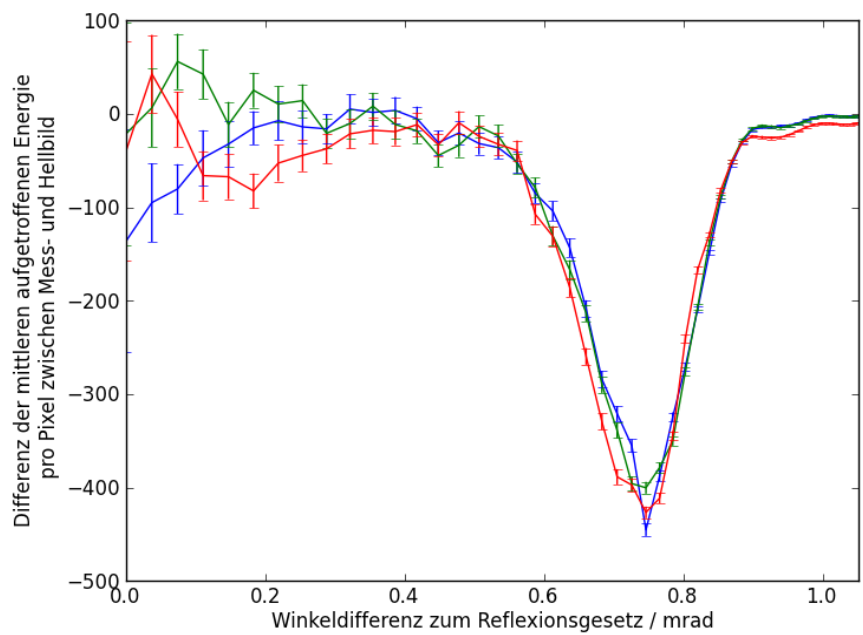

Abbildung 38: Differenz zwischen den Radialverteilungen des Mess- und Hellbildes für drei unterschiedliche Positionen des Hellbildspiegels

Bilden wir die Differenz zwischen normierten Mess- und Hellbild (siehe Abbildung 38), wird sichtbar, dass die Differenz systematisch zum Winkel 0,75mrad hin größer wird, wobei die Differenz zwischen den Bildern leicht schwankt. Es sind hier also systematische Fehler vorhanden, deren Einfluss weit größer als die statistische Schwankung der Kamera ist. Der Einbruch bei dem Winkel 0,75mrad könnte durch die Winkelabhängigkeiten der 200mm-Linse und des CCD-Chips erklärt werden. Aufgrund von mechanischen Beschränkungen konnten beim Hellbild Lichtquelle, Drehpunkt des Probenhalters und CCD-Chip nicht exakt auf eine Achse gebracht werden. Daher trifft der Strahl beim Messbild unter einem leicht anderen Winkel auf die Linse und den CCD als beim Hellbild.

Wichtigste Schlussfolgerung ist, dass eine mögliche Faltung aufgrund einer Strahlaufweitung durch den Spiegel schmaler als die hier erreichte Auflösung ist und daher keinen nennenswerten Einfluss auf das Ergebnis am RMP hat. Wir können also wie erwartet die Strahlaufweitung der Kalibrier-, Mess- und Umlenkspiegel vernachlässigen.

### 4.4.4. Integrale Reflektivität

Um zu überprüfen, ob die Reflektivität der Spiegel den Angaben des Herstellers entspricht, benutzen wir dieselben Aufnahmen wie im letzten Abschnitt. Es wird die Summe der Grauwerte über den CCD beim Messbild durch die Summe beim Hellbild geteilt. Unter der Annahme, dass nur ein vernachlässigbar geringer Anteil des auftreffenden Lichtes neben den CCD gestreut wird, entspricht dies dem Bruch (siehe Gleichung 13)

$$\bar{R} = \frac{\int LQ^{Goniometer}(\lambda) \cdot \eta(\lambda) \cdot R(\lambda) d\lambda}{\int LQ^{Goniometer}(\lambda) \cdot \eta(\lambda) \, d\lambda} \qquad (21)$$

wobei $LQ^{Goniometer}(\lambda)$ das Spektrum der Lichtquelle ist. Da die Reflektivität der Spiegel von der Wellenlänge abhängt und wir durch die Benutzung einer LED als Lichtquelle nicht monochromatisch arbeiten, entspricht der so erhaltene Wert einem mit dem Spektrum der LED gewichteten Mittel. Es handelt sich dabei um den Reflexionswert, der konstant über die Wellenlängen im Bereich des Spektrums der Lichtquelle die gleiche Summe der Grauwerte für das Messbild ergeben würde. Zum Vergleich wurde das Spektrum der LED mit einem Spektrometer aufgenommen, womit wir $const \cdot LQ^{Goniometer}(\lambda)$ bekommen. Anhand des vom Hersteller angegebenen theoretischen Verlaufs der Reflektivität und der Quanteneffizienz wird obiger Bruch berechnet. Aus dieser Berechnung ergeben sich die Werte $\bar{R}^{theo}_{Umlenkspiegel} = \bar{R}^{theo}_{Kalibrierspiegel} = 95,84\%$ und $\bar{R}^{theo}_{Hellbildspiegel} = 97,51\%$. Die Ergebnisse aus den experimentellen Messungen sind im Folgenden zusammengefasst.

| | |
|---|---|
| Umlenkspiegel : | 95,56 ± 0,06 % |
| Hellbildspiegel an Position 1 : | 98,54 ± 0,05 % |
| Hellbildspiegel an Position 2 : | 98,14 ± 0,05 % |
| Hellbildspiegel an Position 3, 1. Messung : | 97,43 ± 0,05 % |
| Hellbildspiegel an Position 3, 2. Messung : | 97,57 ± 0,05 % |

Die darin angegebenen Fehler entsprechen der statistischen Schwankung der Kamera. Jedoch sind auch hier diese nicht entscheidend für die Genauigkeit der Messung. Viel eher sind die Schwankungen der Intensität der Lichtquelle zwischen dem Mess- und dem Hellbild sowie unterschiedliche Mengen an Streulicht auf dem CCD zwischen Aufnahme des Bildes und Aufnahme des Untergrunds von Bedeutung. Letzteres entsteht daher, dass bei den Untergrundaufnahmen der direkte Strahlengang abgeblockt werden muss. Dabei werden zwangsläufig auch manche Streulichtquellen abgeschattet.

Einen Eindruck der Größenordnung des Einflusses des Streulichts bekommen wir, wenn wir, anstatt das Verhältnis der Summen über den gesamten Chip zu bilden, die Summen über einen Ausschnitt um den Spot bilden. Daraus resultieren im Schnitt um ca. 0,1% niedrigere Reflexionsgrade.

Die Schwankung der Lichtquelle wurde extra vermessen und analog zu Kapitel 4.3.1 ausgewertet. Daraus resultiert eine statistische Schwankung um 0,02%. Der Drift der Lichtquelle beträgt in einem Zeitraum von 19 Minuten ca. 0,05% und ist damit im Zeitraum zwischen zwei Messungen nicht von Bedeutung.

Dazu kommt der in Kapitel 4.4.3 angesprochene Effekt der unterschiedlichen Sensibilität für verschiedene Einfallswinkel auf die Linse und auf den CCD. Bei hypothetischen 3°[18] Einfallswinkel sinkt die Sensibilität des CCDs laut Herstellerangaben um 0,12% bezüglich des senkrechten Einfalls, womit der Einfluss des CCDs in derselben Größenordnung wie eine Schwankung des Streulichtes liegen würde. Die Fresnel-Reflexionen an der Linse ändern sich zwischen 0° und 5° Einfallswinkel um 0,005%.

Letztendlich sei angemerkt, dass die vom Hersteller angegebenen Reflektivitäten theoretische Verläufe sind. Auch hier entstehen, zum Beispiel durch die Dicke der Beschichtung, weitere Abweichungen.

Auch wenn die hier durchgeführten Messungen aufgrund der Messungenauigkeit nicht erlauben die Reflexionsgrade genauer als die theoretischen Werte zu bestimmen, haben sie erlaubt die Herstellerangaben zu bestätigen. Letztere können demnach für Berechnungen für den RMP eingesetzt werden.

## 4.5. Glasplatte

Wie der Reflexionsgrad der Spiegel ist auch der Transmissionsgrad der Glasplatte wellenlängenabhängig. Da damit wieder eine Beeinflussung der Reflektivität der Spiegel, des Transmissionsgrades des Diffusors und der Quanteneffizienz der Kamera einhergeht, muss auch dieser vermessen werden um die Wellenlängenabhängigkeit des RMPs zu bekommen. Es wird dabei ähnlich wie für die Spiegel (siehe Kapitel 4.4.2) vorgegangen.

Mit einer Lichtquelle mit breitem Wellenlängenspektrum wird in eine Ulbrichtkugel geleuchtet. An dem Eingang dieser kann eine semitransparente Probe platziert werden. Ein Teil des Lichts, das am Diffusor der Innenwand der Ulbrichtkugel gestreut wird, wird von einem Spektrometer analysiert. Der Aufbau ist in Abbildung 39 zu sehen. Um den Transmissionsgrad der Glasplatte zu bekommen, werden zwei Aufnahmen durchgeführt: Eine mit und eine ohne Glasplatte am Eingang der Ulbrichtkugel. Gemessen wird bei der ersten Aufnahme das Spektrum der Beleuchtung, inklusive Streulicht, gewichtet mit dem Transmissionsgrad der Glasplatte sowie mit der Reflektivität des Diffusors. Bei der zweiten Aufnahme wird selbiges ohne Gewichtung mit dem Transmissionsgrad der Glasplatte gemessen. Teilen wir die erste Aufnahme durch die zweite, bekommen wir den Transmissionsgrad der Glasplatte in Abhängigkeit der Wellenlänge. Er ist in Abbildung 40 zu sehen.

---

[18] $1° = \pi/180$ rad ≈ 0,01745 rad

Aufgrund möglicher Schwankungen in der Intensität der Lichtquelle kann dieser Transmissionsgrad, wie auch schon die Reflektivitäten der Spiegel, mit dem Strahlungsleistungsverhältnis zwischen den zwei Aufnahmen skaliert sein. In dem für uns relevanten Wellenlängenbereich ist die Dispersion des Transmissionsgrades der Glasplatte circa halb so groß wie die Änderungen der Reflektivität der Umlenk- und Kalibrierspiegel.

Diese Messung wird bei der Berechnung der Konzentrationsmatrix des Kalibrierspiegels, bzw. der Probe, eingesetzt. Genaueres dazu ist in Kapitel 5.2 nachzulesen.

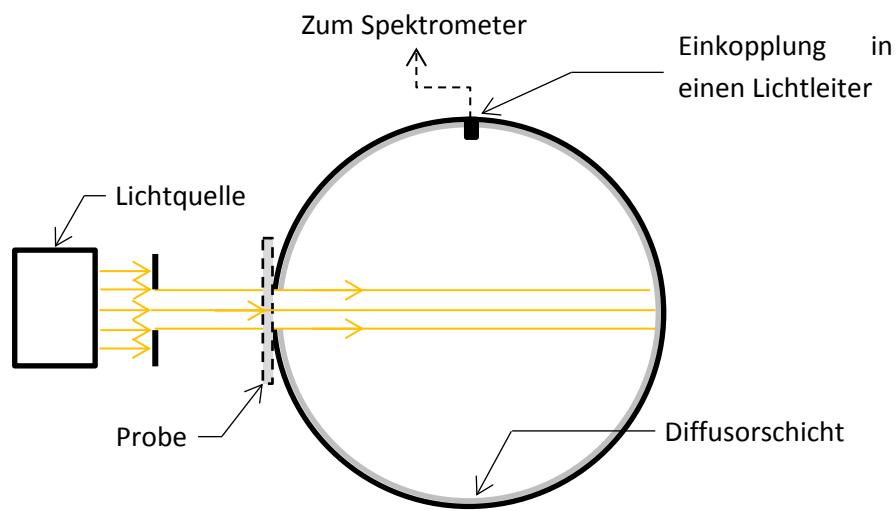

**Abbildung 39: Aufbau zur Messung des wellenlängenabhängigen Transmissionsgrades**

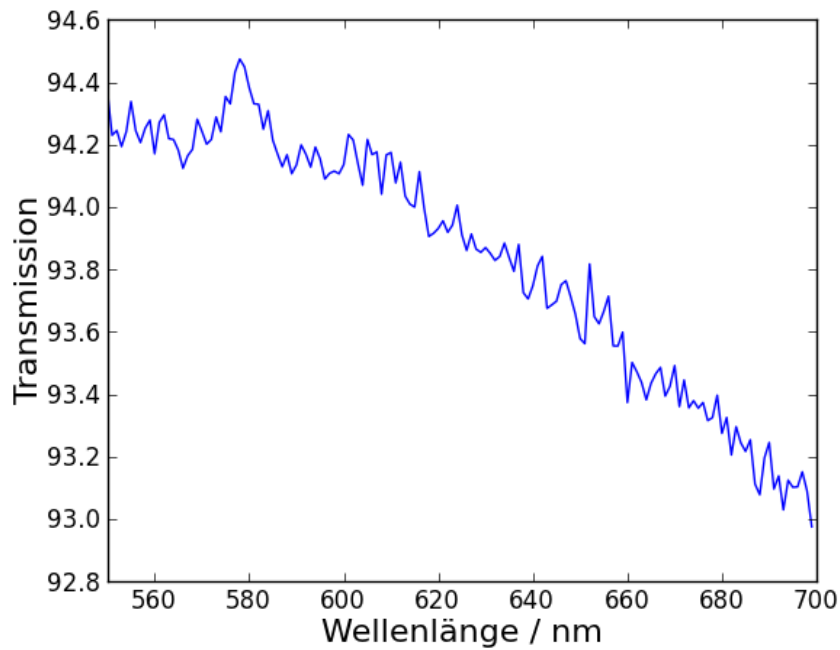

**Abbildung 40: Dispersion des Transmissionsgrades der Glasplatte**

## 4.6. Diffusor

Der Diffusor ist vermutlich, trotz seiner relativen Einfachheit, das Element des Aufbaus dessen Einfluss auf das resultierende Messbild am schlechtesten verstanden ist. Die Rolle des Diffusors im Aufbau ist es, von jedem beleuchteten Punkt aus in Richtungen zu streuen, die das Objektiv einfangen kann, wobei die Strahlungsleistung proportional zur einfallenden Intensität an diesem Punkt sein soll. So kann die Intensitätsverteilung am Ort des Diffusors betrachtet werden. Leider kann jedoch kein physikalischer Streuer diese Anforderung ohne unerwünschte Nebeneffekte erfüllen. Es sollen hier die drei großen, für den RMP interessanten, nicht idealen Verhaltenseigenschaften des eingesetzten Diffusors betrachtet werden: Die beschränkte Ortsauflösung, das Transmissions- und das Reflexionsverhalten. Da nur wenig über den Diffusor bekannt ist [LiQ08], ist dabei die Beschreibung größtenteils empirisch.

### 4.6.1. Strahlaufweitung

*Qualitative Diskussion*

Der Diffusor ist eine 116±2 µm (Messung mit einer Mikrometerschraube) dicke Folie aus gesintertem PTFE. Durch die endliche Dicke führt eine Streuung im Material dazu, dass ein beim Eintritt in den Diffusor schmales Strahlenbündel beim Austritt breiter ist (siehe Abbildung 41 oben). Diese Strahlaufweitung, mathematisch durch eine Faltung der Intensitätsverteilung der Eintrittsfläche beschrieben, verursacht somit eine Verschmierung des Messbildes. Die Streueigenschaften des Diffusors hängen also von dessen Dicke ab, bzw. genauer von der Weglänge über die das Licht gestreut wird.

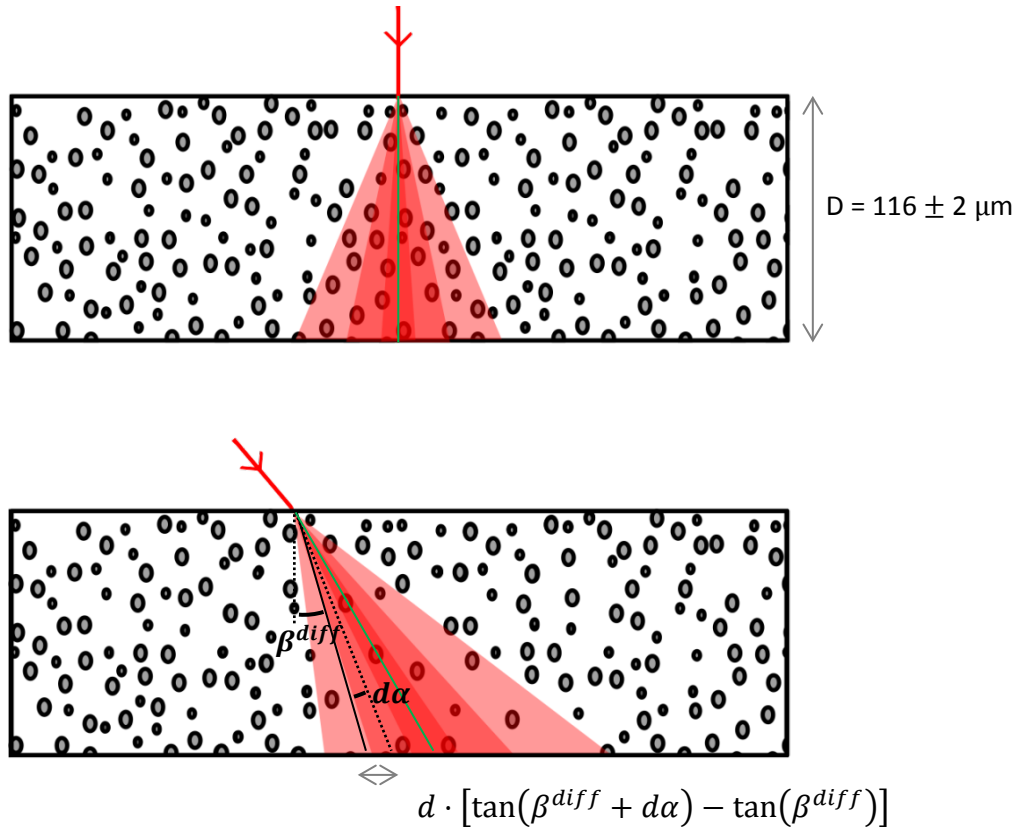

**Abbildung 41: Strahlaufweitung im Diffusor bei senkrechten und schrägen Einfall**

Einen weitereren Parameter der Strahlaufweitung stellt der Einfallswinkel der Beleuchtung dar. Eine von T. Schmid durchgeführte Messung der Intensitätsverteilung über die Austrittswinkel hat gezeigt, dass der Winkel zwischen Ausbreitungsrichtung im Diffusor und Oberflächennormalen mit dem Einfallswinkel steigt. Erstens ist damit der Weg im Material höher und wir erwarten eine stärkere Aufweitung als bei senkrechtem Einfall. Zweitens bedeutet dies, dass der Austrittspunkt aus dem Diffusor bei schrägem Einfall bezüglich des Eintrittspunktes leicht in Einfallsrichtung versetzt ist. Am RMP wird der Diffusor mit einer konzentrierenden Optik beleuchtet, womit Licht aus einem breiten Winkelspektrum auf diesen trifft. Der Abstand zwischen Diffusor und Optik muss demnach so gewählt werden, dass die Austrittspunkte der verschiedenen Richtungen übereinander liegen. Die Brennweite des konzentrierenden Spiegels und der Abstand zwischen Spiegel und Eintrittsfläche des Diffusors unterscheiden sich dadurch leicht.

Weiter könnte eine schräge Beleuchtung eine Schiefe der Intensitätsverteilung auf der Austrittsfläche verursachen. Denn wird angenommen, dass die Strahlaufweitung durch den Diffusor zu der in Abbildung 41 grünen Hauptausbreitungsrichtung symmetrisch verläuft, kommen zwei Effekte zusammen.

Erstens müssen Strahlen, die sich durch die Strahlaufweitung in größeren Winkeln zur Diffusornormalen ausbreiten als diejenigen der Hauptrichtung (in Abbildung 41 unten, rechts der grünen Linie), einen größeren Weg zurücklegen und werden demnach häufiger gestreut, was zu einer größeren Verbreiterung führt.

Zweitens treffen die Strahlen je nach Streurichtung unter unterschiedlichen Winkeln auf die Austrittsfläche. Wird die Ausbreitung von Strahlen im Diffusor vereinfacht als im Mittel geradlinig modelliert, so ist der Winkel $\beta^{diff}$ zwischen Strahl und Oberflächennormale für die in Abbildung 41 unten, linken Strahlen kleiner als für die rechten. Ein Strahlenbündel mit Öffnungswinkel $d\alpha$ und Winkel $\beta^{diff}$ führt mit steigendem $\beta^{diff}$ zu einer im Vergleich zum senkrechten Einfall zunehmend gestreckten Schnittfläche mit der Austrittsfläche des Diffusors. Dieselbe Strahlungsstärke wird demnach auf unterschiedliche Flächen verteilt. Beide Effekte würden, sofern die zugrunde liegenden Annahmen gerechtfertigt sind, zu einem steileren Verlauf der Intensitätsverteilung auf der Seite des Einfallpunktes und einem flacheren Verlauf auf entgegengesetzter Seite führen.

Letztendlich hängen die Streueigenschaften des Diffusors von der Wellenlänge des einfallenden Lichtes ab, was sich ebenfalls auf die Strahlaufweitung auswirken könnte.

### *Modellierung der Strahlaufweitung*

Die Strahlaufweitung des Diffusors bei senkrechtem Einfall wurde bereits durch M. Kiss untersucht [Kis11]. Dabei wurde ein Modell für die Aufweitung durch den Diffusor entwickelt, das besagt, dass diese Intensität bei Beleuchtung eines kleinen Flächenelements der Eintrittsfläche an der Austrittsfläche einem $\cos^4 \alpha$-Verlauf folgt, wobei $\alpha$ der Winkel zwischen Oberflächennormale und dem Ort auf der Austrittsfläche mit Scheitelpunkt am Eintrittspunkt ist. Es gilt also $I(\alpha) = I(0°) \cdot \cos^4 \alpha$. Dieses Modell wird hier näher erklärt und für die folgende Auswertung übernommen.

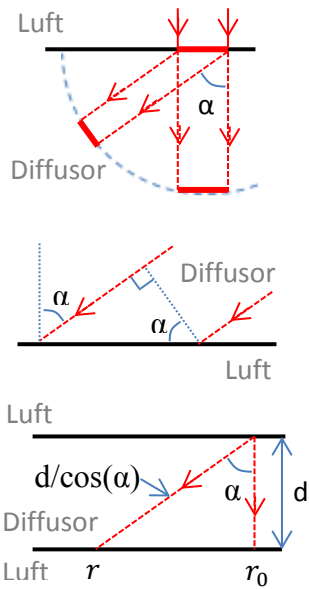

Ein Flächenelement der Eintrittsfläche des Diffusors, das unter dem Winkel $\alpha$ zur Flächennormale betrachtet wird, wirkt aufgrund der Perspektive um den Faktor $\cos\alpha$ gestaucht (siehe Abbildung 42 oben). Wenn eine Fläche Lambert'sch abstrahlt, die von der Fläche emittierte Strahlung also in jede Richtung dieselbe Flussdichte aufweist, ist die Strahlungsleistung unter dem Winkel $\alpha$ ebenfalls um den Faktor $\cos\alpha$ geringer als in Richtung der Flächennormale.

Wird angenommen, dass die Strahlen sich im Diffusor geradlinig ausbreiten, also nur an der Eintritts- und Austrittsfläche gestreut werden, treffen unter dem Winkel $\alpha$ ausgesendete Strahlen ebenfalls unter dem Winkel $\alpha$ auf die Austrittsfläche des Diffusors (siehe Abbildung 42 Mitte). Somit wird der Strahlungsleistung auf eine um den Faktor $1/\cos\alpha$ größere Fläche verteilt. Die Bestrahlungsstärke sinkt somit um den Faktor $\cos\alpha$.

**Abbildung 42: Ursprung der vier Cosinus-Terme im cos⁴-Gesetz**

Letztendlich verläuft die Intensität einer Kugelwelle invers proportional zum Quadrat des Abstands zur Quelle. Eine Kugelwelle die ihren Ursprung in dem Flächenelement der Eintrittsfläche hat, schneidet zunächst die Austrittsfläche in Richtung der Oberflächennormale. Der Weg zur Austrittsfläche unter dem Winkel $\alpha$ ist um den Faktor $1/\cos\alpha$ länger (siehe Abbildung 42 unten). Damit haben wir nun die 4 Cosinus-Terme und es gilt

$$I(\alpha) = \underbrace{\cos\alpha}_{\substack{Projektion\ der\\ Eintrittsfläche}} \cdot \underbrace{\cos\alpha}_{\substack{Projektion\ der\\ Austrittsfläche}} \cdot \underbrace{\cos^2\alpha}_{Kugelwelle} \cdot I_0 \tag{22}$$

Dabei ist $I_0$ die Intensität bei senkrechtem Einfall. Wir können dieses Modell noch umparametrisieren, sodass die Intensitätsverteilung anstatt von dem Winkel $\alpha$ von dem Abstand $r-r_0$ des betrachteten Punktes zu dem Schnittpunkt der Austrittsfläche mit der Oberflächennormale durch den Eintrittspunkt (siehe Abbildung 42 unten) sowie von der Dicke d des Diffusors abhängt. Jeder senkrecht auf die Eintrittsfläche treffender Strahl verursacht somit auf der Austrittsfläche eine Intensitätsverteilung

$$I_0 \cdot \frac{d^4}{[(r-r_0)^2 + d^2]^2} \tag{23}$$

Somit bekommen wir, bei einer gegebenen Intensitätsverteilung auf der Eintrittsfläche, dieselbe Verteilung gefaltet mit der obigen Funktion als Intensitätsverteilung auf der Austrittsfläche.

Es sei hervorgehoben, dass die nötigen Annahmen zur Erklärung dieses Verlaufes zum Teil physikalisch nicht begründet sind. Insbesondere handelt es sich bei dem Diffusor um einen Volumenstreuer, bei dem die mittlere freie Weglänge geringer ist als die Dicke des Diffusors, womit die ungestörte Propagation einer Lichtwelle zwischen Eintrittsfläche und Austrittsfläche nicht gegeben ist. Es handelt sich also hier um ein durch Messungen bei senkrechtem Einfall empirisch entstandenes Modell, das die tatsächliche Physik stark vereinfacht, um eine praktikable Beschreibung des Diffusors zu bekommen. Der darin vorkommende Parameter der Dicke ist nicht zwangsläufig die physikalische Dicke des Diffusors. Er wird daher im Folgenden effektive Dicke genannt.

Dieses Modell wird im Folgenden für die Beschreibung der Strahlaufweitung für senkrechte sowie für schräge Beleuchtung verwendet. Asymmetrien der Strahlaufweitung bei schräger Beleuchtung wirken sich damit in Form einer asymmetrischen Abweichung des Fits, mit der Faltungsfunktion aus Gleichung 23, von den experimentellen Daten aus.

### *Messung*

Im Rahmen dieser Arbeit wurde die Strahlaufweitung durch den Diffusor in Abhängigkeit des Einfallswinkels vermessen. Die Dispersion dieses Effektes kann nicht umfassend berücksichtigt werden. Zwar wurde zur Vermessung der Strahlaufweitung eine LED mit selben Wellenlängenspektrum wie am RMP eingesetzt, doch wird am RMP dieses Spektrum durch Komponenten der Lichtquelle und Spiegel leicht verändert. Um diese leichten Änderungen zu berücksichtigen müsste die Strahlaufweitung wellenlängenaufgelöst vermessen werden, ohne dabei bezüglich weiterer Faktoren, wie zum Beispiel der transversalen Kohärenz, Unterschiede zum RMP zu schaffen. Dieser Aufwand ist hinsichtlich des vermutlich geringen Einflusses dieser Korrektur nicht vertretbar.

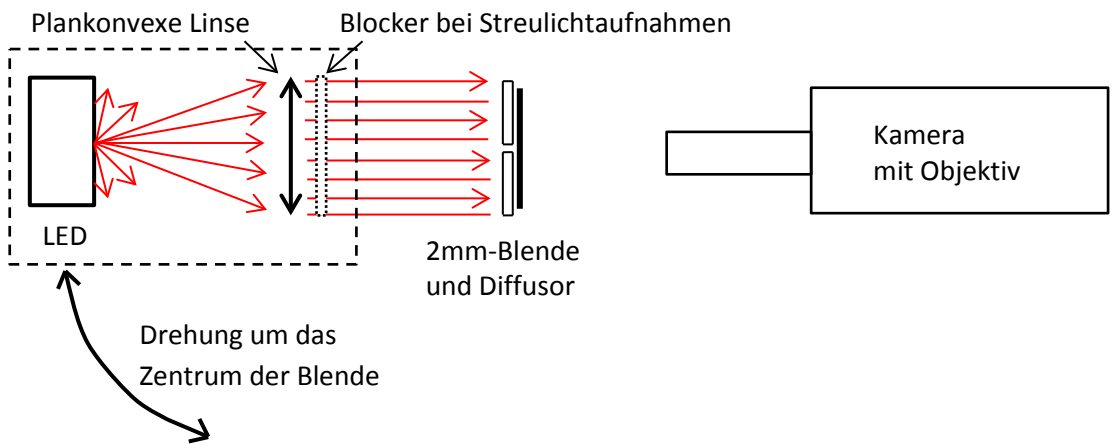

**Abbildung 43: Aufbau am Goniometer zur Charakterisierung der einfallswinkelabhängigen Ortsauflösung des Diffusors**

Zur Charakterisierung der einfallswinkelabhängigen Aufweitung der einfallenden Strahlen durch den Diffusor, wurde das in Abbildung 43 skizziert Experiment aufgebaut, wobei der Einfallswinkel computergesteuert eingestellt werden kann.

Wir erwarten für das Messbild, nach den Korrekturen der Einflüsse der Kamera (siehe Kapitel 4.1), folgende Zusammensetzung:

$$\text{Messbild}(r) = \left[\left(\text{Beleuchtung}(r) \cdot \cos(\beta^{in}) \cdot \delta_{\text{Ausschnitt Blende}}(\beta^{in}) \right.\right. \\ \left.\left. \cdot \text{Diffusorstruktur}(r, \beta^{in})\right) * \text{Aufweitung durch Diffusor}(\beta^{in}) * PSF \right] \quad (24) \\ \cdot \text{Randlichtabfall}(r)$$

Dabei ist r der Ort auf der Blende beziehungsweise auf dem Diffusor und $\beta^{in}$ der Winkel zwischen der Beleuchtungsrichtung und der Oberflächennormalen des Diffusors. Der Cosinus-Term resultiert aus der mit steigendem Einfallswinkel fallenden Beleuchtungsstärke. $\delta_{\text{Ausschnitt Blende}}(\beta^{in})$ ist eine Funktion die innerhalb der Öffnung der Blende den Wert 1, außerhalb den Wert 0 annimmt. Die Diffusorstruktur ist eine Variation des Transmissionsgrades des Diffusors aufgrund von örtlichen Schwankungen der Dicke und Inhomogenitäten im Material. Die aus einem schräg einfallenden Strahl resultierenden gestreuten Strahlen haben einen anderen Weg im Diffusor als diejenigen, die aus einem am selben Ort senkrecht auftreffenden Strahl resultieren. Sie erfahren daher unterschiedliche Dichteschwankungen, wodurch die Diffusorstruktur einfallswinkelabhängig ist.

Die Strahlaufweitung durch den Diffusor wird mathematisch durch eine Faltung beschrieben und wird am Rand der Blende sichtbar. Dies kann ausnutzt werden indem Aufnahmen mit und ohne Diffusor verglichen werden. Aus beiden Bildern kann bei bekanntem Randlichtabfall dieser herausgeteilt werden. Dank der Vertauschbarkeit von Faltungen, besteht der Unterschied zwischen beiden Aufnahmen, nach dem Herausteilen des Randlichtabfalls des Objektivs, nur noch in einer finalen Faltung mit dem Diffusoreinfluss.

Es wurde oben ein Modell für die Strahlaufweitung vorgestellt. Allerdings haben wir dort gesehen, dass der Parameter d nicht unbedingt die physikalische Dicke darstellt. Die effektive Dicke ist somit unbekannt und kann sich mit $\beta^{in}$ ändern. Daher werden für mehrere $\beta^{in}$ Fits durchgeführt. Bei jeder Iteration wird die effektive Dicke variiert, das Bild ohne Diffusor mit der Strahlaufweitung gefaltet und mit der entsprechenden Aufnahme mit Diffusor verglichen. Aufnahmen ohne Diffusor sind jedoch nur bei senkrechtem Einfall möglich, da der Strahl für andere Winkel an der Kamera vorbei führt und somit nicht auf den CCD-Chip trifft. Für diese Winkel muss daher das Bild ohne Diffusor aus der Aufnahme bei senkrechtem Einfall berechnet werden.

Um die benötigten Messbilder zu bekommen, wurden mit dem vorgestellten Aufbau in 8°-Schritten zwischen 0° und 40°, drei durch die Blende beleuchtete Diffusoren je drei Mal fotografiert, um das statistische Rauschen der Kamera und die Diffusorstruktur zu reduzieren. Weiter wurde dieselbe Anzahl an Untergrundsaufnahmen gemacht. Bei senkrechtem Einfall wurden einerseits 5 Aufnahmen ohne Diffusor und ohne Blende gemacht (mit zugehörigen Untergrundbildern), andererseits 5 Aufnahmen ohne Diffusor mit Blende (ebenfalls mit Untergrundaufnahme).

Da die Blende direkt den Diffusor berührt[19], ist es nicht möglich den Diffusor zu wechseln ohne die Blende leicht zu verschieben. Die Bilder müssen also zur Mittelung relativ zueinander neu Positioniert werden. Dies wird getan, indem die quadratische pixelweise Differenz zwischen je zwei Aufnahmen verschiedener Diffusoren durch Verschiebung eines der Bilder minimiert wird. Anschließend werden diese gemittelt. Doch bleibt nach dieser Mittelung eine Gewisse Diffusorstruktur übrig. Diese wird als zusätzlicher Fehler zum statistischen Rauschen der Kamera im direkt beleuchteten Bereich behandelt.

Um zu überprüfen, dass die Diffusorstruktur als statistischer Fehler behandelt werden darf, wurden 3 Aufnahmen von verschiedenen homogen beleuchteten Diffusorflächen aufgenommen. Aus jeder Aufnahme wurde die Standardabweichung der 40000 Grauwerte untereinander berechnet, die aufgrund der homogenen Beleuchtung in guter Näherung nur auf die Diffusorstruktur und Kamerastatistik zurückzuführen ist. Anschließend wurden diese Bilder gemittelt und für das gemittelte Bild ebenfalls die Standardabweichung berechnet. Daraus ergibt sich für die Einzelaufnahmen eine mittlere relative Standardabweichung von 8,22% und für die Standardabweichung des gemittelten Bildes eine Standardabweichung von 4,71%. Wird der zweite Wert mit $\sqrt{3}$ multipliziert (siehe Gleichungen 10 und 11), wobei 3 die Anzahl der gemittelten

---

[19] Das muss sie, denn ansonsten würde der beleuchtete Ort auf dem Diffusor zu sehr mit dem Einfallswinkel wandern und nicht mehr für alle Winkel auf dem CCD sichtbar sein.

Aufnahmen ist, ergibt dies 8,16%. Diese Abweichung kann nicht vollständig aus dem Fehler aufgrund der Berechnung der Standardabweichung aus einer Stichprobe erklärt werden [Squ01]. Dennoch ist die Abweichung zwischen beiden Werten so gering, dass kein großer Fehler gemacht wird, wenn angenommen wird, dass die Diffusorstruktur sich statistisch verhält. Dies gilt allerdings nur sofern die Diffusorstruktur schmal verglichen mit dem betrachteten Bereich ist, da Löcher und dichte Stellen im Diffusor sich über mehrere Pixel erstrecken und daher Nachbarpixel korrellieren.

Die nötige Korrektur des Randlichtabfalls erwies sich jedoch als Problem, denn wie in Kapitel 4.2.1 erläutert wurde, ist diese von der Divergenz der Bestrahlung abhängig. Zur Bestimmung des Randlichtabfalls reicht es prinzipiell eine homogene Lichtquelle direkt zu fotografieren. Doch, im Gegensatz zu diffuser Strahlung, bei der der Himmel fotografiert werden kann, ist eine solche Quelle für die hier benutzte Divergenz nicht natürlich vorhanden. Eine im Labor erstellte Beleuchtung wäre bezüglich der Divergenz und der Homogenität fehleranfällig. Es ist daher weniger aufwendig und nicht zwangsweise ungenauer die Lichtquelle des Aufbaus möglichst homogen zu gestalten und als solche anzunehmen. Bei direkter Betrachtung der Beleuchtung, wird dann der Abfall der Intensität zum Rand des CCD-Chips hin als reiner Randlichtabfall interpretiert.

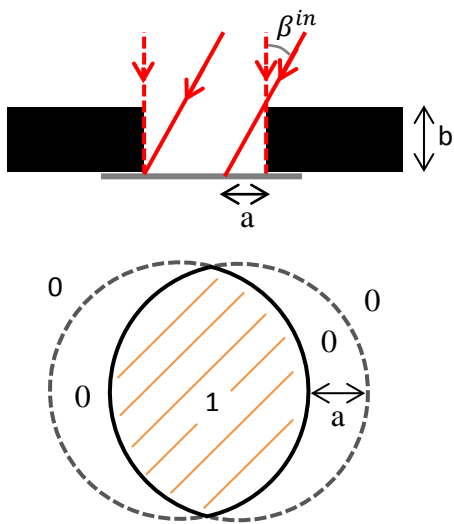

Abbildung 44: Verschattung bei schrägem Lichteinfall

Durch die endliche Dicke der Blende kommt es bei schrägem Lichteinfall zu einer Verschattung der Austrittsapertur durch die Eintrittsapertur, wie in Abbildung 44 zu sehen ist, sodass die beleuchtete Fläche kleiner ist als die Öffnung der Blende. Dabei beträgt der Versatz zwischen den Schattenwürfen der Vorderseite und der Rückseite der Blende $a = b \cdot \tan \beta^{in}$. Um dies zu berücksichtigen, wurde aus den Aufnahmen bei senkrechtem Einfall ohne Diffusor aber mit Blende die Größe der Blende bestimmt und für jeden Winkel die entsprechende Funktion $\delta_{\text{Ausschnitt Blende}}(\beta^{in})$ erstellt, die in Abbildung 44 unten zu sehen ist. Diese enthält als Wert für jeden Pixel, den Anteil der Fläche des Pixels, die innerhalb der Schnittfläche der zwei Schattenwürfe liegt. Anhand dieser Funktion kann nun die gemittelte Aufnahme bei senkrechtem Einfall ohne Diffusor und ohne Blende beschnitten werden um die dem Winkel $\beta^{in}$ entsprechende Beleuchtungsfunktion zu bekommen. Dazu muss noch das Zentrum dieser Funktion an die richtige Stelle gesetzt werden. Theoretisch ist das Zentrum nur um $a = b \cdot \tan(\beta^{in})/2$ gegenüber der Position bei senkrechtem Einfall verschoben. Doch ist ein versehentlicher kleiner Abstand zwischen Blende und Diffusor gegenüber den ca. 120µm Dicke der Blende nicht vernachlässigbar. Dies führt, zusammen mit einer leichten Dezentrierung des Rotationspunktes, dazu, dass das Zentrum der Blendenfunktion nicht berechnet werden kann. Stattdessen wird die Position des Zentrums aus den Aufnahmen mit Diffusor für die jeweiligen Winkel bestimmt.

Die so erhaltenen Bilder werden anschließend mit einer zweidimensionalen Gaußfunktion für die PSF gefaltet. Erwartungsgemäß ändert diese Faltung wenig am Ergebnis. Dies kann man qualitativ verstehen, wenn man den Verlauf der Diffusorstrahlaufweitung grob als Gauß nähert. Die PSF beschreiben wir ohnehin durch einen Gauß. Dann sind zwei konsekutive Faltungen äquivalent zu einer Faltung mit einem Gauß, dessen Varianz gleich der quadratischen Addition der zwei einzelnen Varianzen ist. Wie wir sehen werden, liegt ungefähr ein Faktor 10 zwischen den FWHM der PSF und der Verbreiterung durch die Faltungsfunktion des Diffusors. Bei einem Gauß entspricht die FWHM $2\sqrt{2\ln 2}$ Standardabweichungen. Das Verhältnis der Standardabweichungen ist dasselbe wie das der FWHM. Durch die quadratische Addition ergibt sich dann ein Faktor 100. Die PSF macht daher nur ca. 1% der gemessenen Strahlaufweitung aus. Aufgrund des geringen Einflusses der PSF wurde der Fehler aufgrund der Beschreibung der PSF durch einen Gauß und der Fehler auf die Breite der PSF vernachlässigt.

Die erhaltenen Bilder unterscheiden sich nun nur noch um eine Skalierung und um die gesuchte Faltung von den Messbildern mit Diffusor. Die Skalierung entspricht dem Transmissionsgrad des Diffusors. Um die reine Faltung zu bekommen müssen wir die Aufnahmen noch vom Transmissionsgrad bereinigen. Dies tun wir indem wir die Summe der Grauwerte der Messbilder und nach jedem Iterationsschritt des Fits die Summe der Grauwerte des gefalteten Bildes auf 1 normieren.

Bei dem Fit wird die effektive Dicke des Diffusors angepasst um den Term

$$\chi^2 = \sum_{pxl} \left[\frac{Messbild - Fitfunktion(d)}{\sigma}\right]^2 \qquad (25)$$

zu minimieren, wobei die Fitfunktion das gefaltete berechnete Bild, also

$$\left[\left(Beleuchtung(r) \cdot \delta_{Ausschnitt\;Blende}(\beta^{in})\right) * PSF\right] * \text{Aufweitung durch Diffusor}(\beta^{in}) \qquad (26)$$

ist. Wie oben erläutert enthält der Fehler des Messbildes den statistischen Fehler der Kamera und die Unsicherheit durch die Überreste der Diffusorstruktur. Damit sind jedoch nicht alle Fehlerquellen abgedeckt, denn unsere Fitfunktion enthält ebenfalls Fehler. $\sigma$ würde sich demnach aus $\sqrt{\sigma_{Messbild}^2 + \sigma_{Fitfunktion}^2}$ ergeben.

Bei der Berechnung der Verschattung bei schrägem Lichteinfall spielt die Dicke der Blende herein, die mit einer Mikrometerschraube gemessen wurde und eine Unsicherheit besitzt. Wichtiger ist jedoch, dass der Versatz in Pixel umgerechnet werden muss. Zwar ist die Kantenlänge der Pixel mit 7,4µm sehr genau bekannt, doch spielt ebenfalls der Zoomfaktor des Objektivs in die Umrechnung herein. Dieser ist mit 1,980±0,003 zwar auf 0,15% genau bestimmt worden, wirkt sich aber stark überproportional auf das Ergebnis aus.

Die Behandlung dieser Fehler ist nicht trivial, denn die Gauß'sche Fehlerfortpflanzung ist hier nicht gültig. Die gesamte Gauß'sche Fehlerfortpflanzung beruht auf der Annahme, dass die betrachtete Funktion im Intervall $[x_0, x_0 + \sigma_{x_0}]$ durch ihre Tangente an der Stelle $x_0$ approximiert werden kann. Erstens fließt diese Näherung durch die Taylorentwicklung 1. Ordnung in die allgemeine Formel

$$\sigma_{y(x)} = \sqrt{\sum_i \left(\frac{\partial y(x)}{\partial x_i} \cdot \sigma_{x_i}\right)^2} \qquad (27)$$

ein, die im Prinzip durch eine Taylornäherung höherer Ordnung ersetzt werden könnte. Zweitens und viel entscheidender ist jedoch, dass nur unter dieser Annahme die Gaußkurve, die die Wahrscheinlichkeitsverteilung des x-Werts beschreibt unter Anwendung der betrachteten Funktion $y(x)$ wieder eine Gaußkurve als Beschreibung der Wahrscheinlichkeitsverteilung in $y$ ergibt (siehe Abbildung 45). Dies ist zugleich die Bedingung dafür, dass der Standardabweichung eine Bedeutung zukommt.

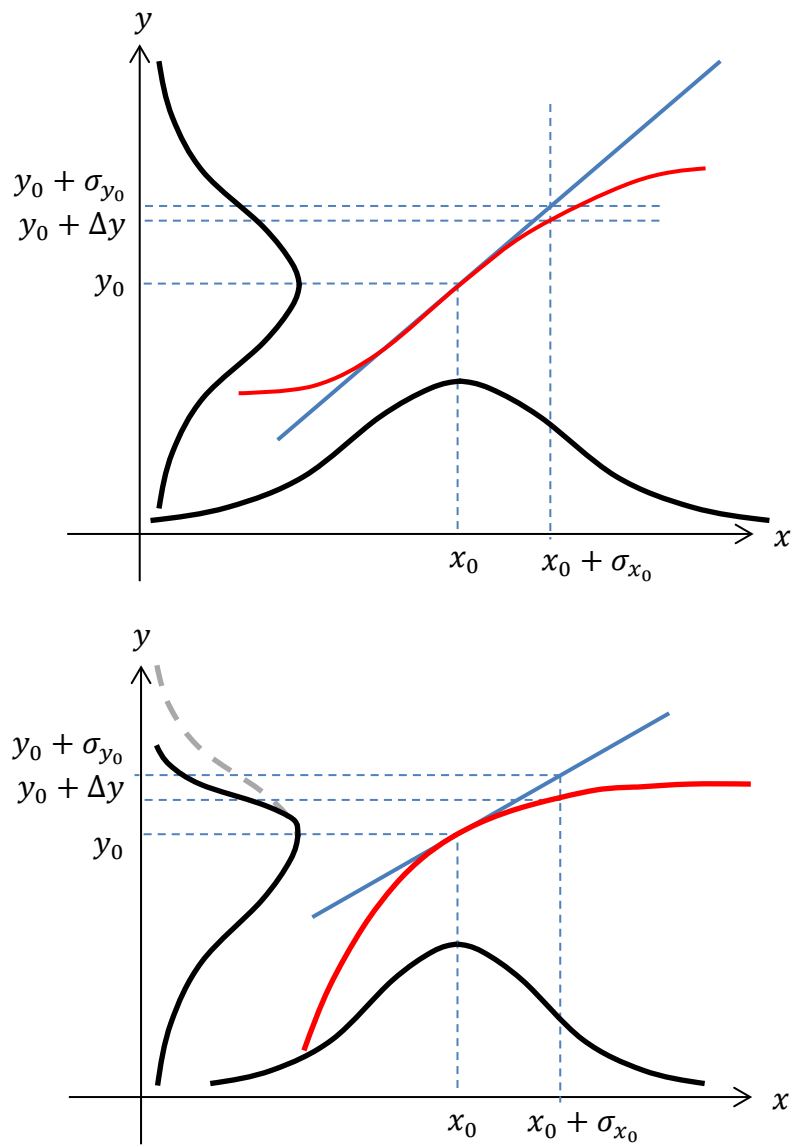

**Abbildung 45: Bedingung für die Gültigkeit der Gauß'schen Fehlerfortpflanzung. Oben: Bei guter lokaler Näherung der Funktion durch ihre Tangente ergibt die Abbildung einer gaußförmigen Wahrscheinlichkeitsverteilung des x-Werts wieder eine Gaußförmige Wahrscheinlichkeitsverteilung für den y-Wert. Hier gilt $\Delta y \approx \sigma_y$. Unten ist dies nicht der Fall.**

Betrachten wir die Funktion $\delta_{\text{Ausschnitt Blende}}(\beta^{in}, b, \text{zoom})$, wobei b die Dicke der Blende war, so fällt auf, dass diese Funktion im Grenzbereich des Schattenwurfs innerhalb von maximal 2 Pixeln[20] in radialer Richtung zwischen den Werten 0 und 1 wechselt und diese Werte dann jeweils konstant annimmt. Dabei hat diese Funktion durch den Fehler auf den Zoomfaktor und den Fehler auf die Dicke der Blende eine Unsicherheit auf die Lage der Stufe in radialer Richtung. Es ist analog ob die Position der Grenze des Schattenwurfs oder die Position des Pixels fehlerbehaftet ist. Um die Analogie zum Allgemeinen Fall von oben herzustellen wird die letztere Betrachtung gewählt. Pixel die weiter als drei Standardabweichungen von der Stufe entfernt liegen haben somit eine vernachlässigbare Wahrscheinlichkeit, dass ihr y-Wert fehlerhaft ist. Kritisch ist der Fehler für Pixel, die aufgrund der Standardabweichung in x eine nicht zu vernachlässigende Wahrscheinlichkeit haben auf der anderen Seite der Stufe zu liegen.

---

[20] Ein Pixel kann auf dem Grenzbereich liegen und den Wert annehmen, der seinem Flächenanteil im Schattenwurf entspricht.

Wie in Abbildung 46 zu sehen, ergibt eine Gaußverteilung als Wahrscheinlichkeitsverteilung der $x$ - Position bei einer Abbildung mit der Funktion $\delta_{\text{Ausschnitt Blende}}(\beta^{in}, b, \text{zoom})$, zwei Delta-Peaks — einen bei 0, einen bei 1 — für die Wahrscheinlichkeitsverteilung der $y$ -Position. Deren Höhen sind durch die Integrale über die Wahrscheinlichkeitsverteilung in x auf der jeweils entsprechenden Seite der Stufe gegeben. Damit ist die Standardabweichung an dieser Stelle kein angebrachtes Maß zur Beschreibung des Fehlers. Es kann daher keine Standardabweichung $\sigma_{Fitfunktion}$ der Fitfunktion berechnet werden um diese beim Fit in die zu minimierende $\chi^2$-Funktion einzusetzen.

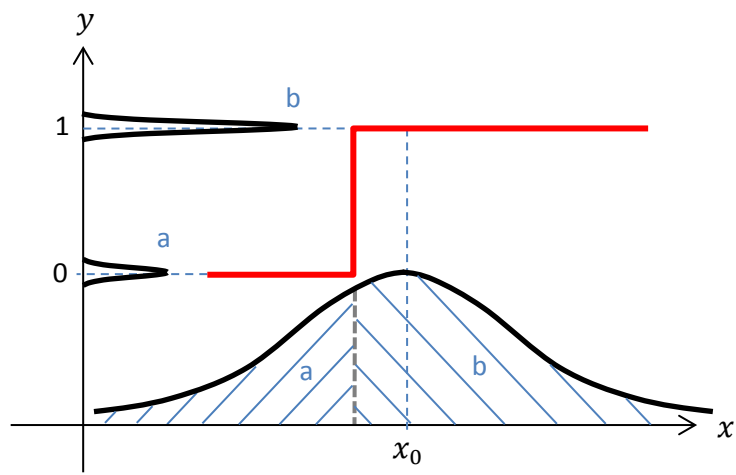

**Abbildung 46: Fortpflanzung eines gaußverteilten Fehlers bei Anwendung einer Stufenfunktion**

Um die Fehler auf den Zoom und die Dicke der Blende trotzdem berücksichtigen zu können wird zu Fehlergrenzen übergegangen. Dazu wird der Fit durchgeführt, wobei für $\sigma$ in obiger Formel für $\chi^2$ nur der Fehler des Messbildes berücksichtigt wird. Dies wird dann zweimal wiederholt, wobei für die Breite der Verschattung aufgrund der Dicke der Blende der berechnete Wert $\pm$ 3 Standardabweichungen dieses Wertes eingesetzt wird. Es werden also für jeden der drei Fits leicht unterschiedliche Funktionen $\delta_{\text{Ausschnitt Blende}}(\beta^{in}, b, \text{zoom})$ gefaltet.

Beispielhaft ist in Abbildung 47, Abbildung 48 und Abbildung 49 ein Schnitt durch die Mitte der zweidimensionalen Fits für 0°, für 16° und 32° zu sehen. Daran ist zu sehen, dass der $\cos^4$-Verlauf nur näherungsweise passt. Diese Abweichungen können durch die bei der Beschreibung des Modells genannten groben Annahmen erklärt werden. Die Anzahl der Mittelungen über die Diffusorstruktur reicht nicht aus um mit hinreichender Genauigkeit anhand der Schnitte Aussagen zu Asymmetrien und Übereinstimmung der Verläufe der experimentellen Kurve und des Fits in Abhängigkeit des Einfallwinkels machen zu können. Die in diesen Abbildungen zu erkennende steigende Abweichung der Messung vom cos⁴-Modell bestätigt sich nicht bei 40° Einfallwinkel und nicht systematisch bei der Wahl anderer Schnitte. Deshalb werden in Kapitel 5.3 die oben gemachten Annahmen und die daraus resultierenden Ergebnisse kritisch hinterfragt.

Wir bekommen aus diesen Fits für jeden Winkel eine effektive Dicke und die zugehörigen Fehlergrenzen. Hierbei sind alle Längen noch in Pixeln, wobei 1 Pixel $7{,}4\,\mu m/Zoomfaktor = 3{,}737 \pm 0{,}005$ µm in der Objektebene entspricht. Die Ergebnisse sind in folgender Tabelle zusammengefasst und in Abbildung 50 links aufgetragen.

| Winkel / ° | Effektive Dicke / Pixel | Untere Grenze der effektiven Dicke / Pixel | Obere Grenze der effektiven Dicke / Pixel |
|---|---|---|---|
| 0 | 32,811 ± 0,008 | 32,811 ± 0,008 | 32,811 ± 0,008 |
| 8 | 33,601 ± 0,009 | 33,433 ± 0,009 | 33,769 ± 0,009 |
| 16 | 34,95 ± 0,01 | 34,60 ± 0,01 | 35,30 ± 0,01 |
| 24 | 35,95 ± 0,01 | 35,40 ± 0,01 | 36,49 ± 0,01 |
| 32 | 37,66 ± 0,01 | 36,89 ± 0,01 | 38,45 ± 0,01 |
| 40 | 38,90 ± 0,02 | 37,86 ± 0,01 | 39,96 ± 0,02 |

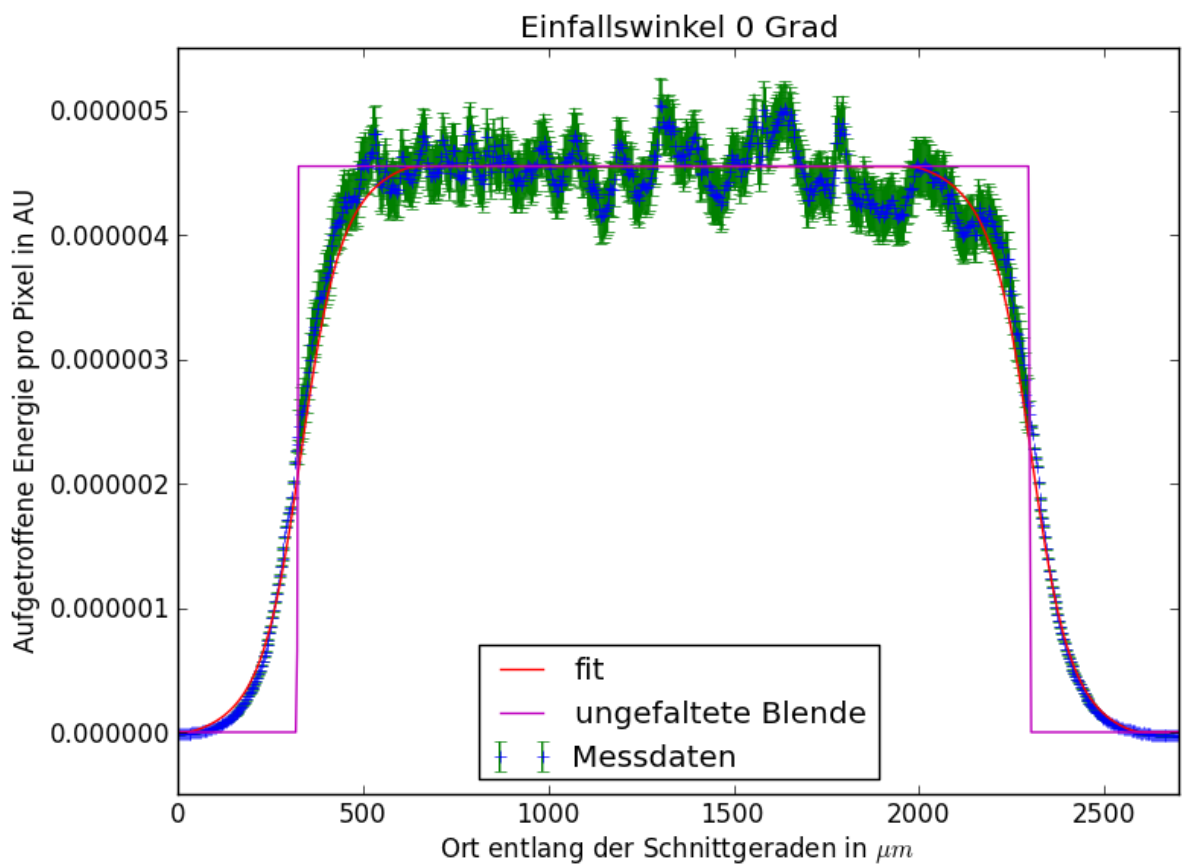

**Abbildung 47: Schnitt durch die Mitte des zweidimensionalen Fit der Diffusorstrahlaufweitung bei senkrechtem Einfall. In blau die Messpunkte, grün die zugehörigen Fehlerbalken, magenta das berechnete Eingangssignal und rot die Fitkurve.**

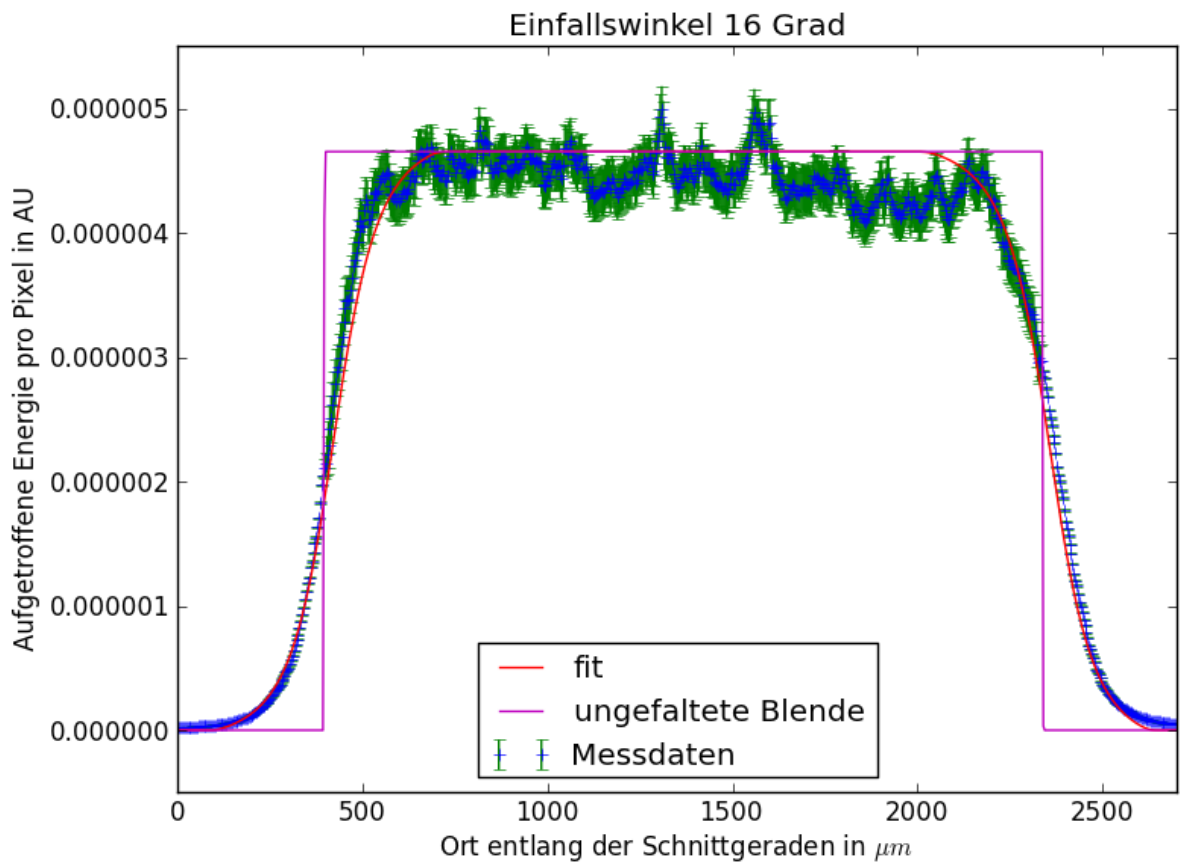

Abbildung 48: Schnitt durch die Mitte des zweidimensionalen Fit der Diffusorstrahlaufweitung bei einem Beleuchtungswinkel von 16°. In blau die Messpunkte, grün die zugehörigen Fehlerbalken, magenta das berechnete Eingangssignal und rot die Fitkurve.

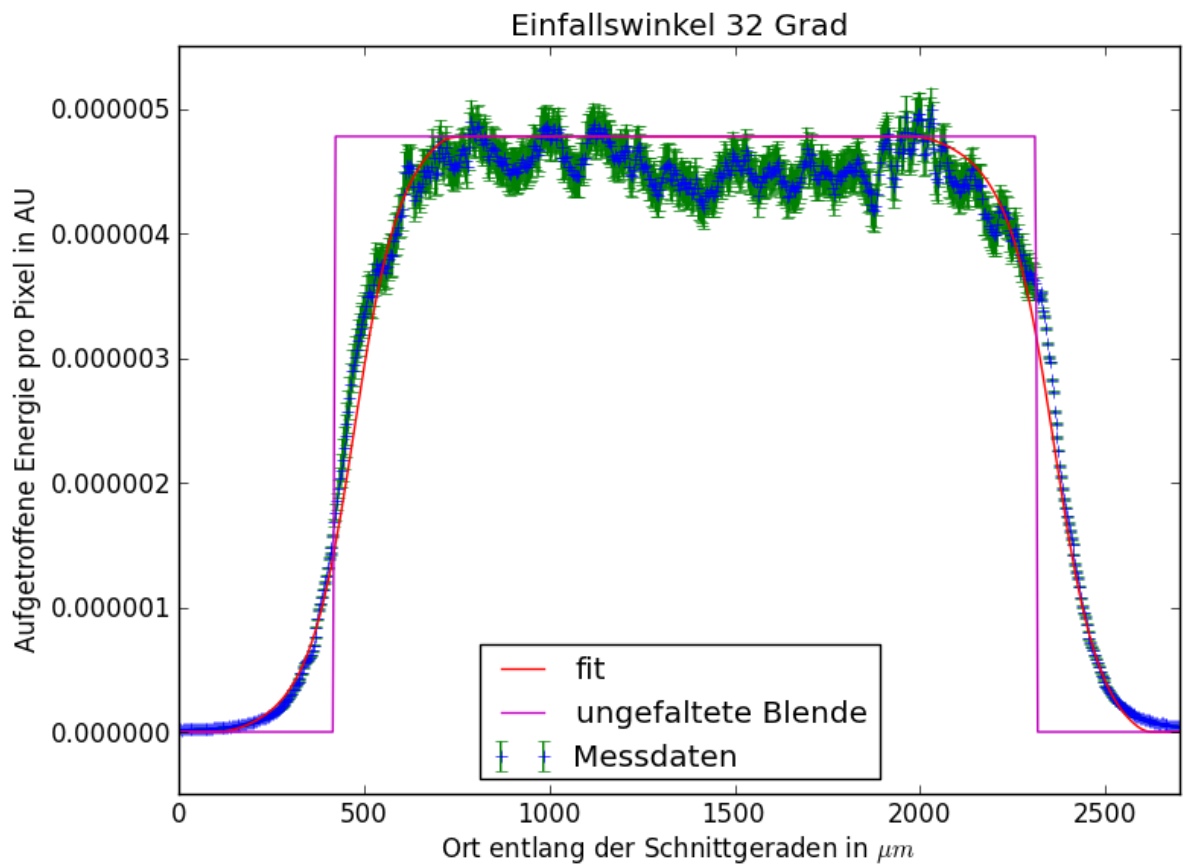

Abbildung 49: Schnitt durch die Mitte des zweidimensionalen Fit der Diffusorstrahlaufweitung bei einem Beleuchtungswinkel von 32°. In blau die Messpunkte, grün die zugehörigen Fehlerbalken, magenta das berechnete Eingangssignal und rot die Fitkurve.

Wir bekommen aber im RMP beliebige Einfallswinkel, die je nach Spiegel bis zu 45° erreichen können und brauchen daher für die Behandlung der Strahlen im Raytracing für jeden beliebigen Einfallswinkel eine zugehörige effektive Dicke. Da das Modell für die Diffusorstahlaufweitung keine genau physikalische Beschreibung darstellt ist es auch nicht möglich daraus einen physikalisch fundierten Verlauf zwischen den Stützstellen abzuleiten. Der geometrische Ansatz

$$d = \frac{d_0}{\cos\left[\sin^{-1}\left(\sin\beta^{in}/n_{PTFE}\right)\right]} \quad (28)$$

wobei $d_0$ die effektive Dicke für senkrechten Einfall und $n_{PTFE}$ der Brechungsindex von PTFE ist, liefert keine brauchbaren Ergebnisse. Auch ein Polynomialfit höherer Ordnung wäre aufgrund der geringen Anzahl an Stützstellen willkürlich. Daher wurde der Weg gewählt die Erhöhung der effektiven Dicke des Diffusors mit dem Einfallswinkel durch einen linearen Fit zu beschreiben.

Da bei senkrechtem Einfall kein Versatz vorhanden ist, sind dort untere und obere Grenze, solange sie in Pixeln ausgedrückt werden, gleich der effektiven Dicke. Der Achsenabschnitt der Geraden die die effektive Dicke und die jeweiligen Grenzen beschreiben muss demnach ebenfalls gleich sein. Daher wurden die drei geraden gekoppelt, das heißt mit einem selben Parameter für den Achsenabschnitt und unter Minimierung der Summe der $\chi^2$ der einzelnen Geraden (abzüglich des doppelten Beitrages des ersten Punktes, der ansonsten dreifach gewichtet würde) gefittet.

Bei der Umrechnung der effektiven Dicken in µm gilt wieder $1\ Pixel = 7,4/Zoomfaktor$. Der Zoomfaktor ist wie vorher fehlerbehaftet. Jedoch haben wir durch die vorhergehende Rechnungen schon festgelegt welcher Zoomfaktor für welche Gerade eingesetzt werden muss: Ein höherer Zoomfaktor verursacht eine größere in Pixel gemessene Verschattung durch die Dicke der Blende, also eine schmalere $\delta_{\text{Ausschnitt Blende}}(\beta^{in})$-Funktion, was durch eine höhere effektive Dicke der Faltungsfunktion ausgeglichen werden muss. Der höhere Zoomfaktor gehört also zur oberen Grenze der effektiven Dicke (siehe Abbildung 50 rechts).

Die Fitparameter sind in folgender Tabelle zusammengefasst.

|  | Achsenabschnitt / µm | Steigung / $\mu m/°$ |
|---|---|---|
| Ermittelter Zoom | 121 | 0,60 |
| Untere Grenze des Zooms | 121,5 | 0,51 |
| Obere Grenze des Zooms | 120,5 | 0,69 |

Die Werte für 0° können mit den Ergebnissen der Messung aus [Kis11] verglichen werden. Die Messung ergab hier $d \approx 126{,}1\mu m$ [Kis11 p. 49]. Da kein Fehler auf diesen Wert angegeben wurde kann über eine Übereinstimmung der Messungen nicht geurteilt werden.

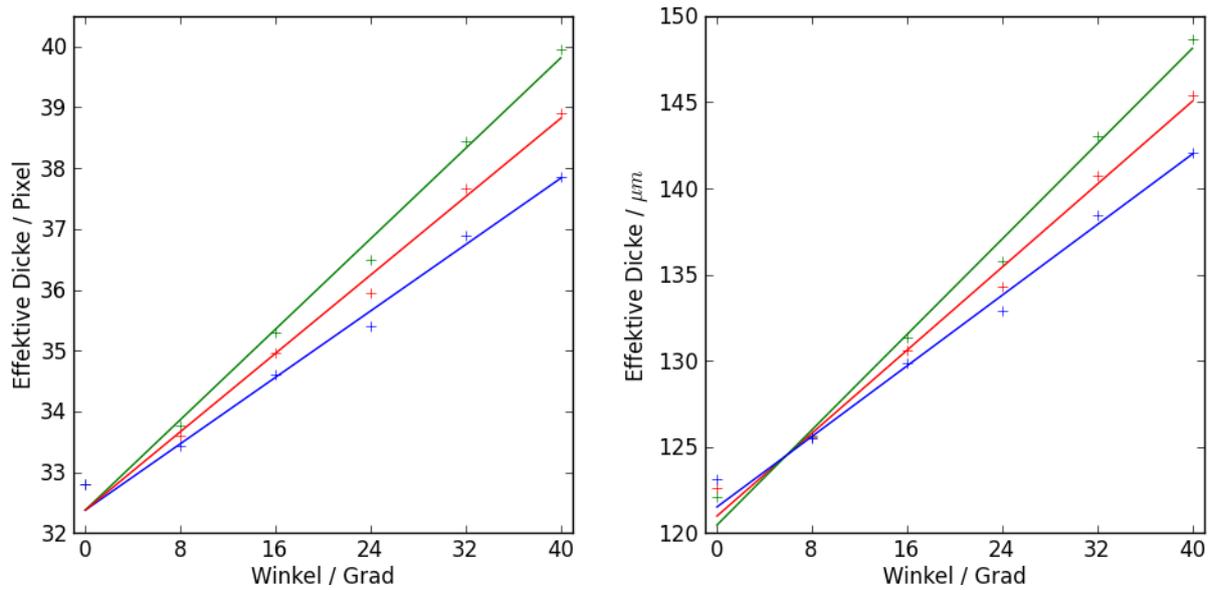

**Abbildung 50:** Effektive Dicke des Diffusors, links in Pixeln, rechts in Mikrometern. In rot die Fit Ergebnisse, in grün und in blau die oberen und unteren Grenzen.

### 4.6.2. Winkelabhängigkeit des Transmissionsgrades

Neben der Strahlaufweitung sind die Transmissionseigenschaften des Diffusors für die Messungen am RMP von großer Bedeutung. Die Lichtstrahlen treffen am RMP unter stark unterschiedlichen Winkeln auf den Diffusor. Es ist daher wichtig, wie sich der Transmissionsgrad in Abhängigkeit der Einfallswinkel ändert. Am RMP treffen nur Strahlen auf den CCD, deren Winkel beim Auftreffen auf das Objektiv kleiner sind als dessen Toleranzwinkel. Letzterer beträgt nur einige Grad. Daher ist für die Transmission durch den Diffusor nur die Strahlstärke für Austrittswinkel nahe 0°, also nahezu in Richtung der Diffusornormalen interessant. Von T. Schmid wurde daher der Transmissionsgrad in Richtung der Oberflächennormale in Abhängigkeit des Einfallswinkels für eine Wellenlänge von 630nm gemessen. Das Ergebnis ist in Abbildung 51 zu sehen. Da nur für ganze Gradzahlen ein Messwert vorliegt wird bei der Simulation für jeden Einfallswinkel zwischen den zwei nächsten Werten interpoliert.

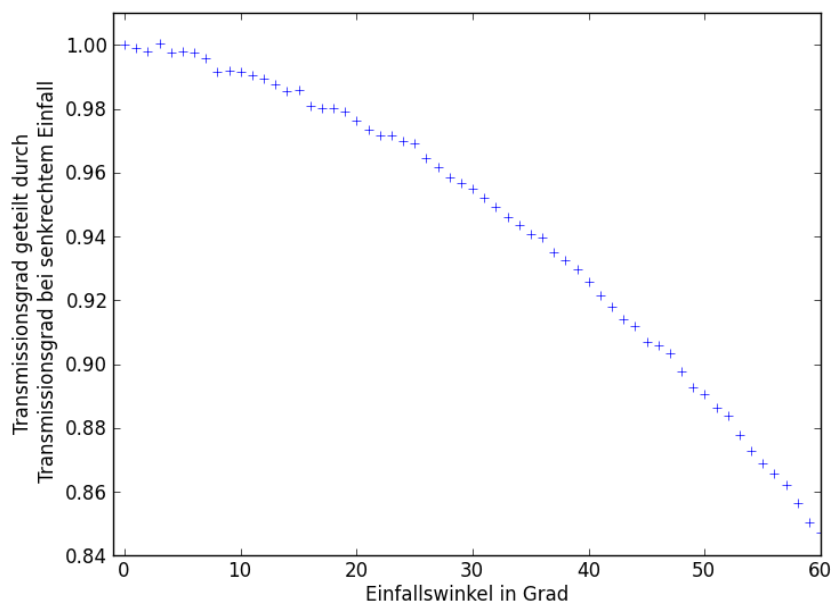

**Abbildung 51:** Einfallswinkelabhängigkeit des Transmissionsgrades des Diffusors

In einer getrennten Messung wurde von T. Schmid die Austrittswinkelabhängigkeit der Transmittivität zwischen -5° und 5° zur Diffusornormalen bei senkrechter Bestrahlung bestimmt. Es konnte im Rahmen der Messgenauigkeit von einem Prozentpunkt keine Austrittswinkelabhängigkeit festgestellt werden. Sie wird daher am RMP vernachlässigt.

### 4.6.3. Wellenlängenabhängigkeit des Transmissionsgrades

Wie schon der Transmissionsgrad der Glasplatte und die Reflektivitäten der Spiegel, ist auch der Transmissionsgrad des Diffusors wellenlängenabhängig. Der Transmissionsgrad bei senkrechter Bestrahlung und hemisphärischer Messung wurde vom Hersteller mitgeteilt und vom T. Schmid an dem in Kapitel 4.5 vorgestellten Aufbau nachgemessen. Am RMP ist nicht der über alle Austrittswinkel integrierte Transmissionsgrad interessant, sondern nur der für Austrittswinkel nahe 0°. Es wird aber angenommen, dass die Wellenlängenabhängigkeit in guter Näherung nicht von der Austrittsrichtung abhängig ist. Daher sind die Ergebnisse bei hemisphärischer Messung proportional zu den gewünschten bei senkrechter Austrittsrichtung. Die Ergebnisse sind in Abbildung 52 zu sehen. Der Ursprung des leichten Unterschieds zwischen den zwei Messungen ist nicht bekannt. Da es sich nicht um eine Skalierung handelt, kann er nicht auf eine Schwankung der Lichtquelle zurückgeführt werden. Da der vom Hersteller vermessene Diffusor zwar dasselbe Produkt ist, jedoch nicht zwangsweise unter denselben Bedingungen wie der am RMP eingesetzte Diffusor produziert wurde, wird im Folgenden die Messung von T. Schmid verwendet.

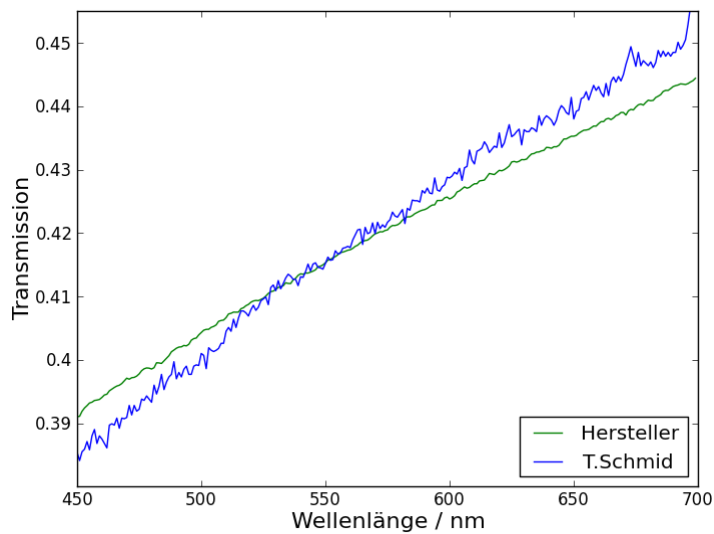

**Abbildung 52: Wellenlängenabhängigkeit der Transmission des Diffusors**

### 4.6.4. Rückstreuung durch den Diffusor

Die Absorption durch den Diffusor ist sehr gering. Die meiste nicht transmittierte Strahlungsleistung wird rückgestreut. Damit beträgt die Reflektivität des Diffusors, je nach Einfallswinkel um die 56%. Die rückgestreuten Lichtstrahlen werden in den Halbraum abgestrahlt und treffen dabei teilweise auf die Probe, den Kalibrier- oder Hellbildspiegel. Der Anteil, der nahezu in Richtung der Diffusornormalen abgestrahlt wird und dadurch direkt zurück auf den Diffusor reflektiert wird sorgt für etwas Streulicht auf dem Diffusor. Der betroffene Raumwinkel bei der Abstrahlung ist jedoch so klein, dass dieses Streulicht nicht bedeutend ist. Durch die Schrägstellung der Glasplatte wurden auch Reflexionen an dieser über die Probe oder den Kalibrierspiegel minimiert (siehe Abbildung 10). Rückreflexionen in die Lichtquelle sind ebenfalls nicht von Bedeutung, da die Innenverkleidung der Lichtquelle nur schwach und diffus reflektiert, womit sich die Strahlungsleistung im Raum verteilt.

In diesem Kapitel wurden die diversen Eigenschaften der Bauteile analysiert. Ihr Einfluss auf das Ergebnis des RMPs kann dabei oft nur durch die gemeinsame Betrachtung mehrerer Effekte abgeschätzt oder berechnet werden. Dies wird in den Kapiteln 5.1 mit der Zusammenfassung der Einflüsse der verschiedenen Bauteile auf die Konzentrationsmatrix und Kapitel 5.2 mit der Berechnung der Wellenlängenabhängigkeit getan.

# 5. Charakterisierung des Gesamtaufbaus

## 5.1. Zusammensetzung der Konzentrationsmatrix

Nun, da die Einflüsse der einzelnen Komponenten des RMPs genauer bekannt sind, wird hier auf die nötigen Annahmen und Rechenschritte eingegangen, die nötig sind um die Konzentrationsmatrix zu berechnen. Dabei wird dem Lauf des Strahlengangs gefolgt. Es sei dabei daran erinnert, dass das Ziel letztendlich nur die Berechnung des Bruchs Messbild durch Hellbild ist, was einige Vereinfachungen mit sich zieht.

Wie im Kapitel 4.3 gesehen, wird derzeit nur eine vorläufige Lichtquelle benutzt. Aufgrund mangelnder Intensität, müssen wir beim Hellbild auf eine zweite Homogenisierung verzichten. Dadurch bekommen wir für das Hellbild und für das Messbild zwei unterschiedliche Beleuchtungsfunktionen, die wir als $LQ^{Hell}(\lambda)$ und $LQ^{Mess}(\lambda)$ bezeichnen wollen. Diese geben die Strahlungsleistung der Lichtquelle für jede Wellenlänge an. Annahme ist dabei, dass die Beleuchtung örtlich homogen ist, worauf wir im Kapitel 5.3 zurückkommen werden. Die zwei Funktionen unterscheiden sich dabei hauptsächlich in ihrer Amplitude, jedoch auch leicht in der Wellenlängenabhängigkeit.

Dieses Licht trifft anschließend auf die Glasplatte. Das Verhalten dieser kann, da uns der Strahlversatz nicht interessiert, gut durch den Transmissionsgrad $T^{Glas}(\lambda)$ beschrieben werden. Die Inhomogenität der Transmission wird vernachlässigt. Auch werden in diesem Modell weitere Reflexionen an der Glasplatte, zum Beispiel nach Streuung am Diffusor oder am Diffusor- und Umlenkspiegelhalter, vernachlässigt.

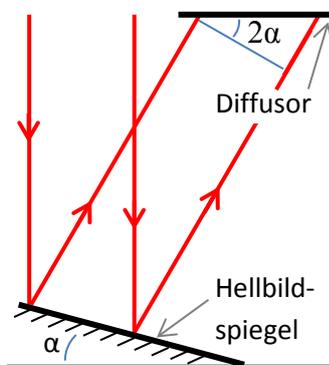

**Abbildung 53: Schwächere Beleuchtungsstärke am Diffusor aufgrund der Verkippung des Hellbildspiegels**

Sowohl bei dem Hellbildspiegel, wie auch bei dem Kalibrierspiegel, ist die Strahlaufweitung vernachlässigbar. Das Verhalten dieser Spiegel kann daher auf ihren Einfluss auf den Strahlengang einerseits, auf die spektral- und winkelabhängigen Reflektivitäten $\tilde{R}^{Hell}(\lambda, \gamma^{in})$ und $\tilde{R}^{Kalib}(\lambda, \gamma^{in})$ andererseits reduziert werden, wobei $\gamma^{in}$ der Einfallswinkel ist. Wir setzen für die Reflektivitäten den Separationsansatz $\tilde{R}(\lambda, \gamma^{in}) = R(\lambda) \cdot R'(\gamma^{in})$ an, was angesichts der Fresnel'schen Formeln eine Näherung ist. Der Einfluss des Kalibrierspiegels auf den Strahlengang wird per Raytracing simuliert und durch die Funktion $f^{Kalib}(\vec{r})$ dargestellt, die die Konzentration je Einfallswinkel an jedem Ort des Diffusors angibt. Jeder Strahl wird dabei je nach Einfallswinkel auf den Spiegel mit dem entsprechenden Wert $R'(\gamma^{in})$ gewichtet. Dieser Wert entspricht dem Reflektivitätsverhältnis zwischen senkrechtem Einfall und Einfall unter dem Winkel $\gamma^{in}$ zur Oberflächennormale des Spiegels am Auftreffpunkt des Lichtstrahls. Der Hellbildspiegel kann, aufgrund der Verschattung durch den Halter nicht waagrecht hingelegt werden. Daher wird er leicht seitlich versetzt unter einem wohldefinierten Winkel $\alpha$ platziert. Hierdurch wird das Licht beim Auftreffen auf den Diffusor auf eine größere Fläche verteilt, wodurch die Intensität des Hellbildes sinkt. Um dies zu korrigieren muss das Hellbild durch $\cos(2\alpha)$ geteilt werden (siehe Abbildung 53). Da die Beleuchtung des Hellbildspiegels als homogen angenommen wird, gilt dies auch für die am Hellbildspiegel reflektierte Beleuchtung des Diffusors.

Komplizierter sieht es aus, wenn anstelle des Kalibrier- oder des Hellbildspiegels eine Probe betrachtet wird. Auch hier gibt es eine wellenlängen- und winkelabhängige Reflektivität sowie eine von der Geometrie des Spiegels abhängige örtliche Verteilung der Beleuchtungsstärke je Einfallswinkel $f^{Probe}(\vec{r})$. Zusätzlich haben die Spiegel hier aufgrund von Formabweichungen von der Sollgeometrie Winkelfehler die über den Spiegel statistisch verteilt sind, sodass jedem einfallenden Strahlenbündel eine Winkelverteilung für den Austrittswinkel zugeordnet wird. Dies ist was uns letztendlich interessiert und wird im Kapitel 5.4 genauer behandelt.

Das durch diese Spiegel reflektierte Licht trifft anschließend auf den Diffusor. Dieser ist nicht trivial zu beschreiben. Zunächst ist die Transmission des Diffusors, aufgrund von Dichteschwankungen im Material und von Löchern ortsabhängig. Wir beschreiben dies durch die Transmissionsfunktion $T(\vec{r})$. Teilt man das Messbild pixelweise durch das Hellbild, sollte sich diese Ortsabhängigkeit herausteilen. Jedoch stimmt dies nur in erster Näherung, denn die Verteilung der Einfallswinkel an jedem Ort ist für Hell- und Messbild nicht dieselbe. Je nach Einfallswinkel ergeben dichtere Stellen und Löcher unterschiedlich breite Projektionen und somit Schwankungen in der Beleuchtungsstärke auf der Austrittsseite des Diffusors. Wir beschreiben diesen Effekt durch die Multiplikation von $T(\vec{r})$ mit einem Korrelationsterm $K(T(\vec{r}), \beta_{in})$. Während $T(\vec{r})$ beim Teilen des Messbildes durch das Hellbild herausgekürzt wird, bleibt dieser Term stehen. Um diese residuelle Struktur abzuschwächen muss mit unterschiedlichen Diffusoren oder an unterschiedlichen Stellen eines selben Diffusors gemessen und über diese Messungen gemittelt werden.

Neben der Diffusorstruktur kommt es zu einer Abhängigkeit der Transmission in Normalrichtung vom Einfallswinkel. Diese wird berücksichtigt, indem ein zusätzlicher multiplikativer Term $T'(\beta_{in})$ aufgenommen wird, der bei senkrechtem Einfall den Wert 1 annimmt und zu höheren Winkeln hin die Abschwächung des Transmissionsgrad wiederspiegelt. Die Transmission ist zusätzlich wellenlängenabhängig, was mit einem multiplikativen Terms $T''(\lambda)$ berücksichtigt wird. Wir werden sehen, dass die Wellenlängenabhängigkeiten einen schwachen Einfluss haben, weshalb auch eventuelle Korrelationen, zum Beispiel zwischen Wellenlängen- und Winkelabhängigkeiten, nicht berücksichtigt werden.

Zu den Transmissionseigenschaften des Diffusors kommt weiter die Strahlaufweitung $g^{Diff}(\beta_{in})$. Für das Hellbild kommt dieser keine Bedeutung zu, da die Beleuchtung des Diffusors homogen ist. Eine Faltung der Beleuchtung hat daher keine Auswirkung. Für das Messbild ist die Strahlaufweitung Einfallswinkelabhängig. Jeder Strahl des simulierten Bildes muss mit der zu seinem Einfallswinkel gehörenden Strahlaufweitung aufgeweitet werden. Dabei wird bei der Normierung berücksichtigt, dass dadurch ein Teil des Lichtes vom Diffusor herunter gestreut werden kann. Eine eventuelle Variation der Aufweitung mit der Wellenlänge wurde nicht untersucht und wird vernachlässigt, da die Strahlaufweitung mit einem ähnlichen Wellenlängenspektrum gemessen wurde. Vor dem Betrieb bei stark abweichenden Wellenlängen muss die Gültigkeit dieses Punktes überprüft werden.

Streng genommen finden jedoch Strahlaufweitung und Transmissionseigenschaften zeitgleich statt. Je nach Tiefe im Diffusor der Dichteschwankungen des Materials, werden also auch die Strukturen, die für die Funktionen $T(\vec{r})$ und $K(T(\vec{r}), \beta_{in})$ verantwortlich sind, mit einer mehr oder weniger breiten Funktion gefaltet (siehe Abbildung 54). Dies ist an sich ein Problem, da wir beim Messbild im Gegensatz zum Hellbild schon bei der Beleuchtung des Diffusors eine örtliche Abhängigkeit haben. Unter Vernachlässigung dieses Effekts indem die Faltung mathematisch vor der Transmission durchgeführt wird, wird beim Hellbild eine homogene Beleuchtung gefaltet. Daher kann die Faltung weggelassen werden. Die erst zum Schluss durchgeführte Multiplikation mit der Transmission kürzt sich dann heraus. Wird der Effekt jedoch nicht vernachlässigt entsteht beim Hellbild aufgrund der Transmission eine örtliche Struktur. Bilden wir den Bruch von Mess- durch Hellbild, bekommen wir nun

$$\frac{\left(f^{Kalib}(\vec{r}) \cdot T(\vec{r})\right) * g^{Diff}(\beta_{in})}{T(\vec{r}) \, * \, g^{Diff}(\beta_{in})} \tag{29}$$

wodurch sich die Diffusorstruktur, im Gegensatz zu dem Fall der Annahme, die Strahlaufweitung trete vor der Diffusorstruktur auf, nicht mehr herauskürzt. Dies scheint jedoch ein unvermeidbarer Fehler zu sein.

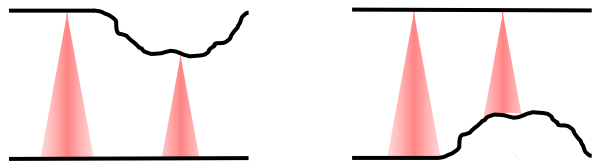

**Abbildung 54: Unterschiedliche Strahlaufweitung bei selber senkrechter Transmission je nach Lage der Struktur (hier ein Loch) im Diffusor.**

Schließlich erreichen für die verschiedenen Punkte des Diffusors nicht dieselben Streuwinkel das Objektiv. Dies sorgt einerseits für unterschiedlich große Raumwinkel in die Licht abgestrahlt werden kann um noch in das Objektiv zu gelangen und dadurch für unterschiedliche Intensitäten $\int T'''(\beta_{out}) \, d\beta_{out}$ der verschiedenen Punkte. $T'''(\beta_{out})$ ist dabei die Strahlstärke in Abhängigkeit des Abstrahlwinkels. Dieser Randlichtabfall kürzt sich jedoch bei der Pixelweisen Division des Mess- und Hellbildes heraus solange die Verteilung der Strahlstärke von jedem Ort zwischen Hell- und Messbild konstant bleibt. Die reine Skalierung dieser Verteilung aufgrund von schräger Beleuchtung wurde bereits durch den Term $T'(\beta_{in})$ berücksichtigt. Eine Änderung im Verlauf der Strahlstärke in Abhängigkeit des Abstrahlwinkels aufgrund der unterschiedlichen Einfallswinkel bei Hell- und Messbild wurde bisher aber nicht berücksichtigt. Der Winkelbereich über den sich $\beta_{out}$ erstreckt ist jedoch kleiner als 0,1 rad. Wie in Kapitel 4.6.2 gesehen konnte bei senkrechter Beleuchtung in diesem Winkelbereich um die Oberflächennormale keine signifikante Variation von $T'''(\beta_{out})$ mit $\beta_{out}$ festgestellt werden. Wir gehen daher davon aus, dass dies auch bei anderen Einfallswinkeln ähnlich ist und somit der Term $T'(\beta_{in})$ eine ausreichende Beschreibung liefert.

Der Diffusor wird also durch die Faltung beim Messbild sowie durch die Skalierung mit

$$T(\vec{r}) \cdot T'(\beta_{in}) \cdot K(T(\vec{r}), \beta_{in}) \cdot T''(\lambda) \cdot \int T'''(\beta_{out}) \, d\beta_{out} \tag{30}$$

beschrieben, wobei $T(\vec{r})$ und $\int T'''(\beta_{out}) \, d\beta_{out}$ sich im Bruch von Mess- und Hellbild herauskürzen und $K(T(\vec{r}), \beta_{in})$ herausgemittelt wird.

Es wurde bereits bei der Beschreibung des Aufbaus erläutert, dass durch die Position des Umlenkspiegels keine Rückspiegelung auf den Diffusor möglich ist. Weiter wurde am Goniometer festgestellt, dass der Spiegel in guter Näherung spekulär reflektiert. Da sowohl das Hell- wie auch das Messbild davon betroffen sind, interessiert uns die Spiegelung des Bildes nicht weiter. Die Winkelabhängigkeit der Reflektion ist mit einer Änderung um ca. 0,0003% über die 0,1mrad auch sehr schwach und wird vernachlässigt. Somit ist die einzige interessante Eigenschaft die wellenlängenabhängige Reflektivität $R^{Umlenk}(\lambda)$.

Wie bereits festgestellt lässt sich der Randlichtabfall zwischen Hell- und Messbild in guter Näherung herauskürzen. Die PSF des Objektivs hat eine Halbwertsbreite von 22,4±0,1µm. Ihre Auswirkung ist eine schmale Faltung, sowohl des Mess- wie auch des Hellbildes, da nun auf dem Hellbild eine Ortsabhängigkeit aufgrund der Diffusorstruktur vorhanden ist. Auch hier tritt wieder ein ähnliches Problem wie bei der Faltung der Diffusorstruktur auf. Während die PSF beim Hellbild die Multiplikation des Randlichtabfalls und der Diffusorstruktur faltet, wird beim Messbild die Multiplikation vom Brennpunkt des Spiegels gefaltet mit der Diffusoraufweitung, dem Randlichtabfall und der Diffusorstruktur gefaltet. Auch hier machen wir wieder einen Fehler, wenn wir die PSF vernachlässigen. Die Alternative die Bilder zu entfalten wurde mit mehreren Fourierverfahren versucht, ergab aber kein verlässliches Ergebnis. Da der Randlichtabfall und das Brennpunktprofil viel schwächere Gradienten aufzeigen als die PSF und die Diffusorstruktur uns nicht interessiert, gehen wir den Weg die PSF zu vernachlässigen.

Letztendlich bleiben die Eigenschaften der Kamera. Wir haben in Kapitel 4.1.2 bereits erläutert welche Korrekturen nötig sind um das Messsignal möglichst weit zu bereinigen. Diese werden auch hier sofern nötig durchgeführt. Eine Korrektur des Smears ist aufgrund der langen Belichtungszeiten (ca. 40s bei dem Messbild und 40min für das Hellbild), bzw. geringen Bestrahlungstärke nicht nötig (siehe Gleichung 6). Die Homogenität des Chips kürzt sich zwischen Hell- und Messbild heraus und die Kamera wird so weit wie möglich im linearen Bereich betrieben. Es bleibt somit, neben dem Belichtungszeitverhältnis das wir hier nicht weiter mitführen wollen aber berücksichtigt wird, der Bruch

$$\frac{\int s_{mess}(\lambda) \cdot \eta(\lambda) \; d\lambda}{\int s_{hell}(\lambda) \cdot \eta(\lambda) \; d\lambda} \tag{31}$$

stehen, wobei $s_{mess}(\lambda)$ und $s_{hell}(\lambda)$ die Zusammenfassungen aller oben genannten Effekte des Messplatzes für das Mess- bzw. das Hellbild sind (siehe Gleichung 13). Wir erhalten somit für das Messbild

$$\int \left[ \left( LQ^{Mess}(\lambda) \cdot T^{Glas}(\lambda) \cdot R^{Kalib}(\lambda) \cdot \tilde{f}^{Mess}(\vec{r}) \cdot T'^{Diff}\left(\tilde{\beta}_{in}^{Mess}(\vec{r})\right) \cdot T''^{Diff}(\lambda) \right) \right. \\ \left. * g^{Diff}\left(\tilde{\beta}_{in}^{Mess}(\vec{r})\right) \right] \cdot R^{Umlenk}(\lambda) \cdot \eta(\lambda) \; d\lambda \tag{32}$$

und für das Hellbild

$$\int LQ^{Hell}(\lambda) \cdot T^{Glas}(\lambda) \cdot R^{Hell}(\lambda) \cdot \cos(2\alpha) \cdot T'^{Diff}(\beta_{in}^{Hell}) \cdot T''^{Diff}(\lambda) \cdot R^{Umlenk}(\lambda) \\ \cdot \eta(\lambda) \; d\lambda \tag{33}$$

wobei $f^{Kalib}(\vec{r})$ bzw. $f^{Probe}(\vec{r})$ aus Kapitel 5.1, Absatz 4 in seine Ortsabhängigkeit $\tilde{f}^{Mess}(\vec{r})$ die die Beleuchtungsstärke und in die Funktion $\tilde{\beta}_{in}^{Mess}(\vec{r})$ die die Winkelverteilung an jedem Ort beschreibt zerlegt wurde. Der Exponent „Mess" steht hier als Überbegriff für „Kalib" und „Probe". Der erste Term beinhaltet in dem Fall der Probe die Strahlaufweitung durch die Oberflächenfehler.

Unter den obigen Annahmen zu den einzelnen Komponenten, insbesondere zum Diffusor, können alle Wellenlängenabhängigkeiten als multiplikative Faktoren behandelt werden. Dadurch können die Wellenlängenabhängigkeiten für das Messbild und für das Hellbild je von den Orts- und Winkelabhängigkeiten desselben Bildes separiert werden. Aufgrund der Aufnahme anhand der Kamera werden die wellenlängenabhängigen Terme mit der Quanteneffizienz gewichtet und integriert.

Da wir uns nur für den Bruch des Mess- durch das Hellbild interessieren, ergibt sich für die Wellenlängenabhängigkeit dieses Bruchs folgender Term:

$$\Gamma(\lambda) := \frac{\int LQ^{Mess}(\lambda) \cdot T^{Glas}(\lambda) \cdot R^{Kalib}(\lambda) \cdot T''^{Diff}(\lambda) \cdot R^{Umlenk}(\lambda) \cdot \eta(\lambda)\, d\lambda}{\int LQ^{Hell}(\lambda) \cdot T^{Glas}(\lambda) \cdot R^{Hell}(\lambda) \cdot T''^{Diff}(\lambda) \cdot R^{Umlenk}(\lambda) \cdot \eta(\lambda)\, d\lambda} \qquad (34)$$

bzw. analog mit $R^{Probe}(\lambda)$.

Damit lässt sich die Konzentrationsmatrix als

$$\Gamma^{Kalib/Probe}(\lambda) \cdot \frac{\left(\tilde{f}^{Kalib/Probe}(\vec{r}) \cdot T'^{Diff}\left(\tilde{\beta}_{in}^{Kalib/Probe}(\vec{r})\right)\right) * g^{Diff}\left(\tilde{\beta}_{in}^{Kalib/Probe}(\vec{r})\right)}{\cos(2\alpha) \cdot T'^{Diff}\left(\beta_{in}^{Hell}\right)} \qquad (35)$$

schreiben, wobei $\Gamma(\lambda)$ eine Konstante die von der konzentrierenden Optik abhängt ist und $T'^{Diff}\left(\beta_{in}^{Hell}\right)$ für alle Strahlen konstant bleibt.

## 5.2. Wellenlängenabhängigkeiten

In diesem Unterkapitel soll der Faktor $\Gamma(\lambda)$ näher betrachtet werden.

Die Transmission der Glasplatte und des Diffusors sowie die Reflektivität des Umlenkspiegels kommen sowohl im Zähler wie auch im Nenner von $\Gamma(\lambda)$ vor. Damit sind Skalierungen dieser Funktionen bedeutungslos und wir können uns mit relativen Verläufen, die wir im Fall der Glasplatte und des Diffusors gemessen haben, zufrieden geben.

Die Reflektivitäten des Umlenk-, Kalibrier- und Hellbildspiegels liegen wellenlängenaufgelöst vor. Es handelt sich dabei um die absoluten Reflektivitäten die von den Herstellern mitgeteilt wurden.

Für die Quanteneffizienz $\eta(\lambda)$ werden ebenfalls die Angaben des Herstellers benutzt.

Aufgrund der Abhängigkeit der Intensität, die durch das Spektrometer angezeigt wird, vom Einfallswinkel auf den zum Spektrometer führenden Lichtleiter, sind die vom Spektrometer gelieferten Spektren nur relativ zu betrachten. Wir bekommen also $A \cdot LQ^{Hell}(\lambda)$ und $B \cdot LQ^{Mess}(\lambda)$. Es wurde jedoch getrennt mit der Kamera das Intensitätsverhältnis zwischen den zwei Konfigurationen bestimmt. Dabei wurde die Kamera mit beiden Konfigurationen der Lichtquelle beleuchtet und das Signal je über den gesamten Chip aufaddiert. Daraus resultiert jeweils $\int LQ(\lambda) \cdot qe(\lambda)\, d\lambda$. Wird nun der Term

$$\underbrace{\frac{\int LQ^{Mess}(\lambda) \cdot \eta(\lambda)\, d\lambda}{\int LQ^{Hell}(\lambda) \cdot \eta(\lambda)\, d\lambda}}_{Messung\ mit\ Kamera} \cdot \underbrace{\frac{\int A \cdot LQ(\lambda)^{Hell} \cdot \eta(\lambda)\, d\lambda}{\int B \cdot LQ^{Mess}(\lambda) \cdot \eta(\lambda)\, d\lambda}}_{\substack{Messung\ mit\\ Spektrometer}} \qquad (36)$$

berechnet, ergibt sich das Verhältnis A/B. Mit diesem Verhältnis kann der Bruch $B \cdot LQ^{Mess}(\lambda)/A \cdot LQ^{Hell}(\lambda)$ korrigiert und das Ergebnis in $\Gamma(\lambda)$ eingesetzt werden.

Damit kann nun der aus den Wellenlängenabhängigkeiten resultierende Skalierungsfaktor der Konzentrationsmatrix $\Gamma^{Kalib}(\lambda)$ für den Kalibrierspiegel bestimmt werden. Dieser beträgt 64,3226. Der Großteil resultiert jedoch aus der Intensitätsdifferenz der Lichtquelle und den unterschiedlichen Reflektivitäten des Hellbildspiegels und des Kalibrierspiegels. Um dies zu sehen wird im Folgenden $\Gamma^{Kalib}(\lambda)$ von diesen Effekten bereinigt, was jedoch nicht für die Benutzung am RMP getan wird. Um diesen Faktor um die Intensitätsdifferenz der Lichtquelle zu korrigieren, werden die Spektren der Lichtquelle so skaliert, dass sie über die Wellenlängen integriert denselben Wert ergeben. Die unterschiedliche Reflektivität des Hellbildspiegels und des Kalibrierspiegels werden herausgerechnet, indem man die mittlere Reflektivität zwischen 595 und 680nm gleich setzt. So reduziert sich der Faktor auf 1,0021. Die Wellenlängenabhängigkeiten beeinflussen das Ergebnis somit entgegen den Erwartungen nur um 2‰.

Letztlich bleibt noch die Frage der Behandlung der Reflektivität der Probe bei der Berechnung des Faktors $\Gamma^{Probe}(\lambda)$ offen. Auch hier wurde der relative Verlauf der wellenlängenabhängigen Reflektivität an der Ulbrichtkugel vermessen, der aufgrund der Instabilität der Lichtquelle skaliert sein kann. Hier kürzt sich weder die Skalierung heraus, noch wurde eine zusätzliche Messung durchgeführt, die erlauben würde diese Skalierung separat zu bestimmen. Dies ist so beabsichtigt, da das Ziel ist, mit möglichst wenigen Messplätzen die Proben vermessen zu können. Die Reflektivität der Probe wird daher so skaliert, dass

$$\int LQ^{Mess}(\lambda) \cdot T^{Glas}(\lambda) \cdot R^{Probe}(\lambda) \cdot T''^{Diff}(\lambda) \cdot R^{Umlenk}(\lambda) \cdot \eta(\lambda) \, d\lambda$$
$$= \int LQ^{Mess}(\lambda) \cdot T^{Glas}(\lambda) \cdot T''^{Diff}(\lambda) \cdot R^{Umlenk}(\lambda) \cdot \eta(\lambda) \, d\lambda \quad (37)$$

gilt. Somit hat die Konzentrationsmatrix die am RMP gemessen wird eine geringere Amplitude als die simulierte. Der Faktor dazwischen gibt dann einen mittleren Reflektivitätsgrad der Probe an, der sich auf das Wellenlängenspektrum am Messplatz und auf die auf die Probe treffenden Winkel bezieht.

## 5.3. Kalibrierung

Mit den vorherigen Überlegungen haben wir nun die Grundlage für die Kalibrierung des Messplatzes gelegt. Der Ablauf ist schematisch in Abbildung 55 skizziert. Vorgehen ist dabei, den Strahlengang für den als glatt und formtreu angenommenen Kalibrierspiegel zu simulieren. Das gesamte aus der Simulation resultierende Bild wird anschließend mit $\Gamma^{Kalib}(\lambda)$, $1/\cos(2\alpha)$ und $1/T'^{Diff}(\beta_{in}^{Hell})$ gewichtet und für jeden Strahl je nach Einfallswinkel mit $T'^{Diff}\left(\tilde{\beta}_{in}^{Kalib/Probe}(\vec{r})\right)$ multipliziert und mit $g^{Diff}\left(\tilde{\beta}_{in}^{Kalib/Probe}(\vec{r})\right)$ gefaltet.

Neben dieser Simulation werden Hell- und Messbild mit dem Kalibrierspiegel am RMP aufgenommen und wie in Kapitel 5.1 erläutert die Korrekturen der Effekte der Kamera durchgeführt. Diese Bilder werden je durch ihre Belichtungszeiten geteilt und pixelweise das Verhältnis der Bilder gebildet.

Von Simulation und Experiment wird anschließend eine Radialverteilung berechnet und die Verläufe verglichen. Ziel ist es durch eine Übereinstimmung der Simulation mit der Messung die Simulation zu bestätigen. Das Ergebnis ist in Abbildung 56 zu sehen.

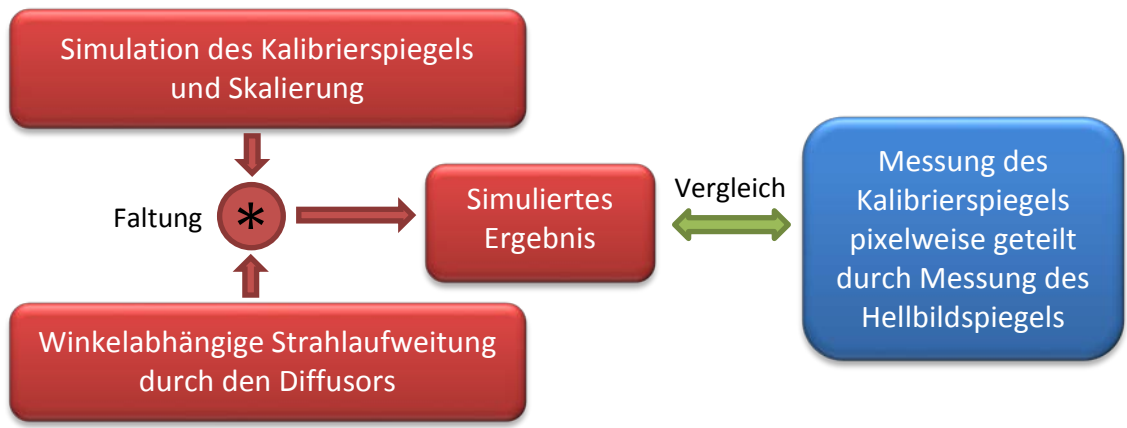

**Abbildung 55: Skizze des Ablaufs der Kalibrierung. Durch Übereinstimmung von Simulation und Messung soll die Simulation bestätigt werden.**

Feststellung dieser Kalibrierung ist zunächst, dass die Effizienz des Kalibrierspiegels keinen physikalischen Wert ergibt.

Jeder Grauwert der Konzentrationsmatrix gibt an, um welchen Faktor die Strahlungsleistung des entsprechenden Pixels höher ist, als die Strahlungsleistung auf einer Fläche derselben Größe an der Eintrittsapertur. Multiplizieren wir diesen Wert mit dem Flächenverhältnis eines Pixels und der Apertur der Optik, bekommen wir für jeden Pixel welcher Anteil der auf die Eintrittsapertur treffende Strahlungsleistung auf diesen Pixel trifft. Summieren wir dies auf, ergibt sich die Effizienz. Bei der Berechnung muss noch der Zoomfaktor des Objektivs berücksichtigt werden.

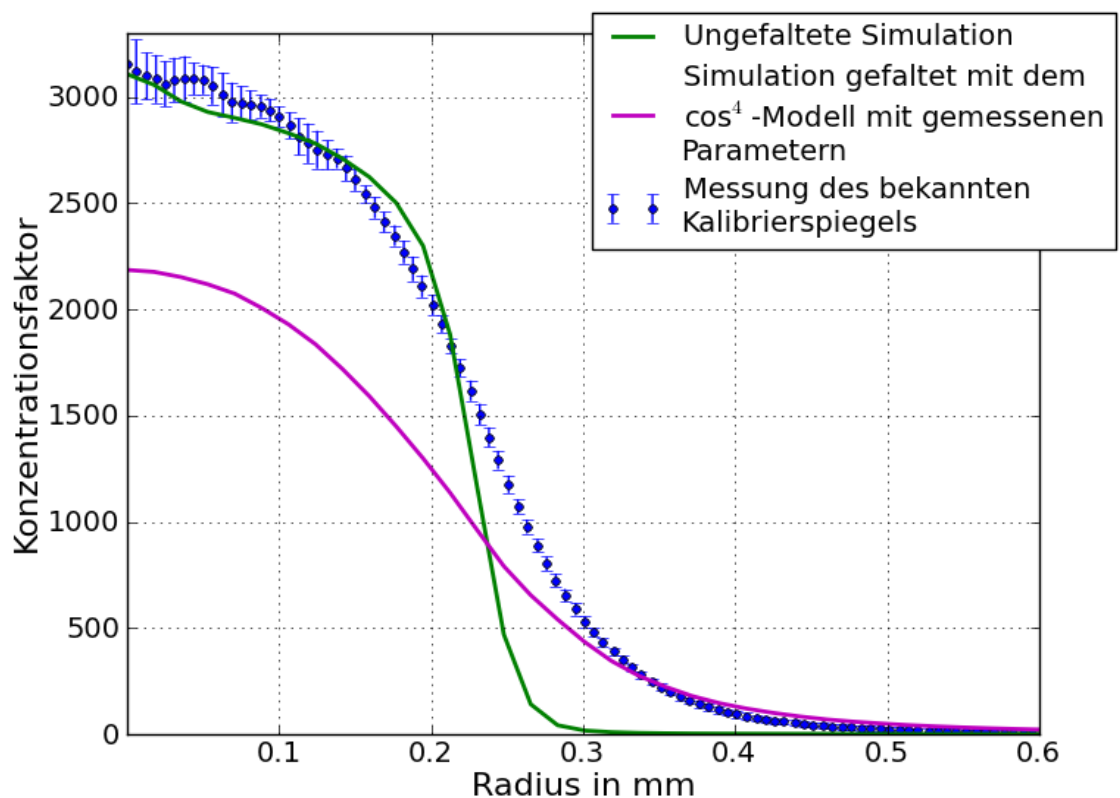

**Abbildung 56: Radialverteilung der simulierten und gemessenen Konzentrationsmatrizen sowie der mit der Diffusorstrahlaufweitung gefalteten simulierten Konzentrationsmatrix des Kalibrierspiegels**

Daraus ergaben sich, je nach Aufnahme, Effizienzen zwischen 125 und 130%. Der Ursprung davon liegt in der örtlichen Inhomogenität der Lichtquelle, die darauf zurück zu führen ist, dass die zweite Homogenisierung der Lichtquelle beim Hellbild ausgebaut wurde. Wie im Zusammenhang mit Abbildung 53 erklärt, ist der Hellbildspiegel verkippt, sodass er nicht durch denselben Bereich der Lichtquelle wie der Kalibrierspiegel beleuchtet wird. Die Intensität der Lichtquelle fällt zum Rand hin ab. Durch die niedrigere Intensität beim Hellbild wird die Effizienz künstlich erhöht.

Da jede weitere Homogenisierung zeitgleich eine Abschwächung der Bestrahlungsstärke bedeutet und damit die Belichtungszeiten nicht mehr praktikabel wären, bleibt dieses Problem bestehen bis eine Leistungsstärkere Lichtquelle vorhanden ist. Hier wird zunächst die Entwicklung der Situation mit der Fertigstellung der neuen Lichtquelle abgewartet. In der Zwischenzeit wird das Problem umgangen, indem auf die Bestimmung der Effizienz vorübergehend verzichtet und die gemessene Konzentrationsmatrix so normiert wird, dass die Summe über die gemessene und über die simulierte Konzentrationsmatrix übereinstimmen. Das Ergebnis ist in Abbildung 57 zu sehen.

Selbst nach der Normierung ist keine Übereinstimmung gegeben. Die aus der Charakterisierung der Strahlaufweitung des Diffusors resultierende effektive Dicke führt zusammen mit dem $\cos^4$-Modell zu einer zu starken Ausschmierung des Brennpunkts. Diese mangelnde Übereinstimmung kann auch nicht behoben werden, indem die effektive Dicke des Diffusors im Rahmen des Fehlers der Charakterisierung variiert wird. Die Benutzung der effektiven Dicken aus der Charakterisierung des Diffusors liefert auch nicht die bestmögliche Übereinstimmung mit der Messung im Rahmen des $\cos^4$-Modells. Wieso dies der Fall ist konnte noch nicht geklärt werden. Damit ist die Schlussfolgerung, dass die Strahlaufweitung durch den Diffusor mit dem $\cos^4$-Modell noch nicht zufriedenstellend beschrieben ist.

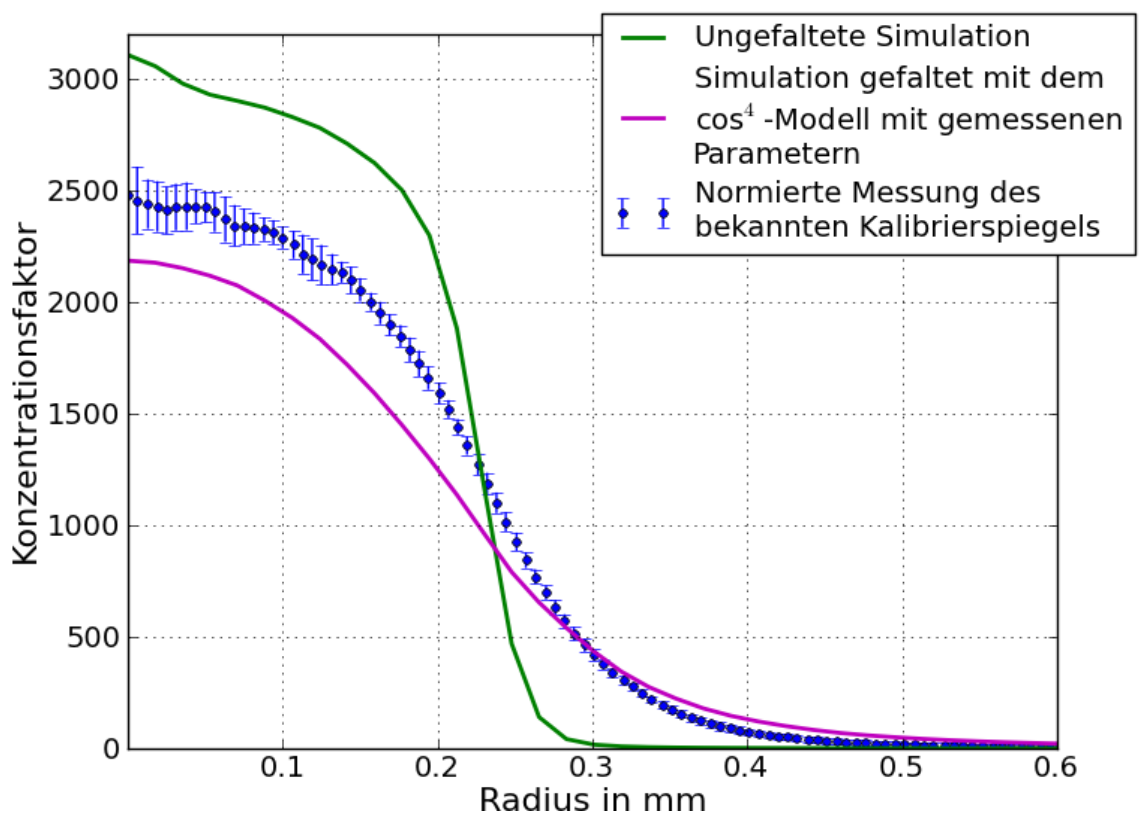

Abbildung 57: Radialverteilung der simulierten und normierten gemessenen Konzentrationsmatrizen sowie der mit der Diffusorstrahlaufweitung gefalteten simulierten Konzentrationsmatrix des Kalibrierspiegels

Wir haben im Kapitel 4.6.1 gesehen, dass das $\cos^4$-Modell auf der Annahme einer Streuung beim Eintritt in den Diffusor und ungestörter Propagation bis zur Austrittsfläche des Diffusors beruht. Da dieses Modell nur durch die Empirie gestützt wurde, war die Propagationsdistanz nicht die physikalische Dicke des Diffusors. Es soll versucht werden dieses empirische Modell etwas zu verfeinern. Die Physik des Diffusors ist jedoch schlecht bekannt [LiQ08] und es ist nicht der Anspruch sie hier korrekt zu analysieren. Viel eher soll das neue Modell für das Raytracing praktikabel sein und muss daher weiterhin die physikalischen Gegebenheiten stark vereinfachen. Die nächsten Absätze sind als ein solcher Versuch aufzufassen.

In einem Volumenstreuer kann das Licht an jeder Stelle des Materials gestreut werden. Im allgemeinen Fall kann es dabei eine Wahrscheinlichkeitsverteilung im Streuwinkel geben. Eine Orientierung des Materials und somit bevorzugte Streurichtungen im Raum unabhängig des Einfallswinkels sollte es in diesem Fall nicht geben, da der Diffusor aus einem Granulat gesintert wird. Weiter kann zwischen den Streuungen ein variabler streuungsfreier Weg zurückgelegt werden, dessen Länge durch eine Wahrscheinlichkeitsverteilung beschrieben wird. Diese könnte mit der Dichte des Materials, also mit der Diffusorstruktur korrelieren. Eine genaue Beschreibung wäre also zu komplex, zumal die hier genannten Größen unbekannt sind.

Das aktuelle Modell wird daher verfeinert, indem der Volumenstreuer als geschichteter Oberflächenstreuer modelliert wird, wobei innerhalb der Schichten das $\cos^4$-Modell gilt.

Beim Eintritt in den Diffusor wird ein auftreffender Strahl wie im vorherigen Modell schon an der Eintrittsfläche gestreut. Nach dieser Streuung und jeder folgenden legt ein Strahl eine freie Weglänge zurück. Es wird angenommen, dass die Wahrscheinlichkeitsverteilung der freien Weglänge gaußförmig mit Maximum bei der mittleren freien Weglänge verläuft.

Nach der ersten Streuung und nach jeder folgenden gibt es ein ganzes Spektrum an Ausbreitungsrichtungen. Interessieren wir uns für die Strecke die von einem bestimmten Strahl entlang der Diffusornormalen z zwischen zwei Streuungen zurückgelegt wird, wird dessen Wahrscheinlichkeitsverteilung immer noch durch einen Gauß beschrieben. Dabei ist jedoch die Breite und die Position des Maximums des auf die z-Achse projizierten Gauß vom Winkel der Ausbreitungsrichtung des Strahls zur Diffusornormalen abhängig. Dies ist in Abbildung 58 links dargestellt. Der gestrichelte orange und der blaue Gauß haben selbe Breite und der Abstand ihres Maximums zum rot dargestellten Streupunkt ist identisch. Die Projektion des orangen gestrichelten Gauß auf die z-Achse verschiebt das Maximum und reduziert die Breite der Gaußkurve. Da mit steigenden Winkeln eine systematische Verschiebung des Maximums zu kleineren z-Werten und eine Verschmälerung der Gaußkurven eintritt, ist die über die Winkel summierte Wahrscheinlichkeitsverteilung streng genommen kein Gauß mehr, sondern besitzt eine Schiefe zu kleineren z-Werten hin. Der genaue Verlauf hängt von der Verteilung der Strahlungsleistung über die Streuwinkel ab. Hier tritt in dem neuen Modell eine Vereinfachung ein: Unabhängig vom Streuwinkel soll die mittlere freie Weglänge und die Breite der Gaußkurve in z-Richtung immer dieselbe sein. Dadurch ist, unabhängig von der Verteilung der Strahlleistung über die Streuwinkel, die Streuwahrscheinlichkeit in z immer durch einen Gauß beschrieben. Das Maximum dieser Kurve entspricht dann nicht mehr der physikalischen mittleren freien Weglänge, sondern ist abhängig von der Streuwinkelverteilung. Dieser Wert wird im Folgenden effektive mittlere freie Weglänge genannt. Analog wird die Komponente der freien Weglänge entlang der z-Achse effektive freie Weglänge genannt.

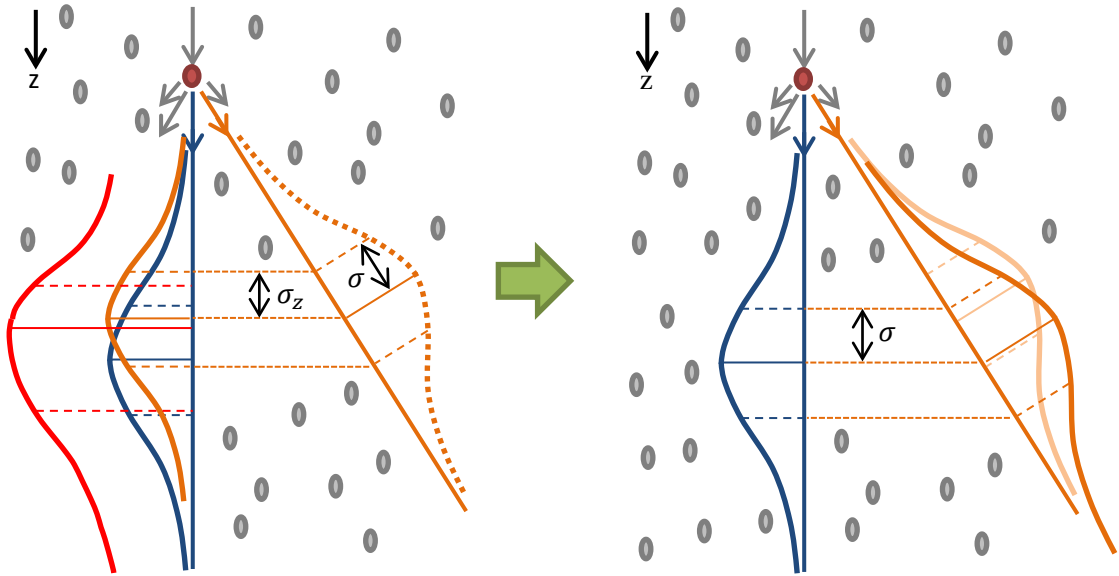

**Abbildung 58: Wahrscheinlichkeitsverteilung der freien Weglänge und ihrer Projektion auf die z-Achse im neuen Modell zur Beschreibung der Diffusorstrahlaufweitung.**

Jede der effektiven freien Weglängen in z ist mit den vorherigen Annahmen durch eine gaußförmige Wahrscheinlichkeitsverteilung beschrieben deren Maximum bei der effektiven mittleren freien Weglänge liegt. Die Wahrscheinlichkeitsverteilung für die Tiefe im Diffusor entlang z der n-ten Streuung ist daher durch die Faltung dieser Gaußfunktionen beschrieben. Das Ergebnis ist wieder ein Gauß dessen Varianz durch die Summe der Varianzen der gefalteten Wahrscheinlichkeitsfunktionen gegeben ist und mit Maximum bei der Summe der z-Werte der Maxima (siehe Abbildung 59).

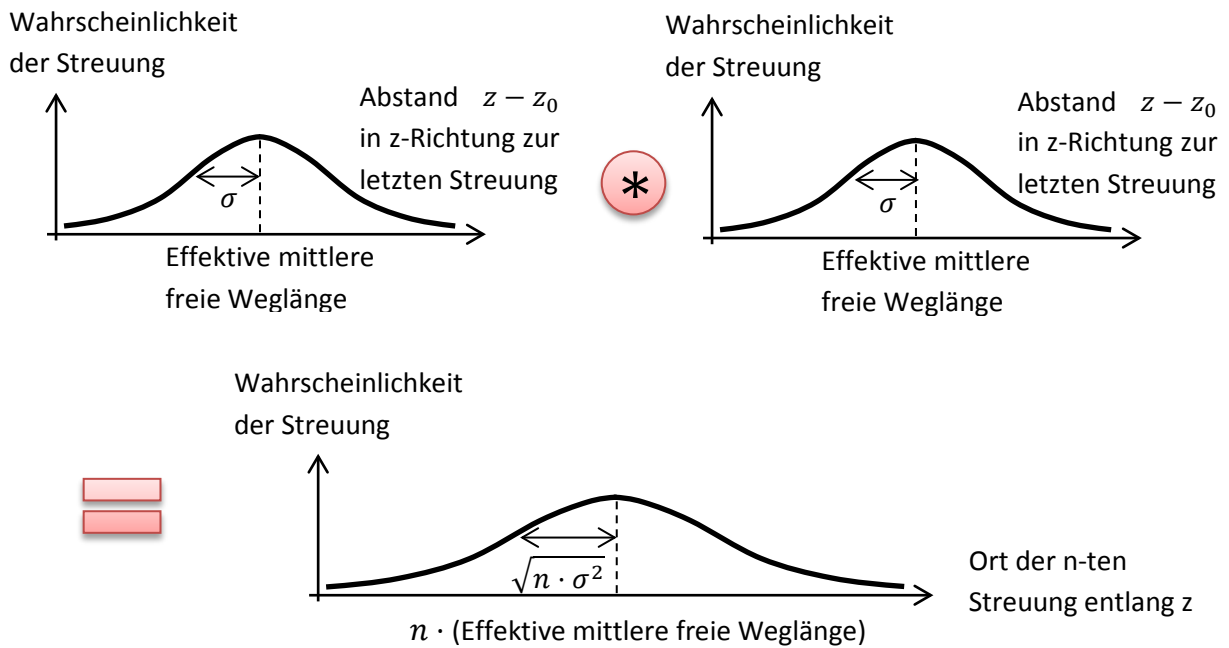

**Abbildung 59: Wahrscheinlichkeitsverteilung der Streutiefe der n-ten Streuung entlang z**

Wir fassen gedanklich alle Strahlen die aus einer Streuung in einem selben kleinen Flächenelement resultieren und nach derselben effektiven freien Weglänge wieder gestreut werden zusammen. Diese Strahlen beschreiben wir wieder durch einen $\cos^4$-Verlauf, der als unabhängig von der Einfallsrichtung des einfallenden Strahls angenommen wird. Dieses Mal ist es fundiert eine freie Ausbreitung der Lichtstrahlen innerhalb dieser Schicht zugrunde zu legen.

Hier kommt eine weitere Vereinfachung ins Spiel: Der Mittelwert eines cos⁴-Verlaufs über eine Schichtdicke $\Delta z_1$ und eines cos⁴-Verlaufs über eine Schichtdicke $\Delta z_2$ wird gleichgesetzt mit einem cos⁴-Verlauf über eine Schichtdicke $(\Delta z_1 + \Delta z_2)/2$. Da der Gauß symmetrisch ist, können paarweise die cos⁴-Verläufe über eine Schichtdicke von (effektiver mittlere freie Weglänge) $- \Delta z$ mit denen über eine Schichtdicke von (effektiver mittlere freie Weglänge) $+ \Delta z$ zusammengefasst werden. Damit mitteln sich alle Strahlaufweitungen zwischen zwei Streuungen zu einer Strahlaufweitung entsprechend einem cos⁴-Verlauf über die effektive mittlere freie Weglänge.

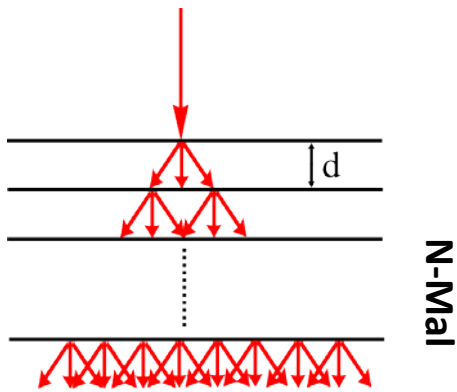

Abbildung 60: Das neue Schichtenmodell für zur Beschreibung der Strahlaufweitung durch den Diffusor: In jeder Schicht der Dicke der effektiven mittleren freien Weglänge d tritt eine Strahlaufweitung nach cos4-Gesetz auf, woraus eine N-fache Faltung der ursprünglichen Intensitätsverteilung resultiert. Quelle: [Sch13]

Dadurch entsteht das neue Schichtenmodell: Über die physikalischen Dicke Diffusors finden in Schichten der Dicke d, die der effektiven mittleren freien Weglänge entspricht, cos⁴-Strahlaufweitungen statt (siehe Abbildung 60). Die Intensitätsverteilung an der Eintrittsfläche wird somit N-fach sukzessiv mit

$$\frac{d_i^4}{(d_i^2 + r^2)^2} \qquad (38)$$

gefaltet. Dabei ist $d_i$ für die N-1 ersten Faltungen gleich (im Folgenden als $d_{1 \to (N-1)}$ bezeichnet), für die N-te Faltung in der Regel geringer, da D kein vielfaches der effektiven mittleren freien Weglänge sein muss.

Anders als beim cos⁴-Modell wird, zumindest vorläufig, kein Unterschied zwischen Strahlen mit unterschiedlichen Einfallswinkeln gemacht, sondern die gesamte simulierte Konzentrationsmatrix mit derselben Funktion mit denselben Parametern gefaltet.

Bisher wurde das neue Schichtenmodell noch nicht anhand der Diffusormessungen aus Kapitel 4.6.1 bestätigt. Dies bedeutet erstens, dass das Modell seine Berechtigung hauptsächlich aus den Ergebnissen am RMP zieht, die in den nächsten Absätzen noch vorgestellt werden. Zweitens ist der Ablauf und das Ziel der Kalibrierung somit ein anderes: Durch einen Fit, bei dem die Simulation mit dem neuen Strahlaufweitungsmodell N-fach gefaltet wird, werden Simulation und Experiment in Übereinstimmung gebracht und dabei der Parameter $\widetilde{N} := (N-1) + d_N/d_{1 \to (N-1)}$ bestimmt. $d_{1 \to (N-1)}$ ergibt sich aus $d_{1 \to (N-1)} = D/\widetilde{N}$ und $d_N$ aus $d_N = D - (N-1) \cdot d_{1 \to (N-1)}$. Somit ergibt sich aus der Kalibrierung eine empirische Messplatzsignatur. Der Ablauf dieser neuen Charakterisierung ist Schematisch in Abbildung 61 skizziert.

Das Ergebnis des Fits ist in Abbildung 62 zu sehen. Als Parameter ergeben sich $\widetilde{N} = 2{,}4 \pm 0{,}1$ [Sch13]. Als Fehler für den Fit wurde die Standardabweichung zwischen drei Messungen sowie Streulicht und Defokussierung angenommen. Diese Fehler wirken sich auf das Fitergebnis aus und werden durch den Fehler auf das Fitergebnis fortgepflanzt.

Der Verlauf des Fits stimmt besser als das cos$^4$-modell mit den experimentellen Daten überein, was dem neuen Modell seine Rechtfertigung zuschreibt.

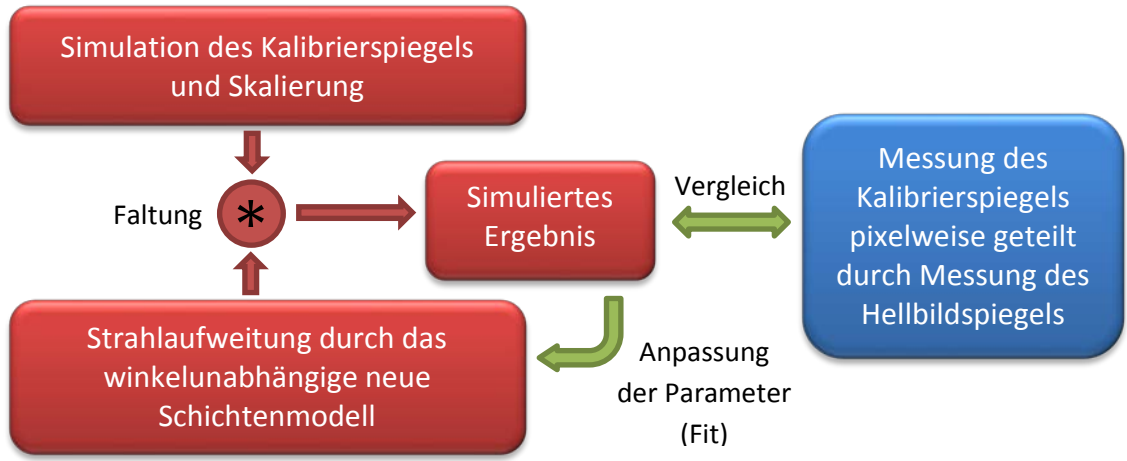

**Abbildung 61: Skizze des Ablaufs der Kalibrierung mit dem neuen Schichtenmodell für die Strahlaufweitung durch den Diffusor. Durch einen Fit der Parameter der Faltungsfunktion des Diffusors soll eine empirische Messplatzsignatur bestimmt werden.**

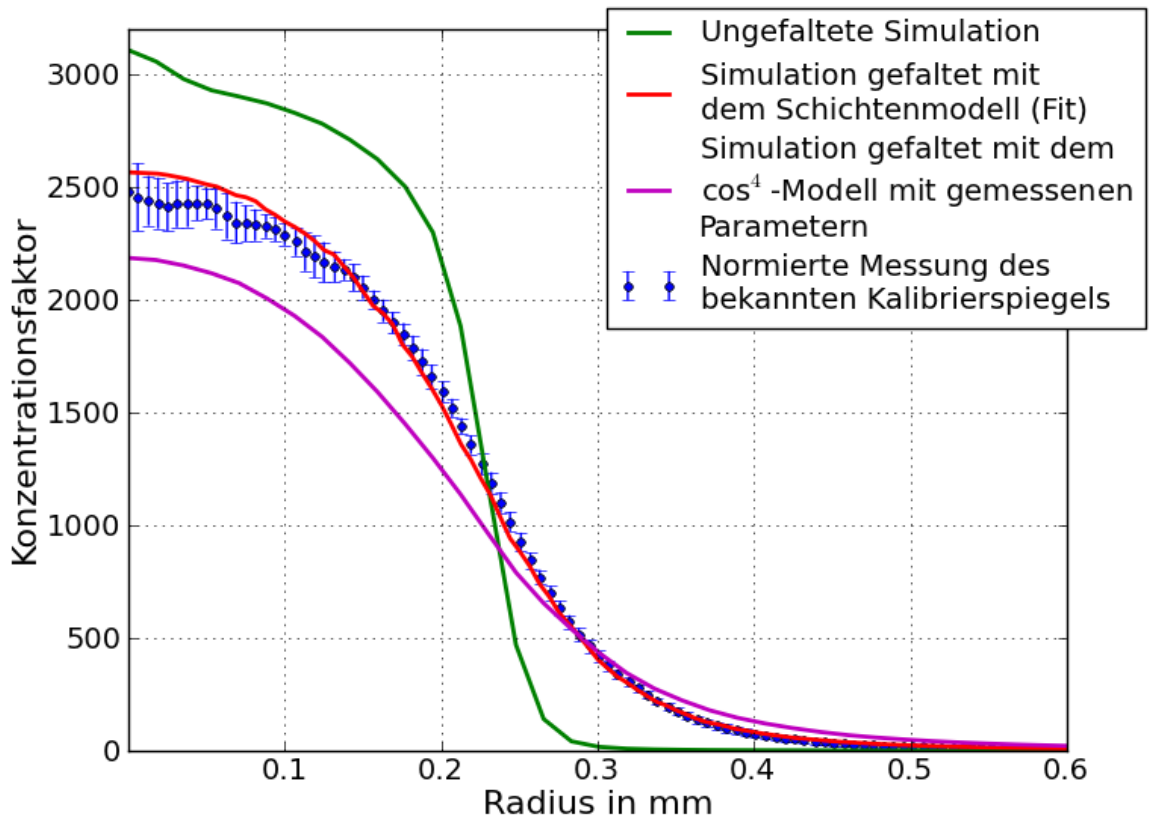

**Abbildung 62: Radialverteilung der normierten gemessenen Konzentrationsmatrix des Kalibrierspiegels (blau) sowie Radialverteilungen der simulierten Konzentrationsmatrix gefaltet mit dem cos$^4$-Modell mit den Parametern aus Kapitel 4.6.1 (grün) und gefaltet mit dem neuen Schichtenmodell, wobei die Parameter aus einem Fit resultieren (rot).**

Jedoch kann auch dieses Modell nicht genau den Verlauf der Strahlaufweitung am RMP beschreiben. Abweichungen des Verlaufs der Strahlaufweitung, die nicht durch eine Variation der Fitparameter gedeckt werden können, werden daher bisher nicht berücksichtigt. Sie sind jedoch klein verglichen mit dem Einfluss der Oberflächenfehler der Probe

Das Ergebnis des Fits mit dem Schichtenmodell kann nun, wie in Kapitel 5.4 beschrieben, als Messplatzsignatur eingesetzt werden, wodurch die Simulation die Effekte des Messplatzes berücksichtigt. Hiermit kann der Einfluss der Oberflächenfehler für eine Probe mit selben Einfallswinkelspektrum wie der Kalibrierspiegel vermessen werden.

## 5.4. Vermessung von Proben

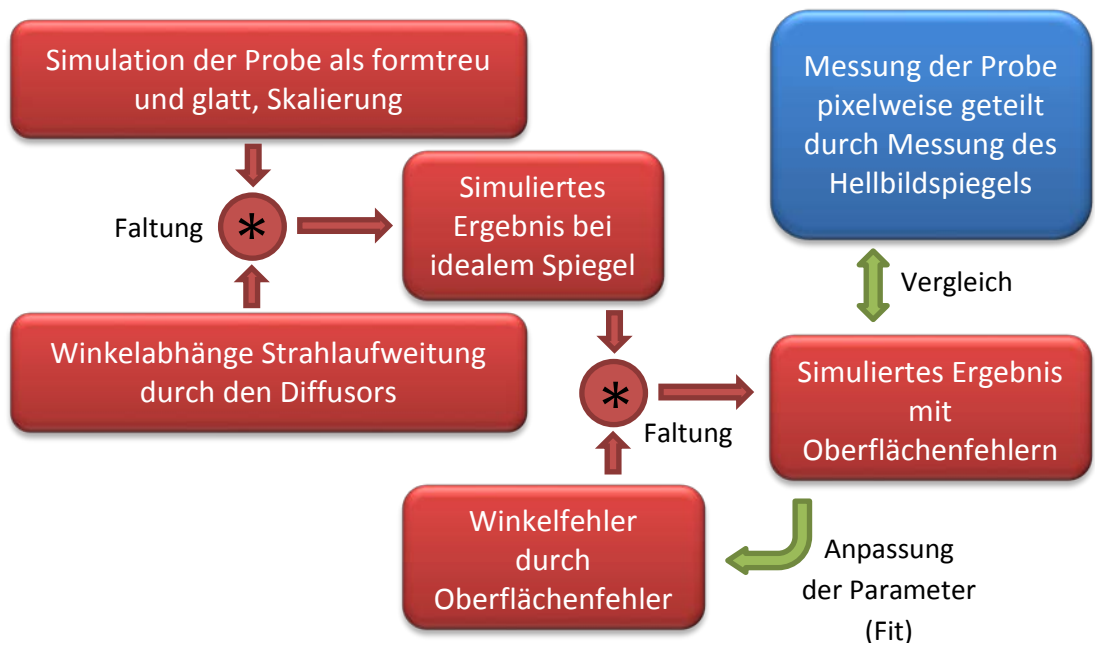

**Abbildung 63: Geplanter Verlauf der Bestimmung der Strahlaufweitung durch die Oberflächenfehler der Proben**

Das Konzept des Messplatzes zur Vermessung der Strahlaufweitung durch die Oberflächenfehler der Proben ist in Abbildung 63 skizziert. Idee ist es die Probe anhand eines Raytracing-Programms mit ihrer Sollgeometrie, also ohne Oberflächenfehler, zu simulieren. Diese Simulation gibt uns an jedem Ort des Diffusors an, wieviele Strahlen und unter welchen Winkeln die Strahlen aufgetroffen sind. Bei genügend Strahlen entspricht die Anzahl der Strahlen der Verteilung der Bestrahlungsstärke. Aus dieser Simulation wird dann, wie in Kapitel 5.1 aufgeführt, die Konzentrationsmatrix unter Berücksichtigung der Messplatzeigenschaften berechnet. Dabei soll für den Auftreffpunkt von jedem Strahl eine Faltung mit der zu seinem Einfallswinkel gehörenden Diffusorstrahlaufweitung durchgeführt werden. Dies ergäbe dann die Konzentrationsmatrix wie sie bei einem idealen Spiegel mit der Sollgeometrie der Probe am RMP gemessen werden würde. Die Abweichung der Simulierten und der gemessenen Konzentrationsmatrizen resultiert dann aus den Oberflächenfehlern der Probe. Die Auswirkung der Winkelfehler aufgrund der Oberflächenfehler der Probe kann, wie wir später noch sehen werden, durch eine Faltung im Ortsraum beschrieben werden. Dadurch können die Faltungen durch den Diffusor und durch den Spiegel vertauschen. Die Breite der Faltungsfunktion aufgrund der Oberflächenfehler kann dann durch einen Fit bestimmt werden, bei dem in jedem Iterationsschritt die simulierte und mit der Diffusorstrahlaufweitung gefaltete Konzentrationsmatrix mit der Faltungsfunktion für die Oberflächenfehler der Probe mit neuen Fitparametern gefaltet wird und die Differenz zur Messung berechnet wird. Ergebnis ist dann diese Faltungsfunktion, bzw. dessen Parameter. Die Breite dieser Funktion ist ein Maß für die Oberflächengüte der Probe.

Die Strahlaufweitung durch den Diffusor in Abhängigkeit des Einfallwinkels ist noch nicht vorhanden. Daher ist dieser Weg noch nicht begehbar. Durch die Kalibrierung mit dem Schichtenmodell die in Kapitel 5.3 beschrieben wurde, ist aber eine empirische Messplatzsignatur vorhanden. Diese kann eingesetzt werden, indem die simulierte normierte Konzentrationsmatrix der Probe mit dieser Messplatzsignatur gefaltet wird. Auch hier beruht der Unterschied zwischen Messung der Probe und gefalteter Simulation, bis auf Ungenauigkeiten der Kalibrierung, nur noch auf geometrischen Formabweichungen und Rauheit der Probe. Dieser Unterschied kann dann wieder mit einer weiteren Faltung beschrieben werden, woraus ein Maß für die Güte der Oberfläche der Probe entsteht. Dieser Verlauf der Bestimmung der Strahlaufweitung durch die Oberflächenfehler der Proben ist in Abbildung 64 skizziert.

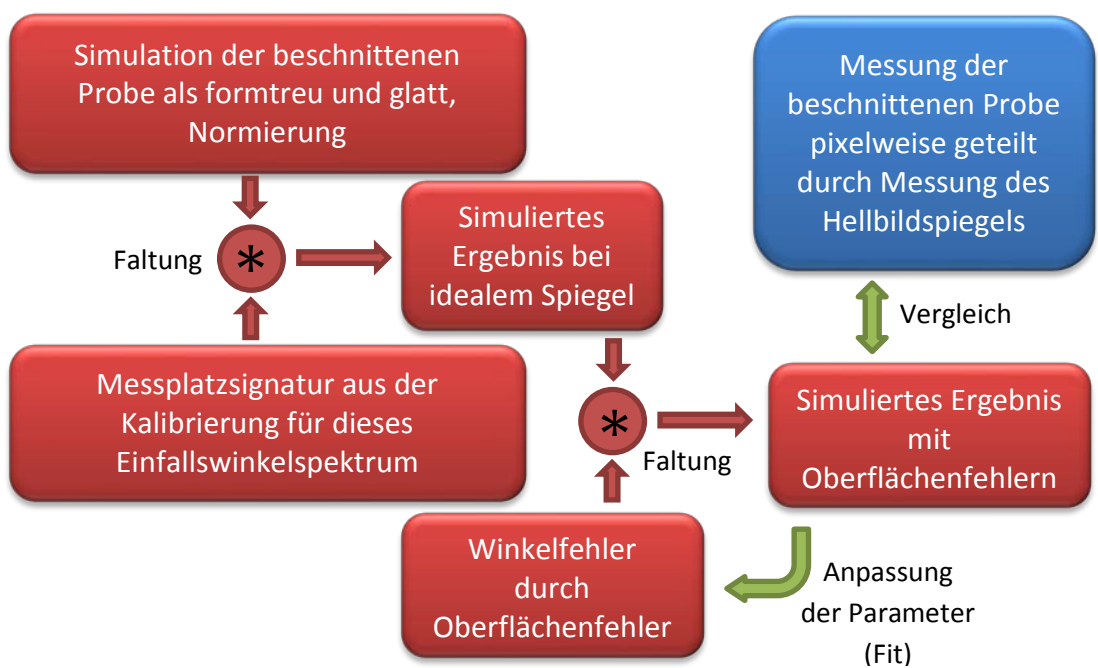

**Abbildung 64: Vorläufiger Verlauf der Bestimmung der Strahlaufweitung durch die Oberflächenfehler der Proben**

Allerdings ist aufgrund der Einfallswinkelabhängigkeit der Diffusorstrahlaufweitung die in Kapitel 5.3 ermittelte Messplatzsignatur nur für Proben gültig, die den Diffusor mit demselben Winkelspektrum wie der Kalibrierspiegel beleuchten. In Kapitel 4.6.1 wurden bei steigenden Einfallswinkeln höhere effektive Dicken des Diffusors ermittelt. Dies muss auch beim neuen Diffusormodell der Fall bleiben. Hat daher eine Probe ein breiteres Einfallswinkelspektrum als der Kalibrierspiegel, so muss auch die Messplatzsignatur breiter werden. Um darauf Rücksicht zu nehmen, können wir zwei unterschiedliche Wege gehen.

Der eine besteht darin, nicht die vollen Proben zu vermessen, sondern mit einer Blende die Probe soweit zu beschneiden, dass der maximale Einfallwinkel auf den Diffusor demjenigen bei der Kalibrierung entspricht. Dies hat den Nachteil, dass damit nicht die Oberfläche des gesamten Spiegels berücksichtigt wird. Die Oberflächengüte der Proben ist nicht zwangsläufig über den ganzen Spiegel homogen. So kann im Scheitelpunkt ein Anspritzpunkt vorhanden sein und sich im Randbereich vermehrt Kratzer befinden. Durch die Blende wird dann der Randbereich vernachlässigt, was sich auf die letztendlich gemessene Oberflächengüte auswirkt.

Der zweite Weg besteht darin, für jede Messung eine eigene Kalibrierung durchzuführen, indem eine Kalibrierprobe eingesetzt wird, dessen maximaler Einfallswinkel größer ist als der der Proben. Der Kalibrierspiegel wird dann durch eine Blende auf das Winkelspektrum der jeweiligen Probe reduziert.

Beide Wege sind nur eine Näherung sobald Kalibrierspiegel und Probe eine unterschiedliche Sollform besitzen, also beispielsweise der Kalibrierspiegel sphärisch, die Probe aber parabolisch ist. Hier ist das Winkelspektrum dann trotz selben maximalen Einfallswinkeln nicht dasselbe. Daher ist eine solche empirische Bestimmung der Messplatzsignatur keine dauerhafte Lösung, sondern wird lediglich übergangsweise angewendet, bis eine zufriedenstellende Charakterisierung der Strahlaufweitung in Abhängigkeit des Einfallswinkels stattgefunden hat. Da derzeit kein größerer Kalibrierspiegel vorhanden ist, wurde der erste der zwei Wege für einen Machbarkeitsnachweis beschritten.

Das Ergebnis der Messung mit beschnittener Probe sowie der zugehörigen Simulation gefaltet mit der Messplatzsignatur ist in Abbildung 65 zu sehen. Hier wurden wieder Radialverteilungen aus den jeweiligen Konzentrationsmatrizen berechnet.

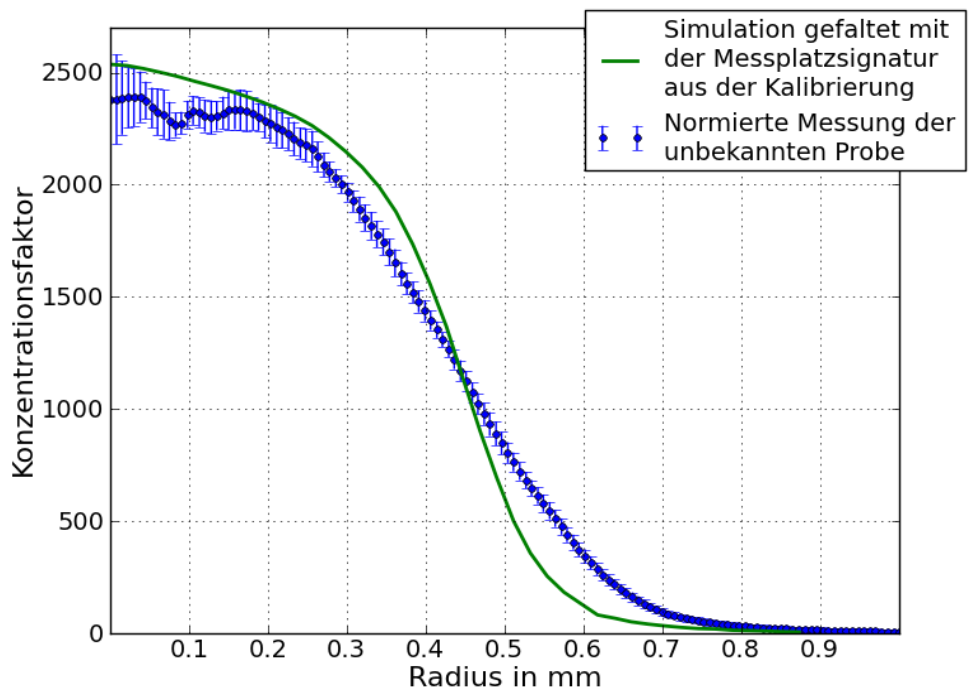

**Abbildung 65: Radialverteilungen der simulierten und mit der Messplatzsignatur aus der Kalibrierung gefalteten Konzentrationsmatrix (grün) und normierte gemessene Konzentrationsmatrix (blau) des Kalibrierspiegels**

Die Abweichung zwischen beiden Radialverteilungen ist deutlich größer als die Fehler, selbst wenn die Ungenauigkeit der Kalibierung mitberücksichtigt wird. Damit ist eine deutliche Auswirkung der Oberflächenfehler der Probe auf die Konzentrationsmatrix der Probe zu sehen.

Die Oberflächenfehler der Spiegel bestehen aus zwei unterschiedlichen Effekten. Erstens kommt es zu Streuung aufgrund der Rauheit der Oberfläche. Wird der Spiegel mit einem infinitesimal schmalen Lichtbündel beleuchtet tritt dieser Effekt immer noch auf. Jeder Punkt des Spiegels reflektiert somit nicht rein spekulär, sondern entsprechend einer um den Winkel nach Reflektionsgesetz zentrierten Intensitätsverteilung. Diese Verteilung wird als rotationssymmetrisch angenommen.

Zweitens weist die Oberfläche des Spiegels Abweichungen von ihrer Sollgeometrie auf. Diese bestehen darin, dass die Spiegeloberfläche lokal einen Winkelfehler besitzt. Ein infinitesimal schmaler Strahl wird hier weiterhin spekulär reflektiert, jedoch mit einem Fehler auf den Ausgangswinkel, der in Kapitel 2.2 erläutert wurde. Er trifft daher mit einem Fehler im Ort auf den Diffusor. Es wird angenommen, dass die Wahrscheinlichkeitsverteilung für die Winkelfehler an jedem Ort des Spiegels bzgl. der Oberflächennormalen an diesem Ort des idealen Spiegels rotationssymmetrisch ist. Die Wahrscheinlichkeitsverteilung soll im Winkel zu dieser Normalen einem Gaußprofil mit Standardabweichung $\sigma_{winkel}$ folgen [Sch13]. Werden die Steigungsfehler über die gesamte Fläche der Probe aufsummiert, so ergeben diese Fehler im Winkel eine Verschmierung der Intensitätsverteilung auf dem Diffusor. Diese kann durch eine Faltung der Intensitätsverteilung auf dem Diffusor mit einem Gauß mit Standardabweichung

$$\sigma_{fit} = \frac{f \cdot \tan(2\sigma_{winkel})}{const} \tag{39}$$

beschrieben werden, wobei f die Brennweite des Spiegels ist [Sch13] [Wen80].

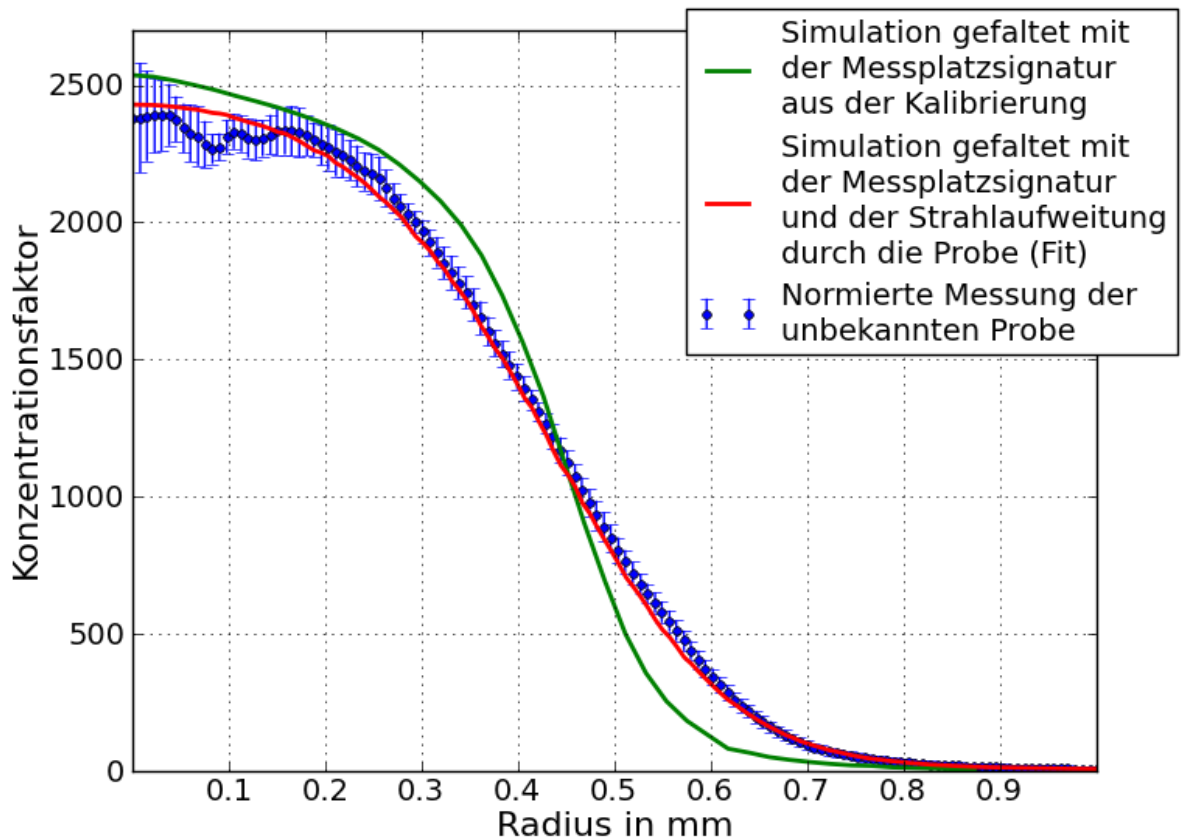

**Abbildung 66: Radialverteilungen der simulierten und mit der Messplatzsignatur aus der Kalibrierung gefalteten Konzentrationsmatrix (grün) und normierte gemessene Konzentrationsmatrix (blau) des Kalibrierspiegels sowie Fit der Strahlaufweitung durch die Formabweichungen der Probe (rot)**

Bei der hier beispielhaft untersuchten Probe kann laut Hersteller die Streuung gegenüber den Formabweichungen vernachlässigt werden. Daher wird die simulierte und mit der Messplatzsignatur gefaltete Konzentrationsmatrix mit einem Gauß gefaltet, wobei die Breite durch einen Fit an die gemessene Konzentrationsmatrix bestimmt wird. Das Ergebnis ist in Abbildung 66 zu sehen. Daraus ergibt sich der Wert $\sigma_{fit} = 0{,}108 \pm 0{,}002$mm, bzw. $\sigma_{winkel} = 0{,}53 \pm 0{,}01$mrad [Sch13]. Bei diesem Fit wurde als Fehler nur der Fehler auf die experimentelle Messung, analog zur Kalibrierung, und auf die Parameter der Faltungsfunktion der Messplatzsignatur berücksichtigt. Der Verlauf der Messplatzsignatur wurde als fehlerfrei behandelt.

Der Toleranzwinkel von HCPV[21]- Systemen beträgt typischerweise um die 10mrad [Sch13]. Es handelt dabei um die gemeinsame Toleranz für Montagefehler, Trackingfehler und Fehler der Optik. Die Oberflächenfehler dieser Probe beanspruchen somit ca. 5% des Toleranzwinkels für das Gesamtsystem.

Damit ist es anhand der provisorischen Methode der Messplatzsignatur gelungen die Strahlaufweitung aufgrund der Formabweichungen der Probe für eine Probe mit selben Einfallswinkelspektrum wie der Kalibrierspiegel zu vermessen.

---

[21] high concentrated photovoltaic

# 6. Zusammenfassung und Ausblick

Ziel dieser Arbeit war die Charakterisierung eines Messplatzes zur Vermessung der Konzentrationsmatrizen im Brennpunkt punktfokussierender reflektiver Optiken für die CPV. Dies ist bis auf einige Einschränkungen gelungen.

Dazu wurden die Einzelteile des Messplatzes charakterisiert. Hier wurden Eigenschaften der Kamera, des Objektivs, der Lichtquelle, der Spiegel, der Glasplatte und des Diffusors analysiert.

Die Charakterisierung der Kamera hat es erlaubt viele ihrer Einflüsse zu verstehen und zu korrigieren. Weiter ist es gelungen den Fehler auf einen von der Kamera gelieferten Grauwert inklusive theoretischer Grundlage zu bestimmen.

Aufgrund der starken Änderung (7,6%) der Quanteneffizienz des CCD-Chips über den Bereich der FWHM des Wellenlängenspektrums der Lichtquelle, wurden für die Spiegel die Wellenlängenabhängigkeiten der Reflektivität sowie für den Diffusor und die Glasplatte die Dispersion des Transmissionsgrades näher betrachtet. Letztendlich hat sich die Wellenlängenabhängigkeit des RMPs mit einem Einfluss von 0,2% als vernachlässigbar erwiesen.

Um sicher zu gehen, dass bis auf die Probe die Spiegel am RMP keine nennenswerte Strahlaufweitung verursachen, wurde auch diese vermessen. Wie erwartet konnte keine bedeutende Strahlaufweitung festgestellt werden.

Eine besonders große Bedeutung kommt dagegen der Strahlaufweitung durch den Diffusor zu. Hier wurde, mit einem von [Kis11] übernommenen Modell, die Einfallswinkelabhängigkeit der Strahlaufweitung analysiert. Eine Einfallswinkelabhängigkeit konnte festgestellt werden. Jedoch hat sich dieses Modell im Laufe der Kalibrierung als nicht ausreichend genau herausgestellt und wurde daher verfeinert. Aus Zeitgründen konnten die Messungen zur Einfallswinkelabhängigkeit der Strahlaufweitung nicht mit dem neuen Modell ausgewertet werden. Dies wird in den kommenden Monaten noch geschehen.

Aus technischen Gründen liefert die Lichtquelle noch eine örtlich stark inhomogene Beleuchtung. Dies hat zur Folge, dass derzeit die Effizienz der Proben noch nicht bestimmt werden kann und die Konzentrationsmatrix normiert werden muss. Hier wird abgewartet, ob dieses Problem durch den aktuellen Umbau der Lichtquelle behoben wird.

Trotz dieser zwei Probleme konnte im Rahmen der Kalibrierung des RMPs durch einen Fit eine empirische Messplatzsignatur ermittelt werden, anhand derer bei den Simulationen der Proben die Effekte des Messplatzes berücksichtigt werden können. So konnte beispielhaft die Brennpunktausschmierung aufgrund geometrischer Formabweichungen einer Probe bestimmt werden.

Die empirische Messplatzsignatur ist jedoch Einfallswinkelabhängig, sodass sie für jede Probe neu ermittelt werden muss oder die Proben nur beschnitten vermessen werden können. Daher stellt diese Methode keine langfristig praktikable Lösung dar.

Daher soll die Einfallswinkelabhängigkeit der Strahlaufweitung durch den Diffusor im Schichtenmodell bestimmt werden. Zusammen mit mehr Strahlungsleistung der Lichtquelle soll dadurch versucht werden die Konzentrationsmatrizen des Kalibrierspiegels korrekt zu simulieren.

Gelingt dies in den nächsten Monaten, kann der RMP einen Beitrag zum Verständnis von Oberflächenfehlern reflektiver Optiken für die CPV liefern und könnte somit langfristig zu einer Senkung der Stromgestehungskosten beitragen.

# Literaturverzeichnis

**[AFP13] AFP. 2013.** Amonix erzielt Weltrekord bei Test zum Wirkungsgrad von PV-Solarmodulen am NREL. [Online] 26. 04 2013. [Zitat vom: 20. 05 2013.] http://www.afp.com/de/profis/partners/business-wire/amonix-erzielt-weltrekord-bei-test-zum-wirkungsgrad-von-pv-solarmodulen-am-nrel.

**[Ant03] Antón, I., Pachón, D. und Sala, G. 2003.** Characterization of optical collectors for concentration photovoltaic applications. *Progress in Photovoltaics.* 2003, Bd. 11, pp.387-405.

**[Art65] Arthurs, A. M. 1967.** *Probability Theory.* Michigan : Routledge & K. Paul, 1967. ISBN 0-7100-4359-7.

**[AZU12] AZURSPACE. 2012.** CPV Solar Cells - Azurspace Power Solar GmbH. [Online] 30. 03 2012. [Zitat vom: 06. 02 2013.] http://www.azurspace.com/index.php/en/products/products-cpv/cpv-solar-cells.

**[Bah00] Bahnmüller, Jochen. 2000.** *Dissertation - Charakterisierung gepulster Laserstrahlung zur Qualitätssteigerung beim Laserbohren.* Universität Stuttgart : Herbert Utz Verlag, 2000. ISBN 3-89675-851-9.

**[Bar75] Barbe, David F. 1975.** Imaging Devices Using the Charge-Coupled Concept. *Proceedings of the IEEE.* 1975. Bde. 63, pp.38-67, No.1.

**[Bey12] Beyerer, Jürgen, Puente León, Fernando und Frese, Christian. 2012.** *Automatische Sichtprüfung: Grundlagen, Methoden und Praxis der Bildgewinnung und Bildauswertung.* s.l. : Springer, 2012. ISBN 978-3-642-23965-6.

**[Bra08] Bradski, Gary und Kaehler, Adrian. 2008.** *Learning OpenCV - Computer Vision with the OpenCV Library.* Sebastopol : O'Reilly, 2008. ISBN 978-0-596-51613-0.

**[Dat] Datenblatt des Herstellers.** Aus Vertraulichkeitsgründen kann leider keine genauere Angabe gemacht werden.

**[Dem09] Demtröder, Wolfgang. 2009.** *Experimentalphysik 2 - Elektrizität und Optik.* Leipzig : Springer, 2009. ISBN 978-3-540-68219-6.

**[ISE12] Fraunhofer ISE. 2012.** Photovoltaics report. [Online] 11. 12 2012. [Zitat vom: 20. 05 2013.] http://www.ise.fraunhofer.de/en/downloads-englisch/pdf-files-englisch/photovoltaics-report.pdf.

**[Fra13] Fraunhofer IWES & Partner.** Kombikraftwerk 2 - Das regenerative Kombikraftwerk. [Online] [Zitat vom: 21. 05 2013.] http://www.kombikraftwerk.de/.

**[Ghe11] Gheţa, Ioana. 2011.** *Dissertation - Fusion multivariater Bildserien am Beispiel eines Kamera-Arrays.* Karlsruher Institut für Technologie : KIT Scientific Publishing, 2011. ISBN 978-3-86644-684-7.

**[Gre03] Green, Martin A., et al. 2003.** Solar Cell Efficiency Tables (Version 21). *Progress in Photovoltaics.* 2003, Bd. 11, pp. 39-45.

**[Gre12] Green, Martin A., et al. 2012.** Solar Cell Efficiency Tables (version 41). *Progress in Photovoltaics.* 2012, Bd. 21, pp.1-11.

**[Gro05] Gross, Herbert. 2005.** *Handbook of Optical Systems - Fundamentals of Technical Optics.* s.l. : WILEY-VCH, 2005. ISBN 3-527-40377-9.

**[Haf03] Haferkorn, Heinz. 2003.** *Optik - Physikalisch-technische Grundlagen und Anwendungen.* Reinheim : Wiley-VCH, 2003. ISBN 3-527-40372-8.

**[Hil03] Hild, Stefan. 2003.** *Diplomarbeit - Thermisch durchstimmbares Signal-Recycling für den Gravitationswellendetektor GEO600.* Universität Hannover : Max-Planck-Institut für Gravitationsphysik, 2003.

**[Hop04] Hopkinson, Gordon R., Goodman, Teresa M. und Prince, Stuart R. 2004.** *A Guide to the Use and Calibration of Detector Array Equipment.* Washington : SPIE - The International Society for Optical Engineering, 2004. ISBN 0-8194-5532-6.

**[alH10] Hornung, T., et al. 2010.** Temperature and wavelength dependent measurement and simulation of Fresnel lenses for concentrating photovoltaics. [Buchverf.] AIP Conference Proceedings. *6th International Conference on Concentrating Photovoltaic Systems: CPV-6.* 2010.

**[Hor13] Hornung, Thorsten. 2013.** *Dissertation - Ein- und mehrstufige optische Konzentratoren für photovoltaische Anwendungen.* Universität Freiburg : s.n., 2013.

**[IPC07] IPCC. 2007.** *Climate Change 2007 - Synthesis Report.* Geneva : WMO - UNEP, 2007.

**[IPC95] —. 1995.** *Second Assessment - Climate Change 1995 - A Report of the intergovernmental panel on climate change.* Geneva : WMO - UNEP, 1995.

**[Jan01] Janesick, James R. 2001.** *Scientific charge-coupled devices.* Washington : SPIE - The international Society for Optical Engineering, 2001. ISBN 0-8194-3698-4.

**[Kir12] Kirshner, Hagai und Sage, Daniel. 2012.** BIG - PSF Generator. [Online] Biomedical Imaging Group at École Polytechnique Fédérale de Lausanne, 27. November 2012. [Zitat vom: 04. März 2013.] http://bigwww.epfl.ch/algorithms/psfgenerator/.

**[Kis11] Kiss, Michael. 2011.** Masterarbeit - Weiterentwicklung eines Sekundärkonzentratorenmessplatz für die konzentrierende Photovoltaik. *Arbeit bis spätestens 08.06.2016 unter Sperrvermerk.* Fachhochschule Vorarlberg : s.n., 2011.

**[Kar04] Kraus, Karl. 2004.** *Photogrammetrie, Band 1: Geometrische Informationen aus Photographien und Laserscanneraufnahmen.* Göttingen : Walter de Gruyter, 2004. ISBN 3-11-017708-0.

**[Küh07] Kühlke, Dietrich. 2007.** *Optik - Grundlagen und Anwendungen.* Frankfurt am Main : Verlag Harri Deutsch, 2007. ISBN 978-3-8171-1741-3.

**[LiQ08] Li, Qinghe. 2008.** Masterthesis - Light Scattering of Semitransparent Media. Georgia Institute of Technology : s.n., 2008.

**[McE] McEvoy, Augustin, Markvart, Tom und Castañer, Luis. 2011.** *Practical Handbook of Photovoltaics: Fundamentals and Applications.* Oxford : Academic Press / Elsevier, 2011. ISBN 978-0-12-385934-1.

**[NAO13] NAOJ.** Subaru Telescope - Reflectivity of the Primary Mirror. [Online] National Astronomical Observatory of Japan. [Zitat vom: 18. 04 2013.] http://www.naoj.org/Observing/Telescope/Parameters/Reflectivity/.

**[NREL] NREL / Scanlon, Bill. 2012.** NREL: News Feature - Award-Winning PV Cell Pushes Efficiency Higher. [Online] 28. 12 2012. [Zitat vom: 05. 02 2013.] http://www.nrel.gov/news/features/feature_detail.cfm/feature_id=2055.

**[NRE13] NREL.** Glossary of Solar Radiation Resource Terms. [Online] [Zitat vom: 20. 05 2013.] http://rredc.nrel.gov/solar/glossary/gloss_g.html.

**[alS96] Ochi, S. 1996.** Charge-Coupled Device Technology. [Buchverf.] T. Lizuka, et al. *Japanese Technology Reviews Vol.30, Section A: Electronics.* s.l. : Gordon and Breach Publishers, 1996.

**[Ped08] Pedrotti, Frank, et al. 2008.** *Optik für Ingenieure.* Leipzig : Springer, 2008. ISBN 978-3-540-73471-0.

**[Sch13] Schmid, Tobias, et al. 2013.** Indoor Charakterization of Reflective Concentrator Optics. [Buchverf.] AIP Conference Proceedings. *9th International Conference on Concentrating Photovoltaic Systems: CPV-9. Veröffentlich voraussichtlich Sommer* 2013.

**[Sol13] SolarJunction.** Solar Junction » Products. [Online] [Zitat vom: 06. 02 2013.] http://www.sj-solar.com/products/.

**[sol13] SolarServer. 2013.** Basiswissen | Photovoltaik: Solarstrom und Solarzellen in Theorie und Praxis - SolarServer. [Online] 2013. [Zitat vom: 06. 02 2013.] http://www.solarserver.de/wissen/basiswissen/photovoltaik.html.

**[Squ01] Squires, Gordon L. 2001.** *Practical Physics.* Cambridge : Cambridge University Press, 2001. ISBN 0-521-77045-9.

**[Til05] Tille, Thomas und Schmitt-Landsiedel, Doris. 2005.** *Mikroelektronik - Halbleiterbauelemente und deren Anwendung in elektronischen Schaltungen.* s.l. : Springer, 2005. ISBN 3-540-20422-9.

**[Wen80] Wen, L., Poon, P. und Carley, W. 1980.** Comparative study of solar optics for paraboloidal concentrators. *Journal of Solar Energy Engineering.* 1980, Bd. 102, pp. 305-315.

# Danksagungen

# Indoor Characterization of Reflective Concentrator Optics

Tobias Schmid, Manuel Frick, Thorsten Hornung and Peter Nitz

*Fraunhofer Institute for Solar Energy Systems ISE, Heidenhofstrasse 2, 79110 Freiburg, Germany*

**Abstract:** We report about the indoor characterization of small point focusing mirrors at Fraunhofer ISE. The goal is to determine the mean slope error of the concentrator. This is achieved by measuring the concentration distribution in the focal plane of such a mirror. We modified and expanded a test site which is used for Fresnel lens characterization [1]. A modified version of the method presented in [2] is employed to measure the concentration distribution. By comparing ray tracing simulation results of the ideal mirror to the measurement, the mean slope error can be deduced.

**Keywords:** point focusing mirror, parabolic mirror, characterization of concentrator optics, optical error
**PACS:** 88.40.FF, 88.40.FC, 88.40.FR, 42.79.EK

## INTRODUCTION

Concentrating mirrors as primary optical elements do not suffer from chromatic aberration and are less prone to temperature effects as it is the case for Fresnel lenses. Thus e.g. a perfect parabolic mirror can achieve a concentration ratio of about 11 000 [3]. For these reasons it is worth studying mirrors as an alternative to Fresnel lenses which currently dominate the HCPV market. On the other hand side any imperfection on a mirror surface that causes the surface normal to deviate by an angle $\theta$ will lead to deviation of the reflected ray by $2\theta$. This is in contrast to refractive optics where surface normal deviations are less severe. This fact has a strong impact on the performance of concentrating mirrors and is thus investigated in this paper.

In this paper we present a measurement method to determine the concentration profile of a point focusing mirror. The method is a modified version of the one presented by Antón et. al. [2]. We exemplify the method by measuring an industrially manufactured sample. From the measurement we calculate the mean slope error of the concentrator. Furthermore we discuss the calibration procedure for the experimental setup.

## MEAN SLOPE ERROR

Real mirrors are subject to various imperfections which may be described by the help of statistics, see [3] for details.

In this paper we focus on the effect of the surface slope deviation. A surface slope deviation is the deviation of the surface normal vector of the real mirror from the designed one at a specific point on the mirror surface (measured e.g. in mrad). It is assumed that the distribution of these errors over the entire mirror surface is rotationally symmetric around the designed surface normal. The surface slope deviations are therefore described by a distribution $E(\theta)$ where $\theta$ is the angle between the designed normal vector and the real one. The probability that the direction of a reflected ray, which falls into the solid angle $d\Omega$, deviates from the designed direction by $2\theta$ is therefore proportional to $E(\theta)d\Omega$. We assume $E(\theta)$ to be Gaussian distributed with zero mean:

$$E(\theta) \sim \exp\left\{-\frac{1}{2}\left(\frac{\theta}{\sigma_{slope}}\right)^2\right\}.$$

The quantity we will determine in this paper is the mean slope error $\sigma_{slope}$.

## EXPERIMENTAL SETUP

The mean slope error may be measured by various techniques, see e.g. [4]. With our setup we measure the concentration distribution in the solar cell plane. Thus, in principle, we can simultaneously determine efficiency of the mirror and the beam deviations.

The measurement configuration is shown in FIGURE 1. Light of solar-like angular divergence hits the mirror and is reflected to the focal plane. A small transmissive diffuser is placed in the focal plane from which a picture is taken of the spot. The mirrors under test have dimensions of the order of 10 cm, therefore any camera positioned behind the focal plane, as proposed in [2], would shade relevant parts of the mirror. Thus we chose a different route and placed a small tilted mirror next to the diffuser. The rays originating from the diffuser are redirected outside the optical path and the diffuser can be imaged with an objective lens. Therefore we are able to obtain a picture of the spot without any shading due to the camera. The shading due to the holder for the tilted mirror and the diffuser does not exceed the typical value of the shading in a real concentrator module due to the solar cell and its cooling or heat spreader.

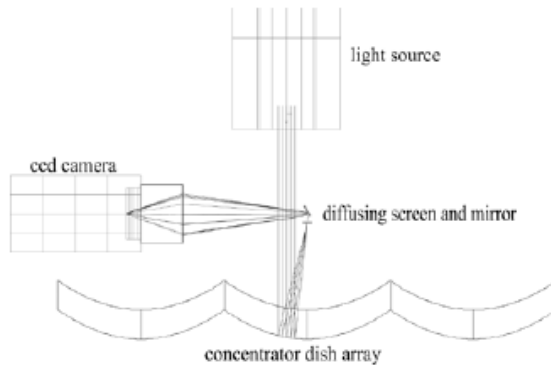

**FIGURE 1.** Schematic configuration of the measurement setup used for characterization of point focusing mirrors.

The characteristics of the setup are the following:

- The light source (LED at wavelength $\lambda$=629 nm) as well as the sample holding stage is described in [1].
- The tilted mirror is placed right above the diffuser. Attention has to be paid to the right size and position of the tilted mirror relative to the diffuser in order not to distort the measurement.
- The common holder for diffuser and tilted mirror is shown in FIGURE 2.
- The objective lens of the camera has to fulfill a range of requirements, the most important being working distance, zoom range and field of view.
- The camera itself is highly linear over at least two orders of magnitude.

A picture of the setup is shown in FIGURE 3.

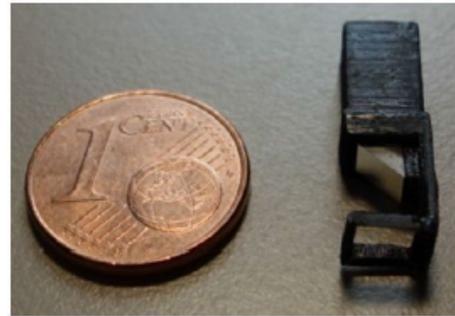

**FIGURE 2.** Holder of the tilted mirror and the diffusing screen (not included in the picture) compared to the size of a one euro cent coin.

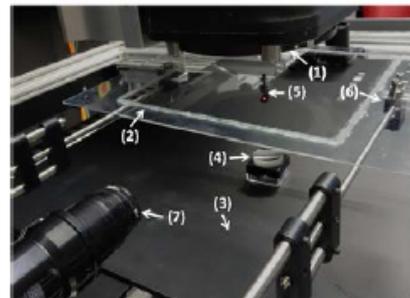

**FIGURE 3:** (1) light source, (2) glass plate, (3) sample holder, (4) sample, (5) diffuser and tilted mirror, (6) adjustment screws, (7) objective lens.

## CALIBRATION

The diffuser is a commercially available PTFE film with thickness $D$=0.116 mm. Its first important property taken into account is the angular dependent transmission in normal direction: The larger the angle of incidence the less the transmission in the direction of the objective lens is. This behavior has been measured at Fraunhofer ISE with a photo goniometer.

The second one is the ray broadening which leads to a distortion of the measurement signal. Any ray bundle hitting the diffuser at the front illuminates an area at the back larger than the cross section of the original ray bundle, see FIGURE 4.

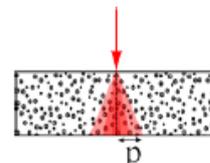

**FIGURE 4:** Schematic representation of the ray broadening in the volume diffuser.

As little is known about the scattering properties of this material, see e.g. [5], we developed a phenomenological model to describe its behavior. The model describes the brightness of any point $p$ on the back of the diffuser depending on the position of the incoming ray.

In the model the diffuser is split into $n$ layers parallel to the boundaries. At each boundary the incident wave is scattered and propagates freely as a spherical wave until it reaches the next boundary. There it is scattered again until it reaches the back of the diffuser. The idea is schematically illustrated in FIGURE 5.

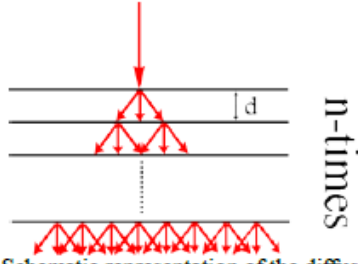

**FIGURE 5:** Schematic representation of the diffuser model.

Mathematically this is modeled by the convolution operator:

$$H = \overset{n}{\underset{i=1}{\otimes}} \frac{d_i^4}{\left(d_i^2 + r^2\right)^2},$$

(the symbol $\overset{n}{\otimes}$ denotes n-fold convolution) where $d_i$ is the thickness of the $i$-th layer and $r$ the lateral distance from the observation point $p$. As the average number of scattering events does not need to be an integer we allow the number of layers to be described by a real positive number $\tilde{n}$. For any number $\tilde{n}$ the $d_i$ are given by $d_i = D/\tilde{n}$, for $0 < i < n-1$ where $(n-1)$ is the integer part of $\tilde{n}$ and $D$ the physical thickness of the diffuser. The thickness of the last layer is given by $d_n = D - (n-1)d_i$.

We simulated and measured a well-known spherical mirror with a high contour accuracy. We fitted the radial distribution obtained by the optical simulation to the measured one by convoluting the concentration matrix of the simulation with $H$. The result is shown in FIGURE 6. The measured curve results from averaging over three independent measurements. The error bars were estimated by taking into account the standard deviation of the three measurements and an estimate of the systematic errors involved. As free parameter for the fit we chose $\tilde{n}$ $(= 2.4 \pm 0.1)$ leading to a fair agreement between the measured and fitted curve.

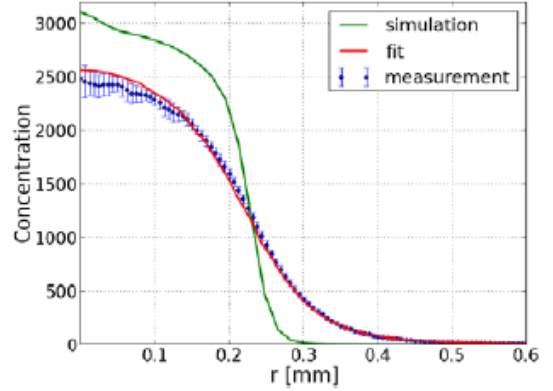

**FIGURE 6.** Resulting radial distributions for the calibration procedure.

## MEASUREMENT PROCEDURE AND ANALYSIS

To demonstrate how the mean slope error can be calculated we measured a parabolic mirror of focal length f=100.1 mm and circular entry aperture of about 5 cm diameter. The mirror was bigger than this but to ensure to have the same incidence angles on the diffuser as in the calibration procedure we shaded parts of it. The sample was an aluminum coated mirror with a protective coating. The substrate was polycarbonate manufactured by injection molding. We simulated the parabolic mirror by treating it as ideal. We then convoluted the signal with the operator $H$ using the value for $\tilde{n}$ found in the calibration. The resulting concentration matrix was further convoluted with a 2-D Gaussian function with zero mean and rotational symmetry using its standard deviation $\sigma_{fit}$ as fit parameter (the dimension of $\sigma_{fit}$ being a length). The result is shown in FIGURE 7. The shown curve of the measured data results from averaging over three independent measurements. The error bars were calculated by taking into account the standard deviation of the three measurements, an estimate of the systematic errors and the uncertainty of $\tilde{n}$ coming from the calibration. A good agreement between the fit and the measurement data can be observed.

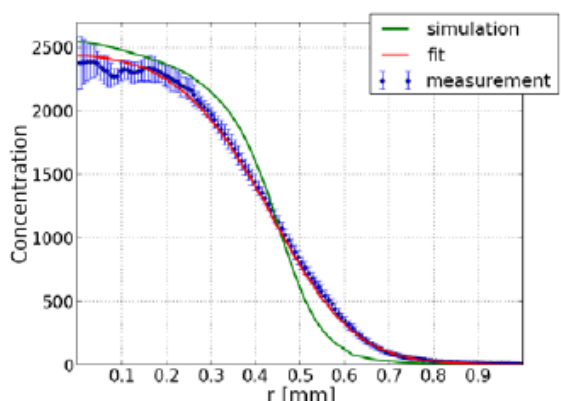

**FIGURE 7.** Measured, simulated (convoluted with *H*) and fitted radial distributions for the parabolic sample.

The mean slope error (dimension: angle) is obtained by

$$\sigma_{slope} = \arctan\left(\frac{\sigma_{fit} \cdot \beta}{f}\right) \cdot \frac{1}{2} = 0.53 \pm 0.01\,\text{mrad}.$$

The factor of 1/2 results from Snell's law of reflection. $\beta$ (=0.986) is a weighting factor according to [6] and depends on the rim angle of the mirror. It results from the fact that rays coming from the rim of the mirror produce a larger spot than those coming from the central part.

## DISCUSSION

The mean slope error was determined to be 0.53 mrad, corresponding to a good optical quality of the sample. For the parabolic mirror this simply means that the sun's image is smeared out and thus the acceptance angle is lowered by about 0.53 mrad. Typical acceptance angles for HCPV systems are of the order 10 mrad. If a secondary optical element or any nonimaging primary optics with this mean optical error is used the situation may change but either the acceptance angle or the concentration ratio will be lower than for the error free mirror regardless of the optical design.

The experimental setup is not limited to parabolic mirrors but the mean slope error for any point focusing mirror can be determined in a similar way. Other slope error models than Gaussian can be used if required.

The overall optical efficiency of the mirror can be determined by integrating the concentration distribution but this needs a further calibration of the setup and is therefore subject to future work.

The entire beam spread (i.e. slope error plus surface scattering) may be wavelength dependent due to surface roughness. Considering that commercially available injection molded optics usually have a rms surface roughness of about 5 nm [7] we expect the slope error to be the dominant effect. Even if this is not the case one can treat the determined value for the mean slope error as upper boundary.

## CONCLUSION

An experimental setup for measuring the concentration distribution in the solar cell plane of a point focusing mirror was developed, assembled and calibrated. An injection molded mirror was measured and the mean slope error was determined to be 0.53 mrad. The main effect of this error is the reduction of the acceptance angle of the CPV system.

## ACKNOWLEDGMENTS

The work presented here was supported by the Federal Ministry for the Environment, Nature Conservation and Nuclear Safety (BMU) under the KomGen project, contract number 0327567A. We thank BECAR s.r.l. (Beghelli Group) for the preparation of the sample used in the research presented here.

## REFERENCES

1. P. Nitz, A. Heller and W.J. Platzer, "Indoor Characterisation of Fresnel Type Concentrator Lenses", *Proceedings of the 4th International Conference on Solar Concentrators for the Generation of Electricity or Hydrogen, San Lorenzo de el Escorial*, 2007, pp. 289-292.
2. I. Antón, D. Pachón and G. Sala, "Characterization of Optical Collectors for Concentration Photovoltaic Applications", *Prog. Photovolt: Res. Appl.* **11**, 387-405 (2003).
3. V.A. Grilikhes, "Transfer and Distribution of Radiant Energy in Concentrating Systems", in *Photovoltaic Conversion of Concentrated Sunlight*, edited by V.M. Andreev, V.A. Grilikhes and V.D. Rumyantsev, Chichester: John Wiley & Sons Ltd., 1997.
4. P. Bendt, H. W. Gaul and A. Rabl, "Determinig the Optical Quality of Focusing Collectors without laser Raytracing", *Technical Report*, Solar Energy Research Institute, Golden, Colorado, 1980.
5. Q. Li, "Light Scattering of Semitransparent Media", Master Thesis, Georgia Institute of Technology, 2008.
6. L. Wen, L. Huang, P. Poon and W. Carley, "Comparative Study of Solar Optics for Paraboloidal Concentrators", *Journal of Solar Energy Engineering* **102**, 305-315 (1980).
7. M. Pfeffer, "Optomechanics of Plastic Optical Components", in *Handbook of Plastic Optics*, edited by S. Bäumer, Weinheim: WILEY-VCH, 2010.

# ANHANG B – Abweichung der Spiegel von der Spekularität

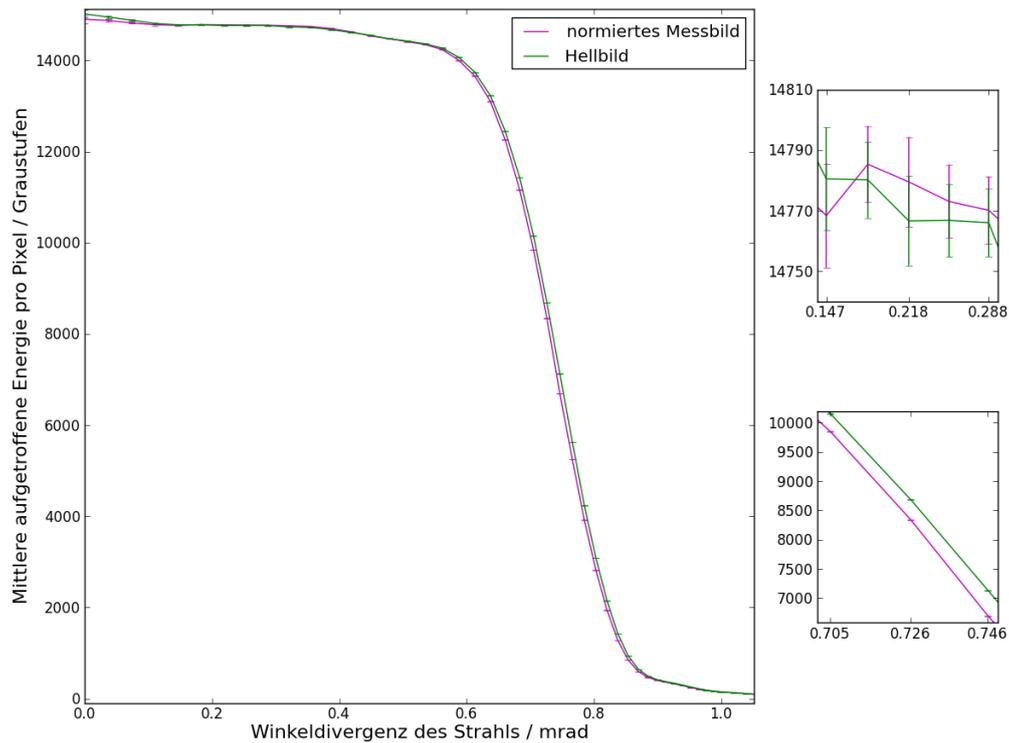

**Abbildung 67:** Radiale Darstellung des Winkelspektrums der reflektierten Strahlen an der zweiten Position des Hellbildspiegels, beim Hellbild aufgrund der Divergenz der Beleuchtung, beim Messbild zusätzlich aufgrund der Strahlaufweitung. Die Fehlerbalken resultieren aus dem statistischen Rauschen der Kamera. Links: Gesamtansicht. Rechts: Nahansichten

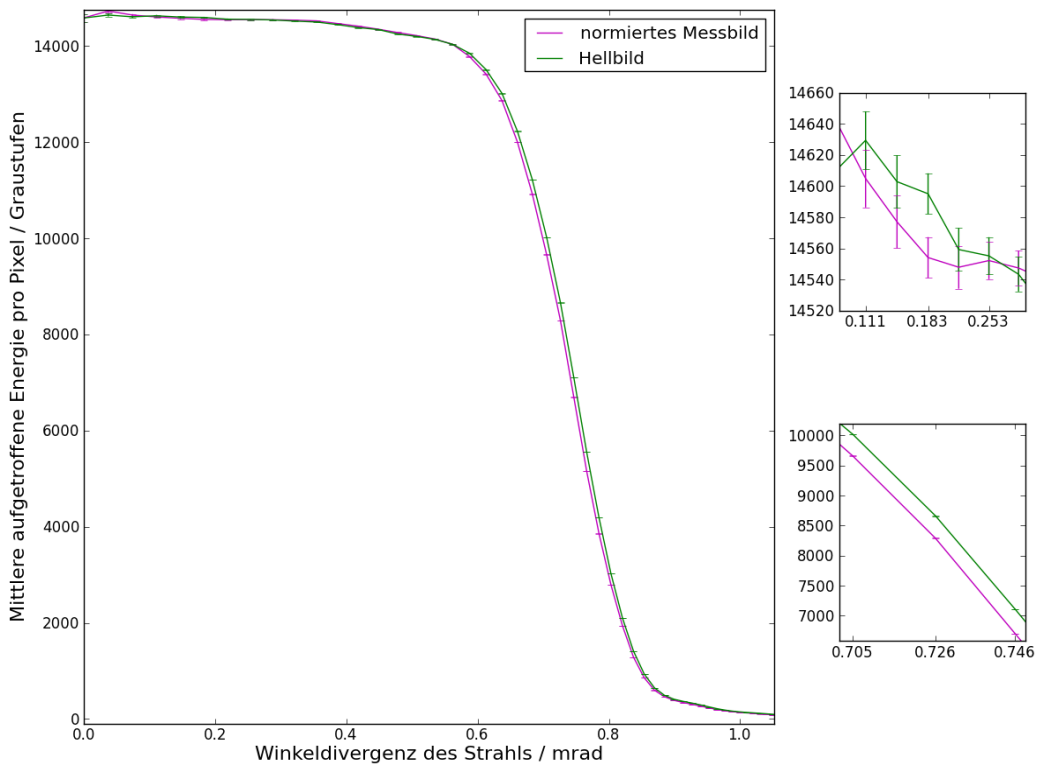

**Abbildung 68:** Radiale Darstellung des Winkelspektrums der reflektierten Strahlen an der dritten Position des Hellbildspiegels bei der ersten Aufnahme, beim Hellbild aufgrund der Divergenz der Beleuchtung, beim Messbild zusätzlich aufgrund der Strahlaufweitung. Die Fehlerbalken resultieren aus dem statistischen Rauschen der Kamera. Links: Gesamtansicht. Rechts: Nahansichten

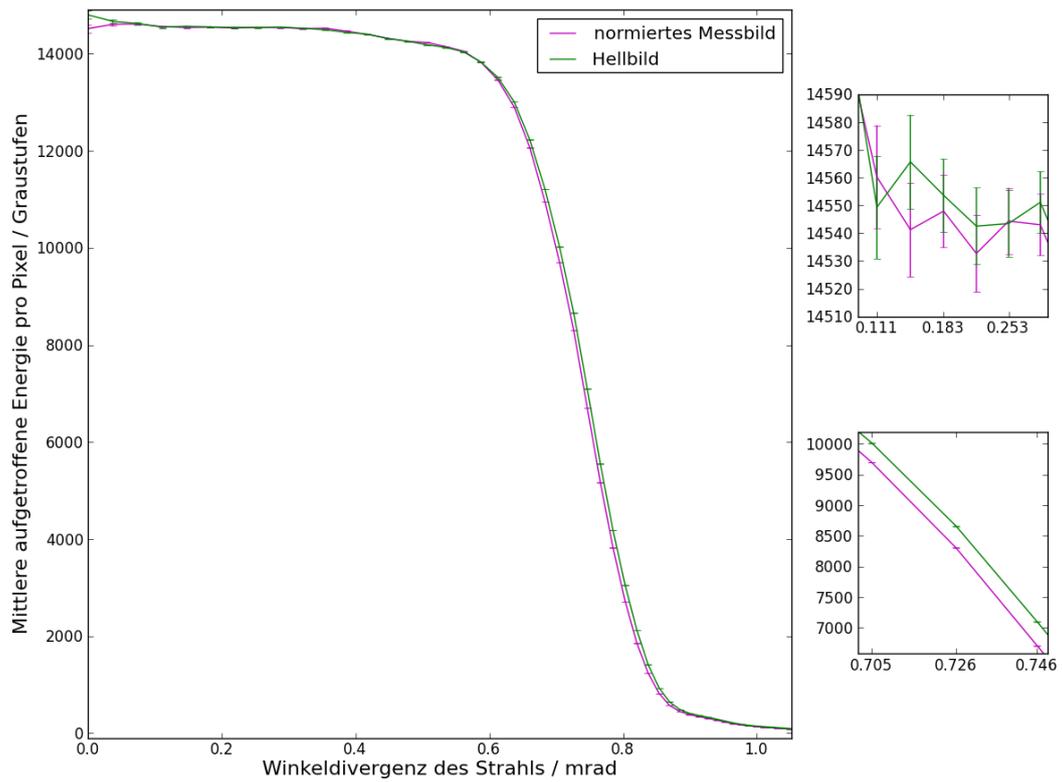

**Abbildung 69: Radiale Darstellung des Winkelspektrums der reflektierten Strahlen an der dritten Position des Hellbildspiegels bei der zweiten Aufnahme, beim Hellbild aufgrund der Divergenz der Beleuchtung, beim Messbild zusätzlich aufgrund der Strahlaufweitung. Die Fehlerbalken resultieren aus dem statistischen Rauschen der Kamera. Links: Gesamtansicht. Rechts: Nahansichten**

# ANHANG C – Fits zur Strahlaufweitung des Diffusors

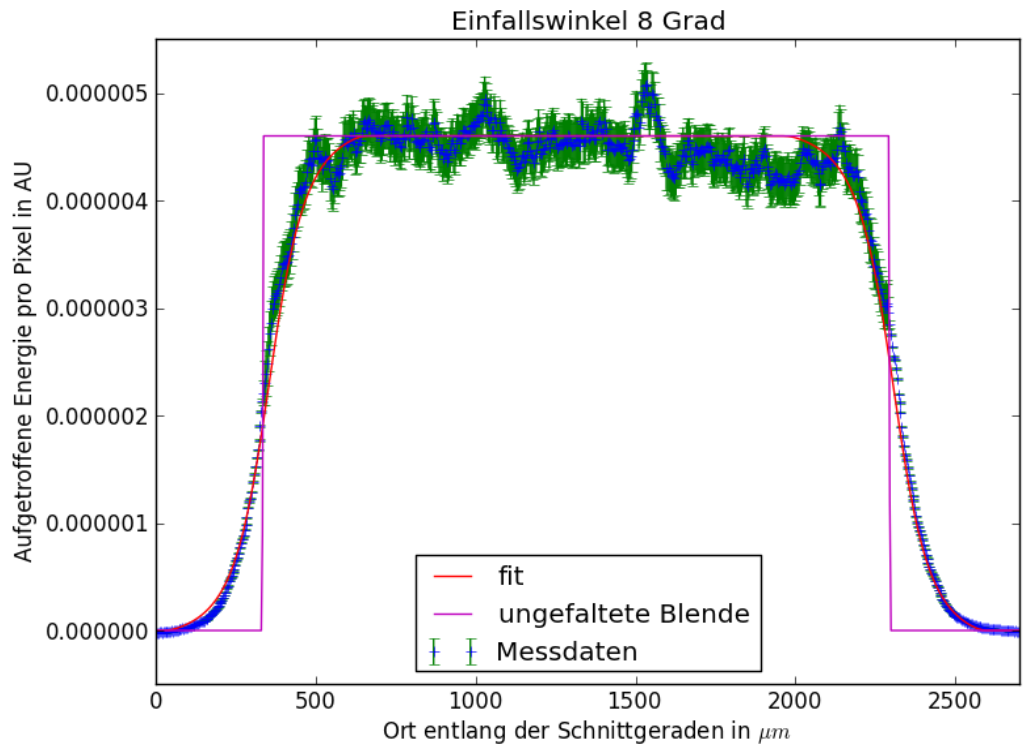

**Abbildung 70:** Schnitt durch die Mitte des zweidimensionalen Fit der Diffusorstrahlaufweitung bei einem Beleuchtungswinkel von 8°. In blau die Messpunkte, grün die zugehörigen Fehlerbalken, magenta das berechnete Eingangssignal und rot die Fitkurve.

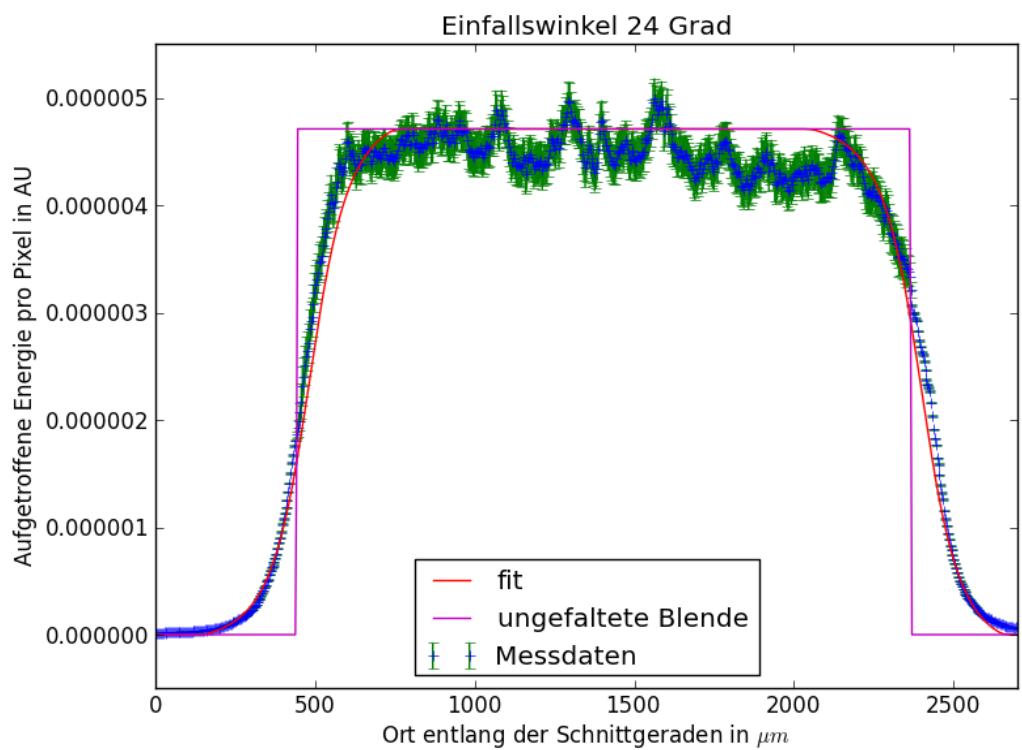

**Abbildung 71:** Schnitt durch die Mitte des zweidimensionalen Fit der Diffusorstrahlaufweitung bei einem Beleuchtungswinkel von 24°. In blau die Messpunkte, grün die zugehörigen Fehlerbalken, magenta das berechnete Eingangssignal und rot die Fitkurve.

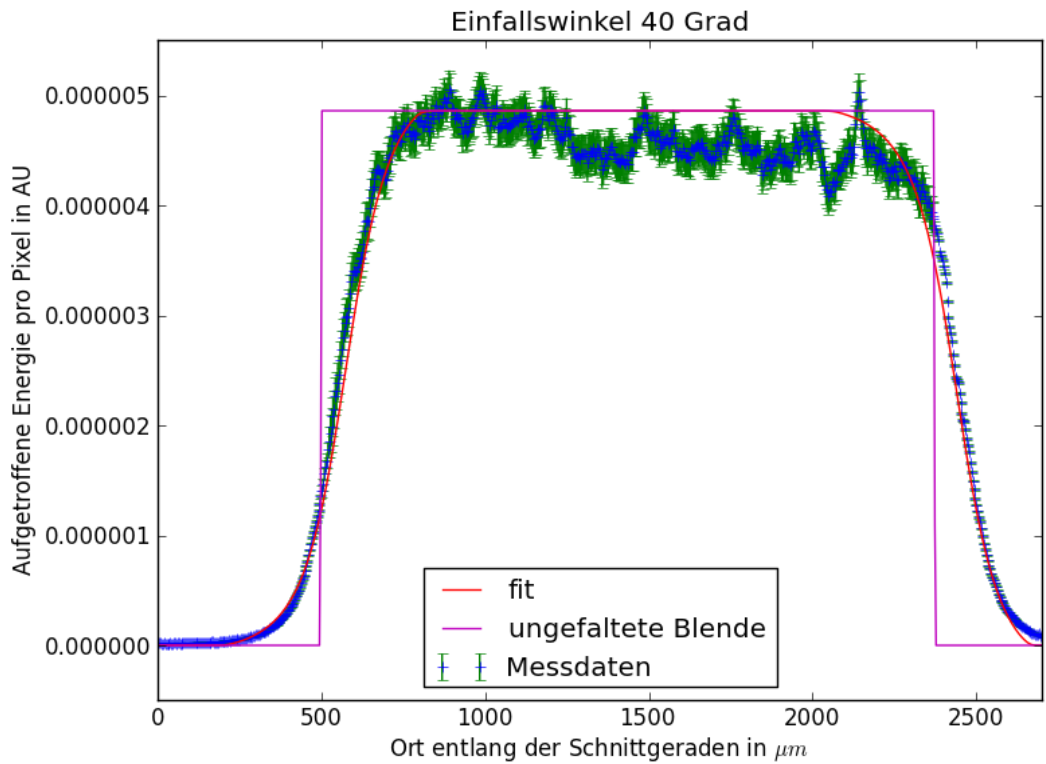

**Abbildung 72:** Schnitt durch die Mitte des zweidimensionalen Fit der Diffusorstrahlaufweitung bei einem Beleuchtungswinkel von 40°. In blau die Messpunkte, grün die zugehörigen Fehlerbalken, magenta das berechnete Eingangssignal und rot die Fitkurve.

Hiermit erkläre ich, dass diese Arbeit von mir selbstständig verfasst und keine anderen als die angegebenen Quellen und Hilfsmittel verwendet wurden.

<div style="text-align: right">Freiburg, den 31.05.2013</div>